\documentclass[twoside,twocolumn]{article}
\pdfoutput=1

\usepackage[hyphens,spaces,obeyspaces]{url}
\usepackage[colorlinks,allcolors=blue]{hyperref}
\usepackage[sort,numbers,compress]{natbib}
\usepackage[T1]{fontenc} 
\usepackage[english]{babel} 

\usepackage[hmarginratio=1:1,top=32mm,columnsep=20pt,left=16mm]{geometry} 

\usepackage{booktabs} 
\usepackage{abstract}

\usepackage{titlesec} 
\usepackage{fancyhdr} 
\usepackage{titling} 
\usepackage{graphicx}
\usepackage{amsmath}
\usepackage{amssymb}
\usepackage[caption=false]{subfig}
\usepackage{xspace}
\usepackage{slashed}
\usepackage{verbatim}
\usepackage{bbm}
\usepackage{amsfonts}
\usepackage{import}
\usepackage{booktabs}
\usepackage{lineno}
\usepackage{multirow}
\usepackage{authblk}
\usepackage[svgnames]{xcolor}

\usepackage{tabularx}
\newcolumntype{L}[1]{>{\raggedright\arraybackslash}p{#1}}
\newcolumntype{C}[1]{>{\centering\arraybackslash}p{#1}}
\newcolumntype{R}[1]{>{\raggedleft\arraybackslash}p{#1}}

\AtBeginDocument{%
  \hypersetup{
    citecolor=blue,
    linkcolor=blue,   
    urlcolor=blue
    }
  }

\pretitle{\begin{center}\Huge\bfseries} 
\posttitle{\end{center}}

\newcommand{\runone}{Run~1\xspace}
\newcommand{\runtwo}{Run~2\xspace}

\newcommand{\fcut}{\ensuremath{f_{\mathrm{cut}}}\xspace}
\newcommand{\zsoftdrop}{\ensuremath{z_{\mathrm{cut}}}\xspace}
\newcommand{\rsub}{\ensuremath{R_{\mathrm{sub}}}\xspace}
\newcommand{\zprune}{\ensuremath{z_{\mathrm{p}}}\xspace}
\newcommand{\dcut}{\ensuremath{d_{\mathrm{p}}}\xspace}

\newcommand{\MeV}{\ensuremath{\,\text{Me\hspace{-.08em}V}}\xspace}
\newcommand{\GeV}{\ensuremath{\,\text{Ge\hspace{-.08em}V}}\xspace}

\newcommand{\TeV}{\ensuremath{\,\text{Te\hspace{-.08em}V}}\xspace}

\newcommand{\fbinv} {\mbox{\ensuremath{\,\text{fb}^\text{$-$1}}}\xspace}

\newcommand{\pt}{\ensuremath{p_{\mathrm{T}}}\xspace}
\newcommand{\kt}{\ensuremath{k_{\mathrm{T}}}\xspace}

\newcommand{\PTm}{\ensuremath{{p}_\mathrm{T}\hspace{-1.02em}/\kern 0.5em}\xspace}

\newcommand{\ETslash}{\ensuremath{E_{\mathrm{T}}\hspace{-1.1em}/\kern0.45em}\xspace}

\newcommand{\ptvecmiss}{\ensuremath{{\vec p}_{\mathrm{T}}^{\kern1pt\text{miss}}}\xspace}
\newcommand{\toptau}{\ensuremath{\tau_{32}}\xspace}

\newcommand{\Mjet}{\ensuremath{M_{\mathrm{jet}}}\xspace}

\newcommand{\ttbar}{\ensuremath{t\overline{t}}\xspace} 
\newcommand{\bbbar}{\ensuremath{b\overline{b}}\xspace} 
\newcommand{\ttH}{\ensuremath{t\overline{t}H}\xspace} 

\newcommand{\antikt}       {anti-\kt}
\newcommand{\antiktfour}    {anti-\kt, \ensuremath{R=0.4}\xspace}
\newcommand{\antiktten}   {anti-\kt, \ensuremath{R=1.0}\xspace}
\newcommand{\antikteight}   {anti-\kt, \ensuremath{R=0.8}\xspace}

\title{Jet Substructure at the Large Hadron Collider: Experimental Review}

\author[a]{Roman Kogler (ed.)\thanks{roman.kogler@uni-hamburg.de}}
\affil[a]{Universit\"at Hamburg, Germany}

\author[b]{Benjamin Nachman (ed.)\thanks{bpnachman@lbl.gov}}
\affil[b]{Lawrence Berkeley National Laboratory, USA}

\author[c]{Alexander Schmidt (ed.)\thanks{alexander.schmidt@physik.rwth-aachen.de}}
\affil[c]{RWTH Aachen University, Germany}

\author[d]{Lily Asquith}
\affil[d]{University of Sussex, UK}

\author[e]{Mario Campanelli}
\affil[e]{University College London, UK}

\author[f]{Chris Delitzsch}
\affil[f]{University of Arizona, USA}

\author[e]{Philip Harris}
\affil[e]{Massachusetts Institute of Technology, USA}

\author[a]{Andreas Hinzmann}

\author[f]{Deepak Kar}
\affil[f]{University of Witwatersrand, South Africa}

\author[g]{Christine McLean}
\affil[g]{University of California, Davis, USA}
 
\author[g]{Justin Pilot}

\author[h]{Yuta Takahashi}
\affil[h]{Universit\"at Z\"urich, Switzerland}

\author[k]{Nhan Tran}
\affil[k]{Fermilab, USA}
\author[k]{Caterina Vernieri}

\author[l]{Marcel Vos}
\affil[l]{IFIC Valencia, Spain}

\author[d]{Emma Winkels}

\date{\today} 
\renewcommand{\maketitlehookd}{%
\begin{abstract}
\noindent 
Jet substructure has emerged to play a central role at the Large Hadron Collider, where it has provided numerous innovative ways to search for new physics and to probe the Standard Model, particularly in extreme regions of phase space. In this article we focus on a review of the development and use of state-of-the-art jet substructure techniques by the ATLAS and CMS experiments. 
\end{abstract}
}

\begin{document}

\maketitle 
 
\tableofcontents{}

\section{Introduction}\label{sec:introduction}

Jets are collimated sprays of particles, produced in abundance in high 
energy particle collisions. They are ubiquitous in particle collider experiments and indespensible to study the underlying 
dynamics and interactions. 
Jets have played a central role in the discovery and property 
measurements of many fundamental particles like the 
gluon ($g$)~\cite{Bartel:1979ut,Berger:1979cj,Barber:1979yr,Brandelik:1979bd} and the 
top quark ($t$)~\cite{Abachi:1995iq, Abe:1995hr}. 
They have provided key insights into the structure of the strong force 
and were indispensable in the study of Higgs boson ($H$) couplings 
to heavy third generation 
quarks~\cite{Aaboud:2018urx, Sirunyan:2018hoz, Aaboud:2018zhk, Sirunyan:2018kst}.
Because of their large production rate at the LHC, 
jets feature prominently in searches for new particles and 
precision measurements of Standard Model (SM) properties. 
However, important information on the underlying particle dynamics 
is not only carried by the total four-momenta of jets, 
but also by their internal structure. 
Investigations of this jet substructure reveal a wealth of physical 
processes and pose interesting theoretical and experimental challenges.
While relatively young, the field of jet substructure has become 
an important field of research over the last decade and 
will gain further importance with the future data taking 
periods at the LHC. 

With the advent of the LHC it was realized that decays of hypothetical, very heavy resonances can lead to 
highly Lorentz-boosted heavy SM particles, $W$, $Z$, $H$ bosons and top 
quarks~\cite{Seymour:1993mx, Butterworth:2002tt, Agashe:2006hk, 
Butterworth:2008iy, Kaplan:2008ie}. 
Since these particles feature the largest branching fractions 
into hadrons, final states with fully-hadronic decays have high sensitivity
in LHC analyses. 
The large boost leads to very collimated decays, where particle masses of $\mathcal{O}(100)$\GeV are not large enough  
for the outgoing quarks to be sufficiently separated relative to each other to be resolved into individual jets. 
It is this small opening angle between the decay products which leads to fully-merged particle decays. 
The following experimental overview describes techniques for measuring jets as proxies for hadronic decays of 
$W$, $Z$, $H$ bosons and top quarks. However, this review is not limited to 
these methods but covers also precision jet substructure measurements and the discrimination of quark and gluon jets, reflecting the versatility of jet substructure. 
The scientific gains from these measurements are manifold, reaching from precision studies of
QCD over the determination of fundamental parameters of the Standard 
Model to searches for new physical phenomena at the highest energy scales. 
A recent review on the theoretical aspects of jet substructure can 
be found in Ref.~\cite{Larkoski:2017jix}.

Since the first evidence for jets in $e^{+}e^{-}$ collisions at SPEAR~\cite{PhysRevLett.35.1609}, jets have had a 
significant impact on the research program of every particle 
collider since DORIS through the LHC, and beyond to the 
design of future colliders.
There is no single, universal definition of a jet -- which particles 
belong to a jet depend on the algorithm used to combine particles into jets.
In the beginning of jets from the mid 1970's, there were no jet clustering algorithms; information from the whole event was used instead of 
localized energy flows. The sphericity tensor~\cite{Bjorken:1969wi} 
was typically used to obtain a jet axis for events with 
a back-to-back dijet topology. 
Quantitative statements about data were obtained from event shapes, like the sphericity or thrust~\cite{Brandt:1964sa, PhysRevLett.39.1587}. 
Sphericity is a measure for the isotropy of the produced particles and 
thrust is a measure of the directed energy flow along an axis that maximises this flow in an event. These event shapes can be used 
to characterize how compatible events are with the assumption of two oppositely directed, collimated jets. 
A clear theoretical advantage of these event shapes is that they are calculable in perturbative Quantum Chromodynamics (pQCD). 
This was realized early on and the calculability ultimately resulted in the confirmation of the parton model and, 
with data from experiments at higher $\sqrt{s}$, the discovery of the gluon in three jet events at 
PETRA~\cite{Bartel:1979ut,Berger:1979cj,Barber:1979yr,Brandelik:1979bd}.

\begin{figure} 
\centering  
\includegraphics[width=0.35\textwidth]{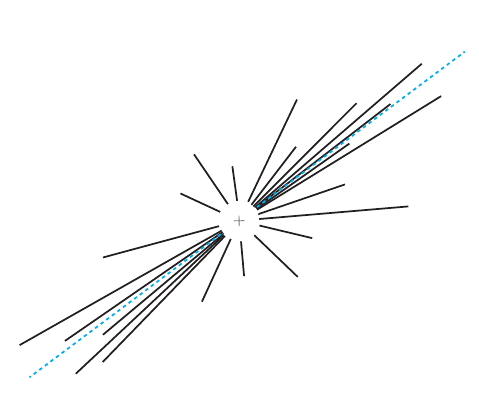}
\caption[Particles from a hard collision]{Schematic drawing of particles emerging from the hard scattering of a high energy particle collision.
The sphericity axis is shown as dashed line.}
\label{fig:schematic_particles_into_jets}
\end{figure}
When studying the dynamics of quark and gluon scattering, it became necessary to perform quantitative analyses 
and calculations that go beyond event shapes. For these to be possible, it was realized that it is mandatory to define a 
deterministic set of rules on how particles are combined into jets. A schematic drawing depicting this problem is shown in 
figure~\ref{fig:schematic_particles_into_jets}. While the sphericity axis is uniquely defined and easily calculable, the direction 
and magnitude of the jet axes depend on which particles should be combined into a given jet, and how the particles are combined to 
obtain the axes. 
An intuitive definition for a jet algorithm consists of summing the momenta of all particles within a cone with fixed size~\cite{Sterman:1977}.   
Naive cone algorithms are not infared and collinear (IRC) safe -- the 
requirement that the resulting jets be insensitive to arbitrarily low energy 
particles and collinear splittings.  IRC safety is a useful theoretical 
requirement for making calculations in pQCD and is also a convenient language
for describing the experimental robustness to noise and detector granularity. 

There exist many variants of cone-type algorithms, developed in the attempt to 
solve the IRC unsafety of naive cone jet algorithms. This stems from the 
necessity of an initial axis, which was eventually solved with 
the formulation of the SISCone algorithm~\cite{Salam:2007xv}. Although this 
algorithm is IRC safe, it is not widely used today because it was found that 
sequential recombination algorithms 
have several advantages over cone-type algorithms. 
First used by the JADE Collaboration~\cite{Bartel1986, Bethke1988}, 
the initial version of a recombination algorithm defined for $e^+e^-$ 
collisions was improved in several steps~\cite{Catani1991, Catani1992}, 
to finally arrive at the 
longitudinally-invariant \kt-clustering algorithm for hadron-hadron collisions~\cite{Catani1993}.
A generalization of this algorithm leads to three classes, distinct 
only by the sign of the exponent of the transverse momentum $p_{T,i}$ 
in the inter-particle distance measure 
\begin{align}
\label{eq:jetdistance}
d_{ij}(p_i,p_j) = \text{min}(p_{T,i}^{2k},p_{T,j}^{2k})\frac{\Delta R^2}{R^2},
\end{align}
where\footnote{
Sometimes the rapidity ($y$) is used and sometimes the pseudorapidity ($\eta$) is used depending on the application.  See Ref.~\cite{Gallicchio:2018elx} for a detailed discussion.
} 
$\Delta R^2=\Delta\phi^2+\Delta y^2$ and $R$ is typically called the \textit{jet radius}.
The original \kt algorithm, with $k=1$ in Eq.~\eqref{eq:jetdistance}, clusters soft and collinear particles 
first, the Cambridge/Aachen algorithm (CA)~\cite{ca1, ca2}, with $k=0$, prioritizes particles in the clustering solely by their
angular proximity, and the anti-\kt algorithm~\cite{Cacciari:2008gp}, with $k=-1$, combines the hardest particles first. 
The proposal of the latter algorithm is also responsible for the disappearance of cone-type algorithms in experimental studies. 
When it was realized that the anti-\kt algorithm results in nearly perfect conical jets the LHC collaborations made a transition to this algorithm. Today, almost all studies involving jets performed at the LHC use this algorithm. Even when analyzing the substructure of jets with advanced grooming or tagging techniques, 
the initial step often consists of building an ensemble of particles that were clustered with the anti-\kt algorithm.

So far, it has not been specified what the term particle refers to when using particles as input to jet clustering. 
In fact, in jet physics, the term particle is often used generically for different sorts of objects, whose ensemble comprises the input to a given 
jet algorithm. Three different ensembles are commonly used. The partonic final state includes all particles resulting from the parton shower before the hadronization starts (which is unphysical). This also include photons when these were
created in the hard interaction or emitted from charged particles during the parton shower. The ensemble on the particle level, 
also called hadron level, consists of hadrons and their decay products, including photons and leptons. 
The detector level input consists of calorimeter clusters, reconstructed particle tracks or combinations thereof. 
Jet algorithms using these different ensembles as input result in parton, particle or detector level jets, respectively. 
Ideally, in any given event, the jets obtained on parton, particle and detector level are as similar as possible. 
Realistically, agreement can not be achieved, but a close correspondence ensures the possibility to study 
the underlying partonic dynamics with the use of jets. It is this correspondence, paired with calculability in pQCD, which makes
jets indispensable tools at high energy particle colliders\footnote{
For a theoretical introduction to jets, we recommend the reviews 
in Refs.~\cite{Salam:2009jx,Ellis:1991qj} as well as the theory companion this experimental review, Ref.~\cite{Larkoski:2017jix}.
}. 

Soon after their discovery, it was realized that not only the kinematics of jets but also their internal structure carry information. 
The parton shower and subsequent hadronization leads to a characteristic multiplicity, as well as angular and momentum distributions 
of hadrons inside jets, which depend on the parton that initiated the shower. For example, the probability of a $q\to qg$ splitting 
is proportional to the color factor $C_F=4/3$ at leading order in QCD, while the probability of $g\to gg$ is proportional to 
$C_A=3$. The larger value of $C_A$ results in a larger multiplicity of hadrons and in broader jets. 
This lead to the suggestion of measuring jet shapes, 
defined as the fractional transverse momentum profile of particles within a concentric inner cone, smaller than the jet cone of 
the original jet, and pointed to their usefulness for distinguishing quark jets from gluon jets~\cite{Ellis:1992qq}. 
Experimental results from LEP~\cite{Alexander:1991ce, Buskulic:1994wy, Abreu:1995hp, Acciarri:1997it}, 
Tevatron~\cite{Abe:1992wv, Abachi:1995zw} 
and HERA~\cite{Breitweg:1997gg, Breitweg:1998gf, Adloff:1998ni} 
confirmed this and can be considered as the starting point of physics with 
jet substructure in particle physics. 

\begin{figure} 
\centering  
\includegraphics[width=0.35\textwidth]{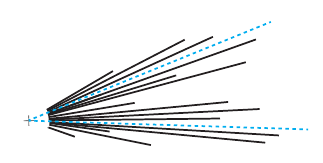}
\caption[Particles from a hard collision]{Schematic drawing of particles clustered into a single jet. Two subjet axes are shown as dashed lines.}
\label{fig:schematic_boosted_particles}
\end{figure}
At the LHC, jet substructure is used to identify highly boosted heavy SM particles in fully hadronic decays. 
An example of a jet with substructure from a two-prong decay is shown schematically in figure~\ref{fig:schematic_boosted_particles}. 
The difficulty lies in identifying the underlying process that led to the final state, for example distinguishing $W\to q\bar{q}^{\prime}$, 
$Z\to q\bar{q}$ or $H\to b\bar{b}$ from QCD splittings like $q\to qg$, $g\to gg$ or $g\to q \bar{q}$. 
Numerous algorithms have been suggested to identify specific decays, which are part of a class of \emph{jet substructure taggers}. The idea behind many of these algorithms is related to event shapes 
in $e^{+}e^{-}$ collisions. By defining $N$ axes within a jet, it is possible to check for the compatibility 
of a fully-merged $N$-prong decay. How these axes are found typically differs from algorithm to algorithm, and some techniques do not even explicitly require axes. 
Popular concepts are an exclusive jet clustering using the particles inside a jet as input, or the maximization of the 
projection of the jet constituents' momenta onto the desired number of axes, as illustrated in figure~\ref{fig:schematic_boosted_particles}. 
Since the opening angle between the quarks depends on the momentum of the parent particle and its mass, larger 
jets ($R\sim 1$) than normally employed in LHC analyses ($R\sim 0.4$) are used to reconstruct boosted heavy particle decays. A larger distance parameter 
is chosen to capture the full kinematics of the decay already at moderate momenta of 200--400\GeV. The drawback of jets 
with large areas is unwanted contributions from the underlying event and from multiple proton-proton 
collisions in a single bunch crossing (pile-up). These lead to a worsening of the resolution in quantities used to identify the substructure of jets, 
like the jet mass. Jet \emph{grooming} and pile-up removal algorithms have been developed to mitigate these 
effects. Grooming algorithms aim at removing soft and wide-angle radiation, therefore not only reducing the effects from 
the underlying event but also reducing the sensitivity to the details of fragmentation. Pile-up removal algorithms 
are designed to identify and subtract contributions from a different interaction vertex, by eliminating uncorrelated radiation from jets. 
A combination of these techniques often leads to the best overall performance and it is an ongoing effort to understand 
the interplay of pile-up removal, grooming and tagging algorithms. 

The theoretical and algorithmic developments are possible due to advances in experimental methods. New technologies, 
like silicon pixel detectors, high-resolution tracking detectors in conjunction with strong magnetic fields,  
highly granular calorimeters with low electronic noise and lightweight materials for detector structures with little dead 
material inside the active detector volume have enabled increasingly precise jet measurements and studies of internal jet structure. 
Modern particle detectors at the LHC are equipped with many layers of high-resolution tracking detectors, 
strong and very homogeneous magnetic fields and finely segmented calorimeters with an excellent 
energy resolution. With these technologies, the ATLAS and CMS detectors\footnote{The ALICE and LHCb detectors are also well-equipped to perform jet substructure studies.  While these experiments do not have access to boosted massive particles due to their data rate (ALICE) or acceptance (LHCb), the are performing many interesting QCD studies with jet substructure.  This review will be focused on ATLAS and CMS, but the future of jet substructure will involve key contributions from all four LHC experiments.} are equipped to track and 
reconstruct individual particles produced in high energy collisions. On average about 60\% 
of a jet's momentum is carried by charged hadrons, photons account for about 25\% of the 
total jet momentum and the remaining 15\% can be attributed to long-lived neutral hadrons~\cite{Khachatryan:2016kdb}. 
With increasing jet energy, the particle multiplicity increases, and also 
the fraction of the jet's momentum carried by soft particles. 
For example, on average 50\% of the momentum of a 50\GeV jet 
is carried by particles with a momentum less than 5\% of the jet's momentum. It is therefore crucial to 
ensure that particles with energies down to $\mathcal{O}(100\MeV)$ can be reconstructed 
in order to retain the full information on a jet's kinematics and internal structure. 

As important as the reconstruction of the total jet energy is 
the measurement of the jet constituent multiplicity and their angular distributions. 
While charged particles can be efficiently reconstructed as tracks, neutral particles develop showers in the 
calorimeters and the possibility to resolve two separate showers depends on the granularity of the calorimeter 
and the lateral shower development. Hence, it becomes more difficult to separate two adjacent particles
in dense environments, such as high momentum jets, and the situation is aggravated by the presence of 
hadronic showers from charged hadrons. Often it is impossible to build one calorimeter cluster per neutral particle. 
A way to improve the angular resolution in jet substructure analyses is to combine 
measurements from the tracking detectors and calorimeters. Using combined detector measurements as input 
to jet algorithms, for example using a particle flow approach, results in improved resolutions of jet 
substructure observables, compared to using only tracks or only calorimeter clusters. 

An important aspect of experimental analyses at the LHC is the calibration of jets, necessitated by the non-compensating 
nature of hadron calorimeters, suppression of electronic noise, tracking inefficiencies, dead material in front 
of calorimeters, the influence of pile-up and other effects. While the calibration of the total jet energy scale 
is an important aspect in all analyses using jets, the precise knowledge of the jet mass scale and the 
detector response to jet substructure observables and jet tagging algorithms is specific to 
jet substructure analyses. Calibrating the jet energy scale results in a change 
of the magnitude of the jet's four-momentum, where the jet mass scale comprises an 
additional degree of freedom that can not be constrained by the typical methods of balancing 
a jet with a well-calibrated reference object. The jet mass scale is usually calibrated 
using jets from fully-merged, highly boosted $W\to q\bar{q}^{\prime}$ decays, 
facilitating a calibration of the peak position in the jet mass distribution. Measurements 
of the jet mass distribution from light quark and gluon jets, as well as from fully-hadronic 
highly-boosted $W$, $Z$ and $t$ decays allow for precise tests of the modelling of perturbative 
and non-perturbative effects in jet production. 
Similar measurements can also be used to study the detector response to jet substructure 
observables and their modelling in simulation. A mis-modelling of variables used for tagging, 
either in the detector simulation or on the level of the underlying physics, can result 
in a wrong estimation of the tagging efficiency or the misidentification rate, with 
important consequences for measurements. 
In order to overcome this limitation, measurements of tagging efficiencies and 
misidentification rates are performed in samples enriched with the particle decays in question. 
While these measurements do not help to understand the cause of the mis-modelling or to 
improve the description of jet substructure distributions, these can be used to correct the 
efficiencies in simulation. It is these measurements that have enabled the use of jet 
substructure taggers in numerous physics analyses since the beginning of data taking at the LHC.
The increased statistics from a data sample corresponding to about 150\fbinv per experiment at a 
centre-of-mass energy of 13\TeV can now be used to improve our understanding of the 
detector response to jet substructure algorithms, the underlying physics and the performance 
differences of taggers. 
These studies and measurements represent the continuation of an exciting physics 
program at the LHC in a field which reached its adolescence in the past few years. 
In the years to come, the field of jet substructure will evolve and mature through precision measurements and the exploration of unknown territory. 

We begin this review with a brief overview of the ATLAS and CMS detectors in 
section~\ref{sec:jetrec-det}, followed by a description of the 
input to jet reconstruction and jet calibration in section~\ref{sec:jetrec}.  
An important aspect of jet reconstruction at the LHC, and jet substructure in particular, 
are algorithms to mitigate the effects of pile-up. 
Recent experimental advancements and algorithms employed in 
ATLAS and CMS analyses are discussed in section~\ref{sec:pile-up}.
In section~\ref{sec:substructuremethodsobservables} we review 
jet grooming techniques in use in experimental analyses and discuss 
their impact on jet substructure observables. A special 
emphasis is given on the jet mass calibration and jet mass 
measurements in different final states. Measurements of other jet 
substructure distributions are described as well. 
One of the key developments within the field of jet substructure 
are tagging algorithms, which are described in detail in 
section~\ref{sec:tagging}. Theoretical and experimental developments 
have resulted in large performance gains of substructure taggers 
in the last years, relevant for a large number of present and future physics analyses. We highlight the main developments and 
improvements and give an overview of relevant experimental studies.
The use of jet substructure taggers in existing cross section measurements is reviewed in section~\ref{sec:crosssections}. 
So far, the major beneficiaries of jet substructure methods 
have been analyses in search for new physical phenomena. 
We review the application of these methods to searches for 
new physics in section~\ref{sec:searches} and conclude 
in section~\ref{sec:conclusions}.

\section{ATLAS and CMS detectors}\label{sec:jetrec-det}

The ATLAS~\cite{atlas-det} and CMS~\cite{cms-det} detectors are designed to observe leptons, photons, and hadrons resulting from LHC $pp$ collisions. The physics of the hard reaction takes place at the point of collision (the primary vertex) within the beam pipe. Beyond the beam pipe\footnote{The LHC collaborations are continuously working to improve the detectors; the numbers given here are for the detectors that operated in 2015-2017.  Before and after this time, the exact values are not the same as reported here.}, at 4.4 cm (3.3 cm) in CMS (ATLAS), the first cylindrical layer of detectors encountered are silicon pixels and strips for identification of charged particles. CMS provides a 3.8\:T magnetic field via a solenoid positioned outside the silicon tracking detector, the Electromagnetic Calorimeter (ECAL) and most of the Hadronic Calorimeter (HCAL). ATLAS has an additional tracking layer composed of straw drift tubes (Transition Radiation Tracking or TRT), with a 2\:T magnetic field encompassing the silicon and TRT detectors, while the ECAL and HCAL are situated outside the solenoidal magnet. The calorimeters are surrounded by muon spectrometers which build the outermost part of the ATLAS and CMS detectors.  Both detectors are nearly hermetic and can therefore measure the missing transverse momentum.

The energy and momentum ranges and resolutions for the barrel
regions\footnote{For example, the ATLAS ECAL barrel covers the pseudorapidity range
  $|\eta|<1.475$, the end-caps cover $1.375 < |\eta| < 3.2$ and the
  forward ECAL layer extends the coverage up to $|\eta| < 4.9$. The
  CMS ECAL barrel covers $|\eta|<1.48$, the end-caps extend the
  coverage up to $|\eta|<3$.} of ATLAS and CMS are shown in table
\ref{tab:det} along with the measurement granularity, which limits the
angular resolution. The better energy resolution of the CMS ECAL is
due to the use of lead tungstate (PbWO$_{4}$) crystals, as opposed to
the Liquid Argon (LAr) used by ATLAS. The differences in the ATLAS and
CMS calorimeter designs are a result of the different ranking of
priorities decided by the two collaborations; ATLAS chose a
radiation-hard technology with sufficient resolution in a fine
sampling LAr calorimeter, while CMS prioritized the excellent
resolution of a total absorption crystal calorimeter (the focus was
Higgs mass reconstruction), and accepted the accompanying limitations
in radiation-hardness associated with this technology. The CMS ECAL
crystal response varies under irradiation, which is partially
recovered in a few hours at room temperature.

The ATLAS ECAL is segmented into three (two) longitudinal layers for $|\eta| < 2.5$ ($|\eta| > 2.5$). The granularity of the ATLAS ECAL in table~\ref{tab:det} refers to its second layer (as most of the electromagnetic energy is deposited there); the first layer has a finer granularity in $\eta$. The multiple layers allow for a finer granularity than the cell size in any of the individual layers, being advantageous over a laterally segmented calorimeter, and additionally provide pointing information. 
The difference between ATLAS and CMS for the HCAL resolution is particularly large at higher energies: a 1~\TeV jet has $\frac{\sigma(E)}{E}\sim 2\%$ in ATLAS, in contrast to $\frac{\sigma(E)}{E}\sim 5\%$ in CMS.  This is one reason why CMS fully adapted a particle flow technique since the beginning of the LHC (see section~\ref{sec:jetrec-rec} below).

\begin{table}[ht]\renewcommand{\arraystretch}{1.5}
\small

\begin{center}
\begin{tabular}{L{2.3cm} | L{2.6cm} | L{2.6cm}}
\hline
\hline
&\textbf{ATLAS} & \textbf{CMS}\\
\hline
\multicolumn{3}{l}{\textbf{Tracking}}\\
\hline
1/\pt resolution         & $0.05\%\times\pt/\GeV\oplus 1\%$~\cite{Aad:2010bx}  &  $0.02\%\times\pt/\GeV\oplus 0.8\%$~\cite{Chatrchyan:2014fea}         \\
$d_{0}$ resolution ($\mu m$)  & $20$~\cite{Abbott:2018ikt}                       & $20$~\cite{Chatrchyan:2014fea}         \\
\hline
\multicolumn{3}{l}{\textbf{ECAL}}\\
\hline
$E$ resolution        & $10\%/\sqrt{E}\oplus0.2\%$~\cite{atlas-det}    & $3\%/\sqrt{E}\oplus12\%/E\oplus0.3\%$~\cite{cms-det} \\
granularity                 & $0.025\times0.025$ & $0.017\times0.017$   \\  
\hline
\multicolumn{3}{l}{\textbf{HCAL}}\\
\hline
$E$ resolution        & $50\%/\sqrt{E} \oplus 5\%$~\cite{atlas-det}     & $100\%/\sqrt{E}\oplus5\%$~\cite{CERN-LHCC-97-031}         \\
granularity                 & $0.1\times 0.1$                  & $0.087\times0.087$ \\
\hline
\hline
\end{tabular}
\caption[ATLAS and CMS resolution and granularity.]{ ATLAS and CMS detectors in the barrel regions.  The granularity is in pseudorapidity and azimuth
 ($\eta \times \phi$) and $d_{0}$ is the transverse impact
  parameter resolution with respect to the beam-line.  The tracker momentum resolution is from muons while the $d_0$ resolution is from generic charged particles (mostly pions) in $t\bar{t}$ events.  The ECAL energy resolution is presented for electrons.   The granularity for the ATLAS calorimeters are for the middle layers only, which collect the largest amount of energy.  For the ATLAS EM calorimeter, the innermost layer has $\Delta\eta = 0.0031$ for $\gamma/\pi^0$ separation.   }
\label{tab:det}
\end{center}
\end{table}

\section{Jet Reconstruction}\label{sec:jetrec}

\subsection{Inputs}\label{sec:jetrec-rec}

Both experiments have dedicated algorithms to reconstruct particle kinematics from calorimeter and tracker information designed to minimize the fake rate, maximize the efficiency, and minimize the bias and resolution of the particle candidate parameters.  As there is no algorithm that can simultaneously optimize all of these objectives, the various approaches trade off optimality under one metric for improvements under another.  ATLAS and CMS have also developed different algorithms that cater to the experiment's hardware as well as the collaboration's goals for the tradeoffs.  By default, CMS combines tracker and calorimeter information into unified particle flow objects as inputs to jet reconstruction~\cite{CMS-PAS-PFT-09-001,CMS-PAS-PFT-10-002,Sirunyan:2017ulk}.  ATLAS has traditionally used calorimeter-only information for jet reconstruction, with tracking information used to augment/enhance the performance.  While ATLAS is current migrating to a variation of particle-flow~\cite{atlas-pflow-13tev}, most of this review will focus on calorimeter-only jets as they are still the most widely used setup.  ATLAS benefits less than CMS from particle flow because of its weaker magnetic field and longitudinally segmented calorimeter. 

ATLAS and CMS combine calorimeter cells using topological clusters~\cite{atlas-topoclusters,Sirunyan:2017ulk}.  These clusters are three dimensional in ATLAS as a result of the longitudinal segmentation.  Cluster seeds are started from highly significant energy (high cell signal to average electronic $\oplus$ pileup noise) deposits which are combined (or split) based on the distribution of the significance of energy in nearby cells.  Calorimeter-cell clusters in CMS are obtained using a Gaussian-mixture model, which results in one or more calorimeter clusters within each topological cluster. HCAL clusters can be split according to the number and energy distribution of associated ECAL clusters.  Cluster splitting is critical to achieve a better estimate of the spatial energy distribution as input to jet substructure algorithms~\cite{ATL-PHYS-PUB-2017-015,CMS-PAS-JME-14-002}.  

The topological clusters are calibrated using simulations to account for the non-compensating calorimeter response to hadrons, signal losses due to energy deposited in inactive detector material and signal losses on cluster boundaries caused by the topological clustering algorithms. In ATLAS, the calibration scheme relies on a classification of clusters as hadronic or electromagnetic in origin based on the energy and position of the cluster, the longitudinal depth ($\lambda_{clus}$) and normalized signal energy density; hadronic showers tend to occur deeper in the calorimeter and be less dense~\cite{atlas-topoclusters}.  Charged and neutral pions are used to derive this classification and calibration, called the Local Cell Weighting (LCW). In CMS, dedicated ECAL (based on photons) and HCAL (based on neutral kaons) calibrations are combined to account for energy and $|\eta|$-dependent non-linearities in the hadron calorimeter response~\cite{Sirunyan:2017ulk}.  Both ATLAS and CMS validate the performance of these calibrations with single particle studies in data~\cite{Aaboud:2016hwh,Sirunyan:2017ulk}.

Different strategies are used by ATLAS and CMS to reconstruct tracks from their inner detectors.  ATLAS focuses first on maintaining a high efficiency with a rather inclusive first pass through inner detector hits.  A second step known as ambiguity solving reduces the fake rate.  In contrast, CMS uses a sequential approach with multiple passes through the remaining inner detector hits.  With each pass, the efficiency increases while maintaining a low fake rate.  Both procedures are effective at identifying about 90\% of charged pions above 1 GeV with a percent-level (or smaller) fake rate.  Lower momentum particles can be reconstructed, at the cost of a higher fake rate and lower efficiency.  Due to its weaker magnetic field, ATLAS is able to reach low track momentum of 100 MeV for physics analysis~\cite{Aaboud:2016itf}, although most jet substructure measurements and searches use a threshold of 500 MeV.  In contrast, the momentum resolution in CMS is excellent up to higher momenta than in ATLAS.  The TRT can be used to improve the momentum resolution of high $p_T$ tracks~\cite{Aaboud:2017odu}, but the weaker magnetic field despite a comparable inner detector radius is a fundamental limitation. 

Both experiments have implemented dedicated strategies for track
reconstruction in high density environments such as the core of high
\pt jets. In such environments, pixel and strip clusters can merge resulting in a loss
in tracking efficiency and degraded resolution.  ATLAS has implemented
a stacked neural network (NN) approach to examine pixel clusters to
identify multi-particle clusters, estimate the position of the
particles passing through the clusters, and also predict the residual
resolution of the position
estimates~\cite{Aad:2014yva,Aaboud:2017all,ATL-PHYS-PUB-2016-007,ATL-PHYS-PUB-2017-016,ATL-PHYS-PUB-2015-044}. 
CMS has introduced a dedicated tracking step in which a
cluster splitting procedure attempts to split merged clusters
exploiting the information of the jet direction, predicting the
expected cluster shape and charge \cite{CMSJetCore}. 

For particle flow in CMS, tracks and calibrated clusters are combined taking the tracking and calorimeter resolutions into account. First, a link is created between tracks in the central tracker and calorimeter clusters. Links are also created between clusters in the ECAL and HCAL, when the cluster position in the ECAL is within the cluster envelope in the less granular HCAL. Tracks with a \pt uncertainty in excess of the calorimetric energy resolution expected for charged hadrons are masked, which allows the rate of misreconstructed tracks at large \pt to be reduced. 

The ECAL and HCAL clusters not linked to any track give rise to photons and neutral hadrons. Charged hadrons are created from the remaining ECAL and HCAL clusters, linked to tracks. If the calibrated calorimetric energy is compatible with the corresponding track momenta under the charged-pion hypothesis, no neutral particles are created. Otherwise, the excess energy is interpreted to originate from photons and neutral hadrons for deposits in the ECAL and HCAL, respectively.  The particle flow algorithm in ATLAS is similar to the one used by CMS and is described in more detail in Ref.~\cite{atlas-pflow-13tev}.

The combination of tracking and calorimetric measurements results in an optimal input for jet substructure measurements, making use of the superior angular resolution from the tracking detector and calibrated calorimeter clusters. Once the calibrated PF objects are clustered into jets, their relative momenta and angular distances are kept constant, and only the total energy response of jets is corrected with factorized JES calibrations (see section~\ref{sec:jetrec-jes}). 

\begin{figure} 
\centering  
\includegraphics[width=0.45\textwidth]{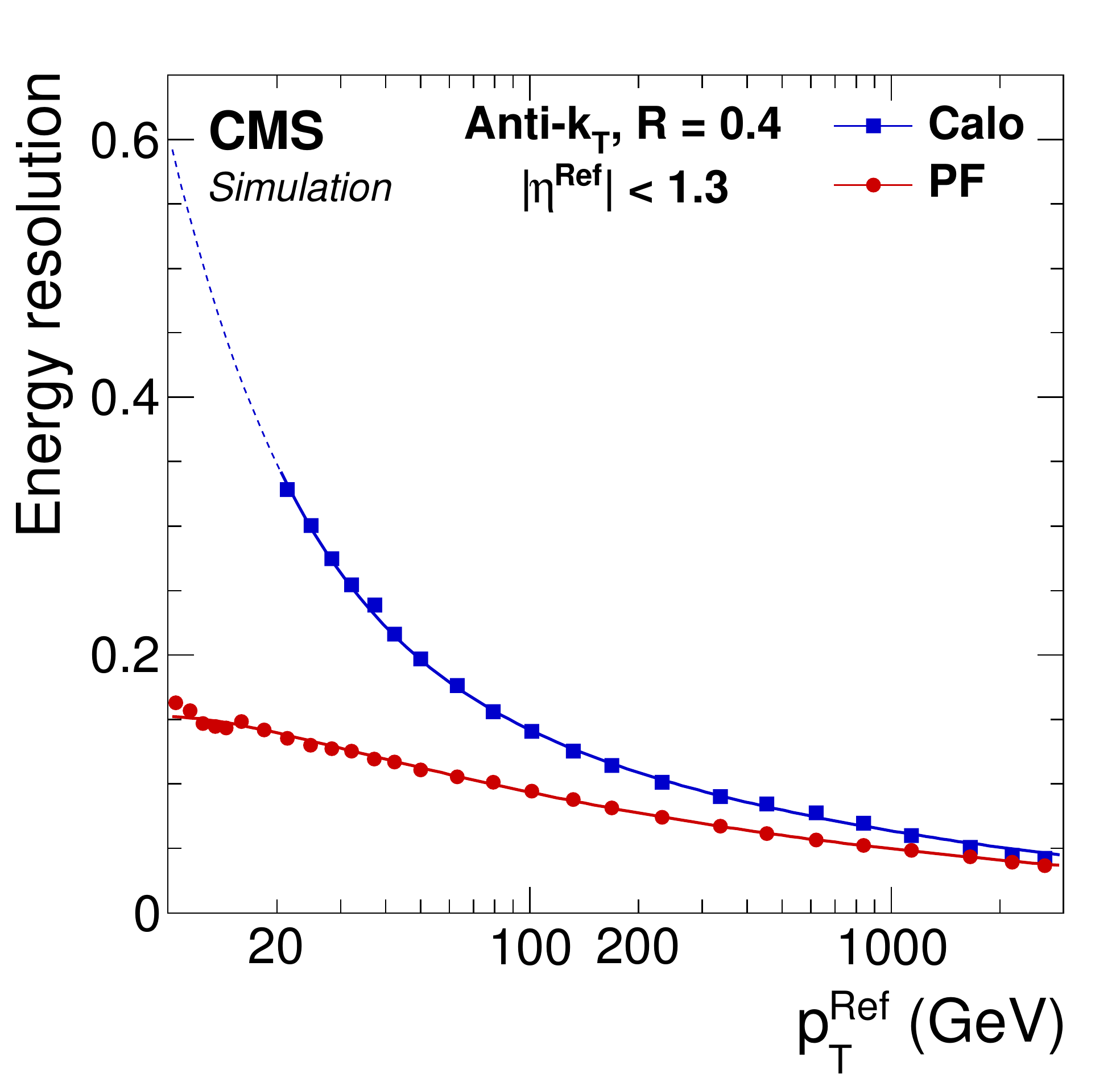}
\caption[CMS jet energy resolution with PF]{Jet energy resolution for particle flow (red, lower line) and calorimeter-only (blue, upper line) jets in the barrel region in CMS simulation, with no pile-up, as a function of the $\pt$ of the reference jet. Taken from~\cite{Sirunyan:2017ulk}. }
\label{fig:ptres_CMS_pFlow}
\end{figure}

The particle flow algorithm improves the energy resolution as shown in figure~\ref{fig:ptres_CMS_pFlow}. A similar performance gain is observed in ATLAS, but the weaker magnetic field means that the point where calorimetery and tracking are comparable is lower (about 100 GeV).

\subsection{Calibration}\label{sec:jetrec-jes}

The ratio of the measured energy $E_\text{reco}$ to the deposited energy
$E_\text{true}$ is the jet energy \textit{response} which depends on the
energy, pseudorapidity and other features of the jet.
Due to the properties of tracking detectors and calorimeters, the average response is not unity. For example calorimeter jets in ATLAS with $E_\text{true}=30$\GeV may have
responses below 0.3, while jets of higher energies may have responses
above 0.8.  For this reason, the Jet Energy Scale (JES) is calculated in bins of the particle-level jet energy $E_\text{true}$ and $\eta_\text{det}$ as the mean of a Gaussian fit to the response distribution and a numerical inversion procedure is used to derive calibration factors in bins of the reconstructed jet energy from $E_\text{true}$~\cite{Aad:2014bia,atlas-jes-13tev,Chatrchyan:2011ds,Cukierman:2016dkb}.

\begin{figure}
\centering
\includegraphics[width=0.45\textwidth]{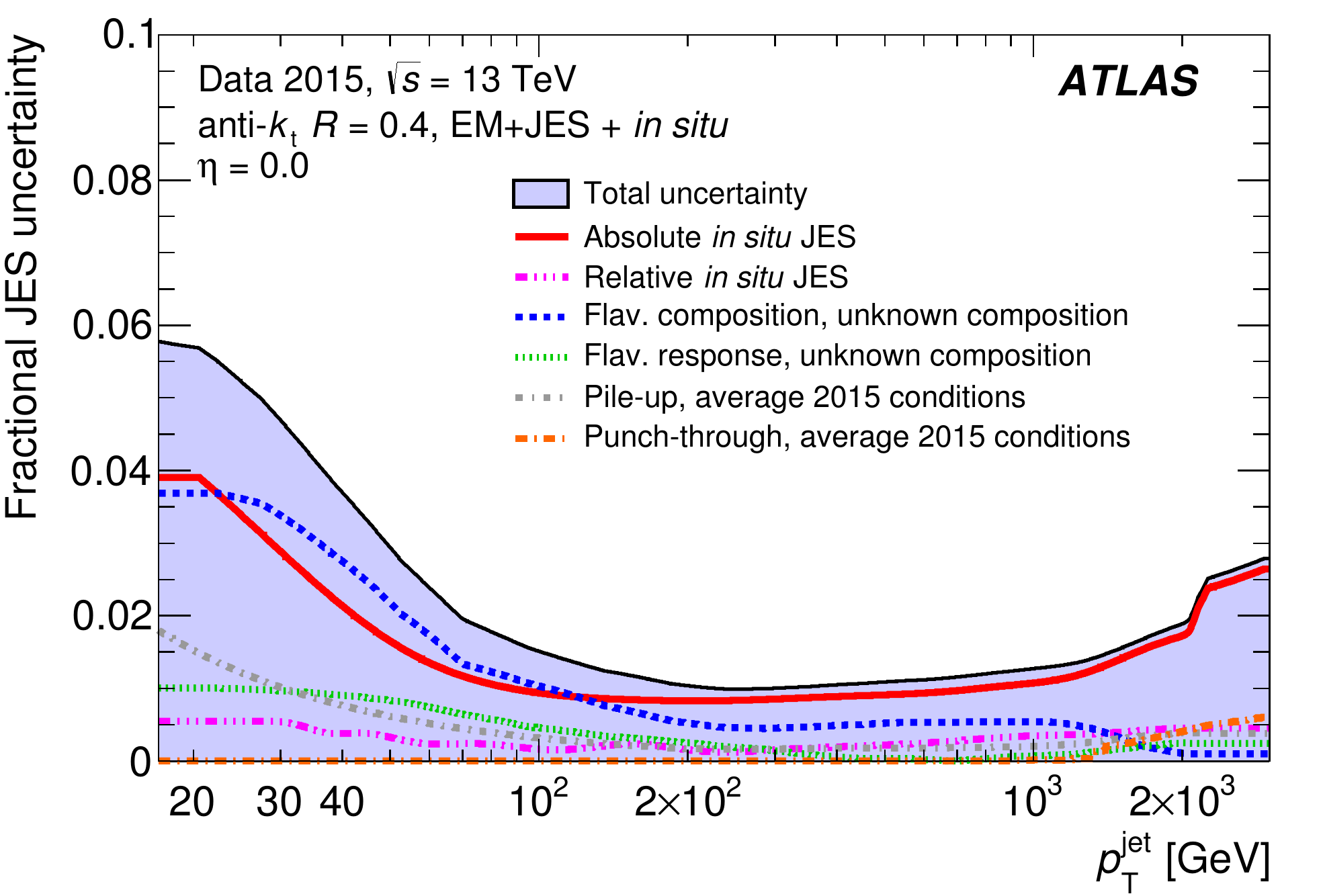}
\caption[ATLAS JES ]{ATLAS jet energy scale uncertainty. Adapted from~\cite{atlas-jes-13tev}.}
\label{fig:atlas-jes}
\end{figure}

In ATLAS, the calibration of the JES is undertaken in several stages, starting from jets either at the electromagnetic (EM) or LCW (built from calibrated inputs) scale. Using calibrated inputs bring the JES to within 10\% of unity for $E = 30$\GeV and $|\eta| < 0.3$~\cite{Aad:2014bia}.
  The Global Sequential
Calibration~\cite{ATLAS-CONF-2015-002,atlas-jes-13tev} was introduced
for \runtwo and reduces the sensitivity to differences in the responses
of quark versus gluon-initiated jets (quark/gluon separation is also
discussed in section \ref{sec:quarkgluon}). This additional calibration
results in a significant jet \pt resolution improvement of up to 35\%
depending on the \pt and $\eta$ of the
jet~\cite{ATLAS-CONF-2015-002}. The JES uncertainty varies between
1-6\% in the central region with $\eta=0$ as shown in
figure~\ref{fig:atlas-jes}~\cite{atlas-jes-13tev}. 

\begin{figure}
\centering
\includegraphics[width=0.45\textwidth]{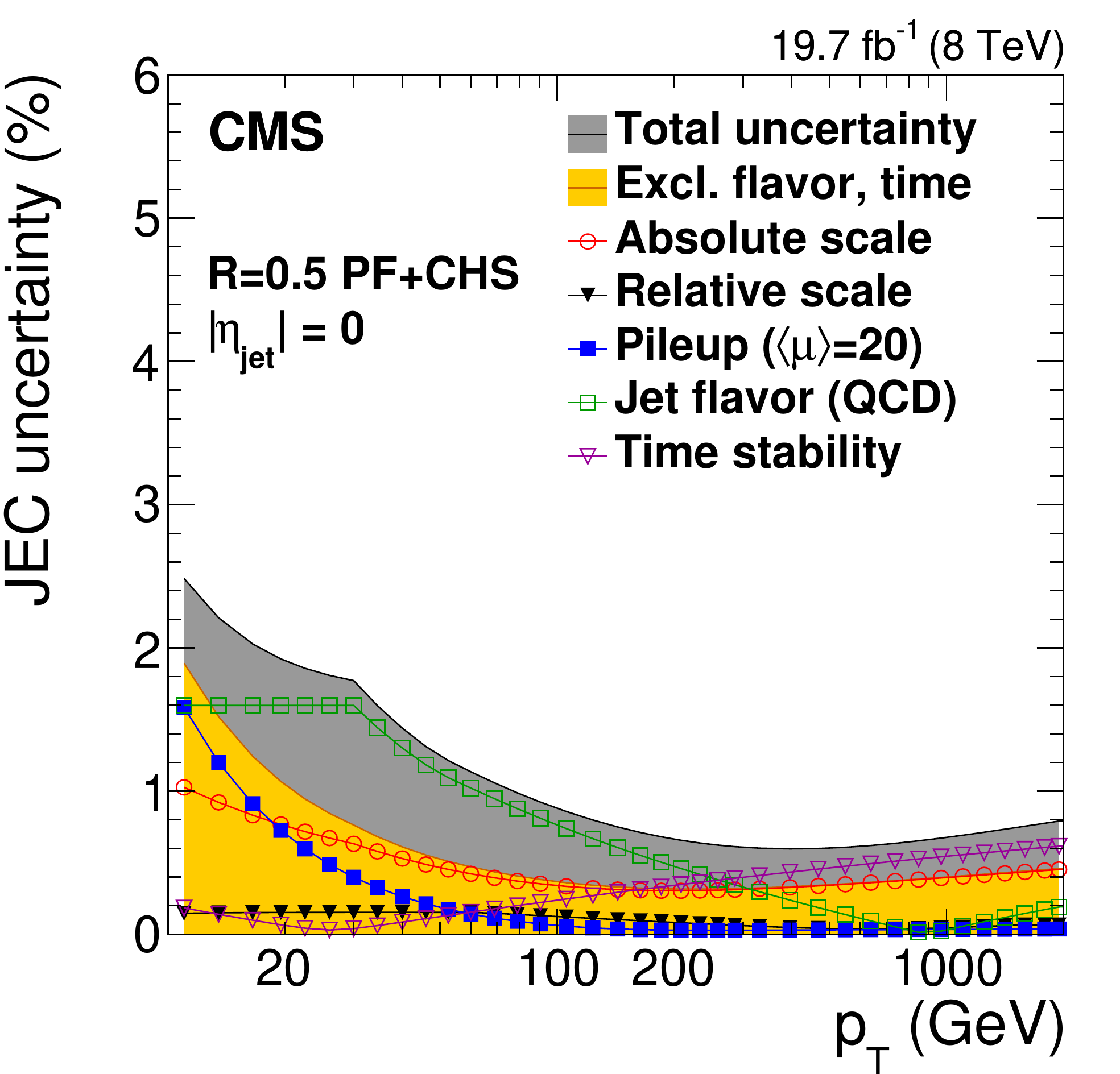}
\caption[CMS JES calibration]{CMS jet energy scale uncertainty, from~\cite{Khachatryan:2016kdb}. JEC means Jet Energy Correction, which has the same meaning as JES.}
\label{fig:cms-jes-uncertainty}
\end{figure}

In CMS, jets are clustered from calibrated particle flow objects, thus the
uncalibrated JES is within 6\% of the expected value of 1 for
central jets with $\eta <0.7$ and $\pt >
30\GeV$~\cite{Khachatryan:2016kdb}.  To account for deviations from unity,
factorized JES calibrations are applied in multiple stages
\cite{CMS-PAS-JME-14-003}  including pile-up corrections, simulation-based response corrections and small residual corrections for tracking inefficiencies and threshold effects, derived \textit{in-situ} from $\gamma$+jet, $Z$+jet
and dijet samples~\cite{Chatrchyan:2011ds}. This additional correction
is not used when jet substructure observables are constructed, but
dedicated corrections are derived as described in
section~\ref{sec:jetrec-mass}. Figure~\ref{fig:cms-jes-uncertainty}
shows the calibrated JES uncertainty obtained in CMS, which is below
1\% for jets with $\pt > 100\GeV$  in the central region with $\eta=0$. Even for jet $\pt$ as low as 10\GeV the uncertainty is below 3\%, owing to the excellent performance of the particle flow reconstruction.  

A detailed discussion of the different approaches for deriving jet energy scale 
uncertainties in ATLAS and CMS can be found in Ref.~\cite{CMS-PAS-JME-14-003}.

\section{Pile-up Mitigation}
\label{sec:pile-up}

\subsection{Definition}
Pile-up originates from simultaneous proton-proton ($pp$) collisions
that occur in addition to a hard scattering collision of interest. The
hard scattering event of interest is referred to  as the Primary
Vertex (PV).  Pile-up is uncorrelated with the PV and typically
consists of an admixture of inelastic, elastic and diffractive $pp$
processes which are separated in the  longitudinal direction.  As the detector response is not instantaneous, pile-up events from both the same (\textit{in-time}) and neighboring (\textit{out-of-time}) bunch crossings can contribute.  This review focuses on the mitigation of in-time pile-up, 
though out-of-time pile-up is also mitigated to differing degrees due 
to the specifics of the ATLAS and CMS detector technologies and reconstruction algorithms.

During the LHC \runone the mean number of pile-up interactions reached $\langle\mu\rangle = 21$, 
and $\mu$ values up to 60 were attained in certain runs of 2017 (\runtwo) with possibly even higher values in Run 3, and culminating at the high luminosity LHC (HL-LHC) reaching up to $\langle\mu\rangle = 140-200$.

Pile-up typically leaves about $0.5\GeV$ of energy in the detector per unit area (${\eta,\phi}$), per pile-up vertex; the effects of this are present in all aspects of LHC physics, from detector design and software performance to the final sensitivity of measurements and searches.

\begin{figure}[t]
\centering
\includegraphics[width=0.45\textwidth]{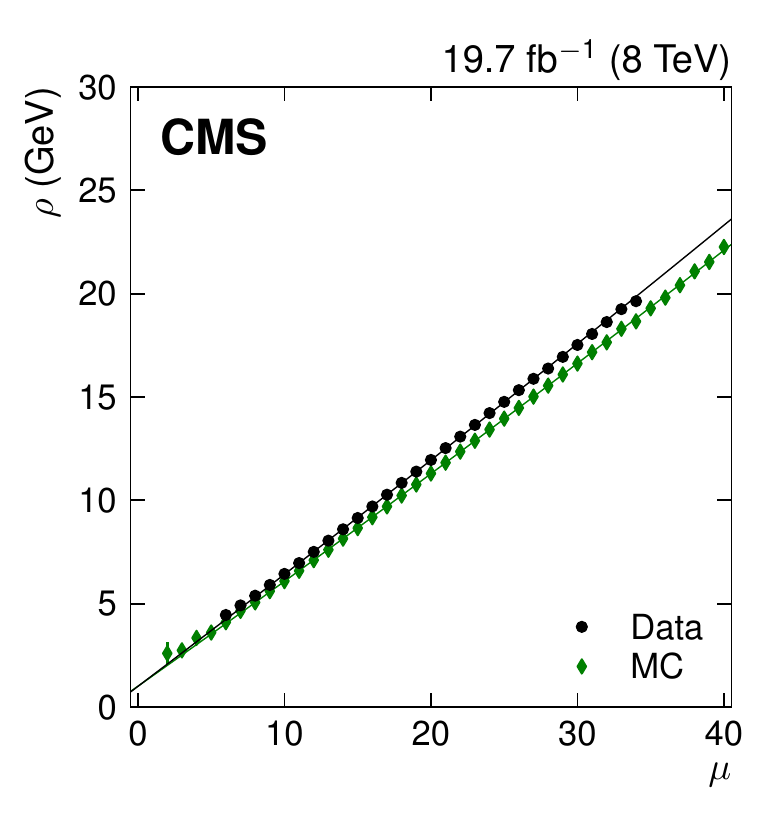}
\caption{\label{fig:pu_offset} Average pile-up contribution to the jet \pt, $\rho$, as a function of the average number of pile-up interactions per bunch crossing, $\mu$, for data (circles) and simulation (diamonds)  at the CMS experiment. Taken from~\cite{Khachatryan:2016kdb}.}
\end{figure}

\subsection{Mitigation Methods}

Properties of pile-up interactions are exploited  to discriminate
pile-up particles from particles originating from the primary vertex,
or to remove energy contributions from pile-up to the individual jet. 

Pile-up can be approximated as a spatially uniform deposition of energy. The so-called \textit{area subtraction} uses a pile-up \pt density per unit area estimator,
$\rho$, and defines a jet catchment area, $A$, to remove energy that
is assumed to originate from pile-up interaction. This approach corrects the jet in the following way: $\pt^\text{corr} = \pt^\text{orig} -\rho A$. 
An example of $\rho$ is shown in
figure~\ref{fig:pu_offset}. There are many subtleties in defining both
$\rho$ and $A$, which are discussed in e.g.
Refs.~\cite{Cacciari:2008gn, CMS-PAS-JME-14-001, Aad:2015ina}. 
An extension to this method is \textit{shape subtraction}~\cite{Soyez:2012hv}, where 
randomly distributed ghost particles are used to calculate a jet shape's 
sensitivity to pile-up, which can then be corrected for non-uniformities in the 
spatial distribution of pile-up particles.

\begin{figure}[t]
\centering
\includegraphics[width=0.45\textwidth]{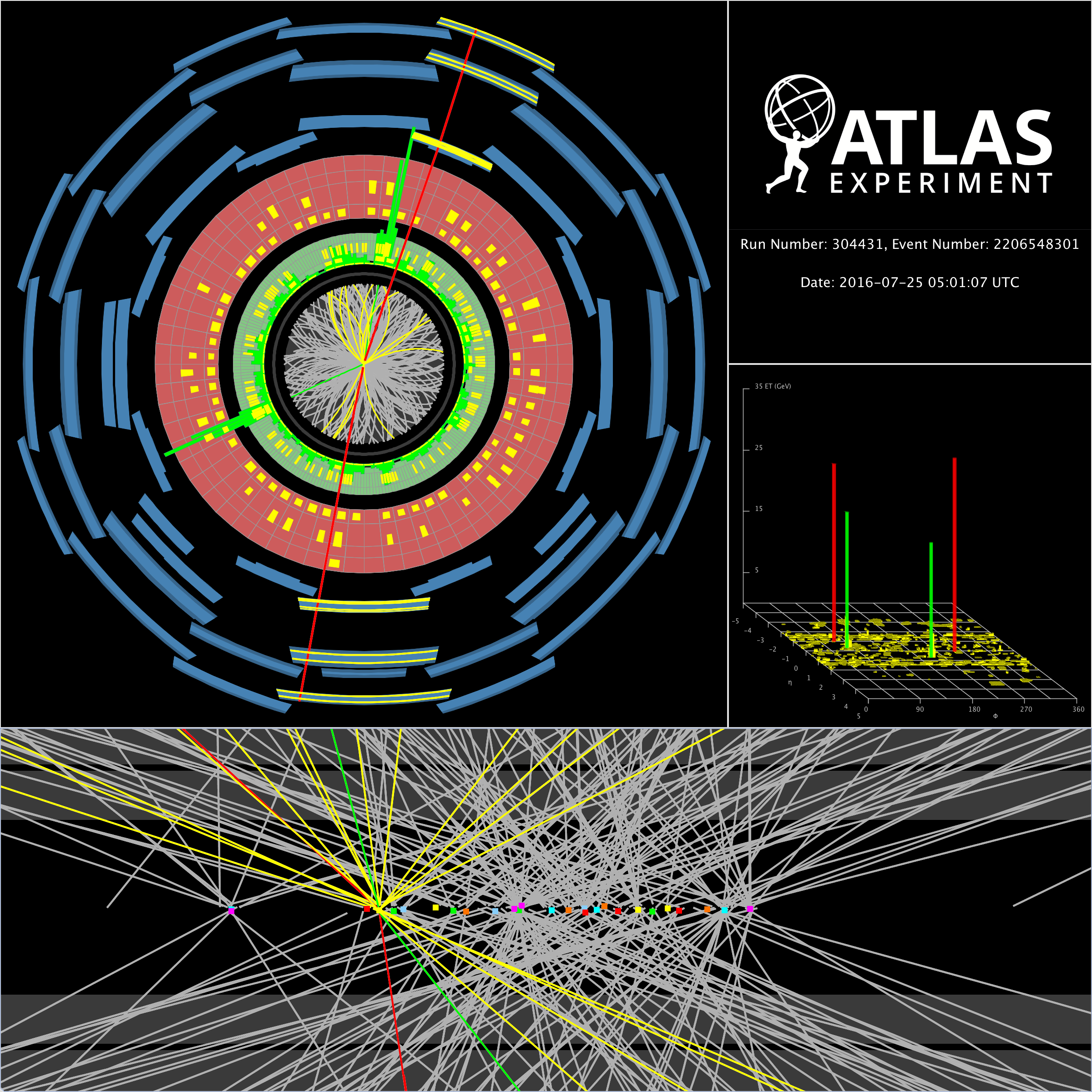}
\caption[Pile-up vertices in z.]{\label{fig:pile-upz} $H\rightarrow 2e2\mu$ candidate event with 25 additional reconstructed vertices recorded in 2016. Taken from~\cite{ATLAS_EVENT_DISPLAYS_RUN2}.}
\end{figure}

Instead of a global, collective, treatment of pile-up for the whole
jet, the individual particles within the jet can be classified to
whether they belong to the actual jet or to the underlying
pile-up. Charged particles leave tracks in high granularity tracking
detectors at the heart of multi-purpose detectors like ATLAS and CMS
and can be separated based on their longitudinal position $\hat{z}$ (along the beamline)
within the luminous region (see
figure~\ref{fig:pile-upz}). The \textit{charged hadron subtraction}
(CHS)~\cite{CMS-PAS-JME-14-001} method identifies each pile-up track individually. Used in
concert with particle flow concepts which attempt to identify each
particle in the event uniquely, CHS can effectively remove all charged
pile-up radiation from the event, including calorimeter signals that
are linked to tracks through the particle flow algorithm. Identification of pile-up jets, formed
predominantly from the energy of one or many pile-up vertices, is
another technique for removing pile-up using charged particles; by
determining the fraction of energy of the jet from the primary vertex,
one can distinguish such pile-up jets from the PV
jets~\cite{CMS-PAS-JME-13-005, Aad:2015ina}.  

The two methods discussed above can be
combined. First the more precise CHS method subtracts the pile-up
contribution from charged particles; in a second step, the remaining
contributions from neutral particles are removed with the area
subtraction method.  

In a more advanced approach, local, topological information is used,
as  QCD radiation from pile-up vertices is often uncorrelated and soft.
It and can thus be removed based on the local energy profile, i.e. if
the radiation is not consistent with hard scattering radiation from
the PV. This can be done in the transverse plane ${\eta,\phi}$ and
also as a function of radiation depth. The jet grooming
 technique is such an example to clean the jet of soft and wide-angle radiation which incidentally removes pile-up radiation. It is
discussed in more detail in section~\ref{sec:jetrec-groom}.  Topoclustering~\cite{atlas-topoclusters}, used by the ATLAS Collaboration, is deployed at the formation of clusters in the calorimeter requiring radiation to have a certain topological profile. In the forward region, where no tracking information is available, jet shapes and topological correlations can be used to identify pile-up~\cite{Aaboud:2017pou}.

While the above methods have been successfully deployed in the LHC experiments, they each have some deficiencies as well; ideally, one would hope to effectively combine all pile-up mitigation handles in order to maximally distinguish pile-up from PV radiation and to remove pile-up at the most granular level possible, i.e. at the particle or constituent level, in order to be as generic as possible. For example, while area subtraction is very effective for correcting the jet $\pt$, it is not used to mitigate the pile-up dependence of jet substructure observables as it is only able to correctly remove pile-up contributions on average. 
In fact, jet substructure variables are among the most difficult to
correct for pile-up because they are so reliant on radiation
profiles. A number of hybrid methods have been proposed operating at
the event constituent level. One example is the
PUPPI~\cite{Bertolini:2014bba} algorithm which is extensively used in
CMS. The PUPPI algorithm uses both event energy density and local
topological information incorporated in an event-by-event
particle-level discriminator to determine if a particle is from
pile-up. The algorithm defines a shape which attempts to distinguish parton shower-like radiation from pile-up-like
radiation. The shape is calculated from  \pt, angular distance to
nearby particles, and other information.    Particle four-vectors are
then weighted proportional to the value of the discriminator value.   Ideally, particles from the hard scatter
would get a weight of one and pile-up particles would get a weight of zero. Almost all pile-up
particles have values within a few standard deviations of the median and are assigned
small weights. Values that deviate far from the charged pile-up are
 indicative of a hard scatter, and these particles are assigned large
 weights. This weighting method allows for experimental information, such as
tracking, vertexing and timing information, to be included.

Other examples of such hybrid methods are Constituent
Subtraction~\cite{Cacciari:2014gra,Berta:2014eza, Bertolini:2014bba}, SoftKiller~\cite{Cacciari:2014gra}
and  PUMML~~\cite{Komiske:2017ubm}. Precursor hybrid methods include \textit{jets without jets}~\cite{Bertolini:2013iqa} and \textit{jet cleansing}~\cite{Krohn:2013lba}.  

\subsection{Performance Studies}

Pile-up removal algorithms are commissioned for use in ATLAS and CMS via detailed studies of jet observables in terms of the resolution and absolute scale, pile-up dependence, and the background rejection versus signal efficiency for boosted heavy particle taggers. 

For observables like jet $\pt$, dependencies on the number of reconstructed vertices and $\mu$ are observed even with area subtraction methods for the pile-up levels currently observed at the LHC, $\langle\mu\rangle \sim 25$. To correct for these effects, an additional residual correction is applied~\cite{atlas-jes-13tev,Khachatryan:2016kdb}. Enhancements are also possible from combining area subtraction methods with e.g. CHS. 

\begin{figure}[t]
\centering
\includegraphics[width=0.45\textwidth]{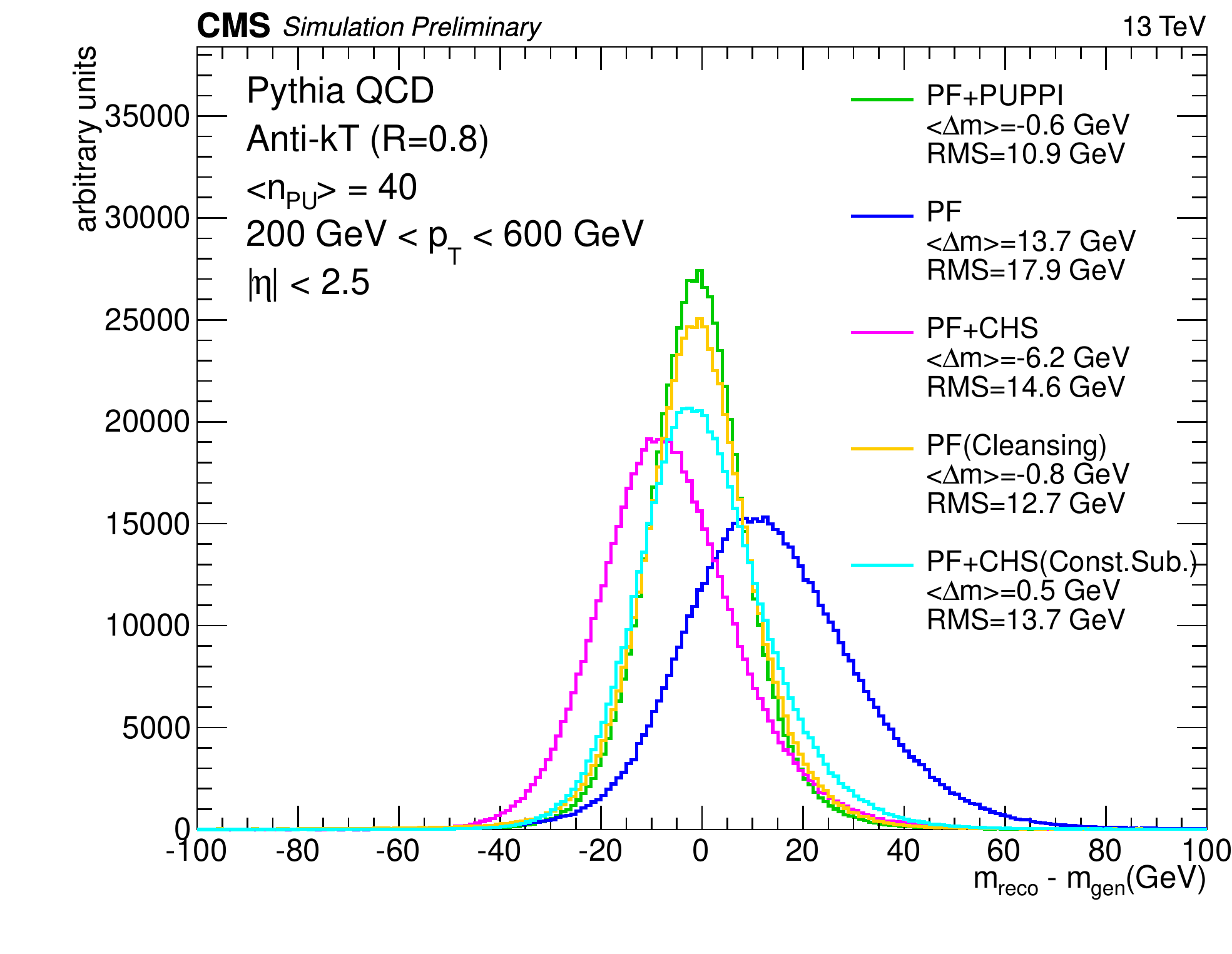}
\caption{\label{fig:pu_algos} Comparison of different pile-up removal algorithms for the leading ungroomed jet mass response in simulated QCD multijet events with CMS. Taken from~\cite{CMS-PAS-JME-14-001}.}
\end{figure}

For jet substructure observables, particle- or constituent-level pile-up mitigation strategies have been shown to improve performance, especially in simulation studies for up to $\langle\mu\rangle \sim 40$. 
An example is given in figure~\ref{fig:pu_algos}, where the ungroomed jet mass of the leading jet in \pt 
in simulated QCD multijet events is corrected with different pile-up removal techniques. 
The jet mass resolution can be improved further when using a grooming algorithm. 
The effect of different pile-up removal techniques on the groomed jet mass depends 
however strongly on the choice of the grooming algorithm as discussed in detail in Refs.~\cite{CMS-PAS-JME-14-001,ATL-PHYS-PUB-2017-020}. 
The improved performance observed in simulation has also been verified in collision 
data~\cite{CMS-PAS-JME-16-003}.

Generally these techniques, particularly those which operate at particle-level, can also be used to improve performance of non-jet objects such as missing transverse energy and lepton isolation. 
In the latter case, where the energy in a small cone around the lepton is 
summed, pile-up mitigation techniques help to reduce the isolation's susceptibility to pile-up. 

Preliminary studies (detector configurations have not yet been finalized) into the application of these advanced hybrid techniques at the higher pile-up levels anticipated at the HL-LHC suggest that they are effective in the $\langle\mu\rangle = 140-200$ range~\cite{CMSCollaboration:2015zni,ATLAS-CONF-2017-065}.


\section{Jet Substructure Methods and Observables
\label{sec:substructuremethodsobservables}}
  
\subsection{Jet Grooming \label{sec:jetrec-groom}}

Jet grooming techniques have seen a particularly high level of interest from the experimental
and theoretical communities alike. 
Jet grooming is an additional `post-processing' treatment of large 
radius jets, an extra step used
to remove unwanted soft radiation and to allow the underlying hard
substructure associated with a two-prong (e.g. $W$ boson)
or three-prong (e.g. top quark) decay to be identified more efficiently. 

In particular, grooming is the systematic removal of radiation from within a jet, 
often targeting soft and wide angle radiation.  There are a variety of techniques 
and each one has tunable parameters which are chosen to suite the particular needs of the application.  
The three main algorithms used by ATLAS and CMS are trimming~\cite{Krohn:2009th}, 
pruning~\cite{Ellis:2009me}, and soft drop~\cite{Larkoski:2014wba}.  
In each of these cases, the constituents of a jet are re-clustered and soft/wide angle 
radiation is rejected in this process. 
For trimming, the \kt algorithm is used to re-cluster and 
the radius parameter of the re-clustering is called $\rsub$.  
Those smaller-radius jets with a momentum fraction $f<\fcut$ are removed to produce the trimmed jet.  
The two other algorithms impose a condition on each $2\rightarrow 1$ clustering step, 
by going backwards in the sequence in which the particles were combined in the 
re-clustering. 
The transverse momentum fraction of the softer particle to the merged system, 
$z=\min(p_{\mathrm{T},1}, p_{\mathrm{T},2}) / (p_{\mathrm{T},1} + p_{\mathrm{T},2})$, 
is a natural choice for determining the scale of the soft radiation, 
and the angular distance $\Delta R$ between the two particles for 
identifying wide-angle radiation. 
The difference between pruning and soft drop lies in the way how particles 
and their combinations get rejected based on the values of $z$ and $\Delta R$. 
For pruning, the softer particle of the $2\rightarrow 1$ clustering step is discarded if 
$z < \zprune$ and $\Delta R < \dcut$. For soft drop, the softer particle is 
discarded if $z< \zsoftdrop(\Delta R/R)^\beta$, where $\zsoftdrop$ and the angular 
exponent $\beta$ are free parameters\footnote{
Most applications of soft drop use $\beta=0$, in which case it is equivalent 
to an earlier algorithm known as modified mass drop tagger (mMDT)~\cite{Dasgupta:2013ihk}.  
Since both collaborations call this soft drop, we also refer to the algorithm by this 
name, but encourage the users to cite the mMDT publication in addition to the soft drop one.
}.

The role of grooming has traditionally satisfied two purposes in ATLAS, 
being the mitigation of pile-up effects on jets, and the removal of soft/wide-angle radiation. 
The particle flow algorithm employed in CMS in conjunction with CHS or PUPPI 
allows for a correction for pile-up effects. This reduces the usefulness of grooming 
for pile-up mitigation, but retains its advantage for the removal of soft/wide-angle radiation.

ATLAS performed a broad study of the relative performance of different grooming techniques for boson-tagging~\cite{Aad:2015rpa,ATL-PHYS-PUB-2015-033,ATL-PHYS-PUB-2017-020}, top-tagging~\cite{Aad:2016pux,ATL-PHYS-PUB-2015-053} and SM measurements~\cite{Aaboud:2017qwh,Aad:2014haa}, using the removal of pile-up-dependence, the jet mass resolution, and the tagging efficiency versus background rejection as performance metrics.  The `standard' grooming procedure adopted by ATLAS is trimming with \fcut = 0.05 for boson tagging in both \runone ($\rsub=0.3$) and \runtwo ($\rsub=0.2$). The trimming algorithm with the same parameters was adopted for top tagging, along with several other techniques (see section~\ref{sec:TopTagging}). Another technique currently in use by ATLAS is the \textit{reclustering} of small-$R$ jets~\cite{Nachman:2014kla}, which uses fully-calibrated \antiktfour jets as inputs to the \antikt algorithm with a larger distance parameter (typically $R=1.0$). This has proven a popular method in ATLAS analyses due to the flexibility of optimizing the jet distance parameter depending on the considered phase-space of the analysis~\cite{Aad:2016eki, Aaboud:2016lwz, Aaboud:2017hdf}. A recent study of \textit{in-situ} measurements~\cite{ATLAS-CONF-2017-062} (including `closeby' effects) confirm that the differences between data and simulation observed with reclustered jets are indeed covered by simply propagating the uncertainties associated with the input \antiktfour jets.

 CMS studied a large number of grooming techniques in the context of
 boosted boson-tagging~\cite{Khachatryan:2014vla,CMS-PAS-JME-14-002},
 top-tagging~\cite{CMS-PAS-JME-13-007, CMS-PAS-JME-15-002} and SM
 measurements~\cite{Chatrchyan:2013vbb, Sirunyan:2018xdh}. During
 \runone the grooming techniques were used together with
 charged-hadron subtraction for pile-up mitigation (see
 section~\ref{sec:pile-up}). All groomers studied showed reasonable or
 good agreement between data and simulation and the
 pruning algorithm ($R=0.8$, \zprune = 0.1 and
 \dcut = 0.5) showed the best performance for boson
 tagging~\cite{Khachatryan:2014vla}. 
 For \runtwo, soft drop (\zsoftdrop = 0.1 and $\beta$ = 0) is used for jets with $R = 0.8$ in jet substructure analyses in CMS together with the pile-up removal algorithm PUPPI~\cite{Bertolini:2014bba} (see section~\ref{sec:pile-up}).  Soft drop jets in combination with PUPPI show a similar performance as pruning when comparing signal efficiency versus background rejection~\cite{CMS-PAS-JME-16-003,CMS-PAS-JME-15-002}, but allow for better theoretical control.
  While grooming techniques were found to improve the performance
  (higher background rejection at fixed signal efficiency) of the jet
  mass, $N$-subjettiness ratios~\cite{Thaler:2010tr, Thaler:2011gf} were found to perform better without
  grooming for boosted boson tagging~\cite{Khachatryan:2014vla}. 
For top-tagging applications, however, soft drop groomed
 $N$-subjettiness ratios improved the performance with respect to
 ungroomed ones for jets with $\pt < 400$~\GeV. For higher \pt jets
 there was no significant gain observed
 with grooming  for $N$-subjettiness ratios~\cite{CMS-PAS-JME-15-002}.

\subsection{Jet Mass}\label{sec:jetrec-mass}

The reconstruction of jet energies mainly relies on the capability of a detector to measure the total energy of all particles deposited in the detector; however, the measurement of jet mass requires detection of the deposited energy with a granularity that is finer than the size of a jet. The mass of a jet can only be estimated if the energy is deposited in at least two detector elements, as it depends on both the energy and opening angle between the jet constituents. For jet substructure techniques that rely on the rejection of soft particles, it is also important to be able to reconstruct particles with low \pt separately from harder particles in a jet.

The jet mass \textit{response} distribution $R_\text{reco}$ is constructed from the calibrated, reconstructed jet mass $M_\text{reco}$ divided by the particle-level jet mass $M_\text{true}$. The mass response distribution is calculated in bins of reconstructed jet $p_{\mathrm{T}, \text{reco}}$ and $\eta_\text{reco}$.
In ATLAS, the Jet Mass Scale (JMS) is defined as the mean of this response distribution. The Jet Mass Resolution (JMR) is then defined as half the 68\% interquartile range (IQnR) of the response distribution, as

\begin{align}
\label{eq:atlas:jmr}
r = 0.5 \times 68\%\; \mathrm{IQnR} (R_\text{reco}).
\end{align}
This is robust to large non-Gaussian tails but, if the distribution is Gaussian, is equal to its 1$\sigma$ width. The fractional JMR is expressed as the JMR divided by the \textit{median} of the response distribution.

\begin{figure}[tb]
\includegraphics[width=0.45\textwidth]{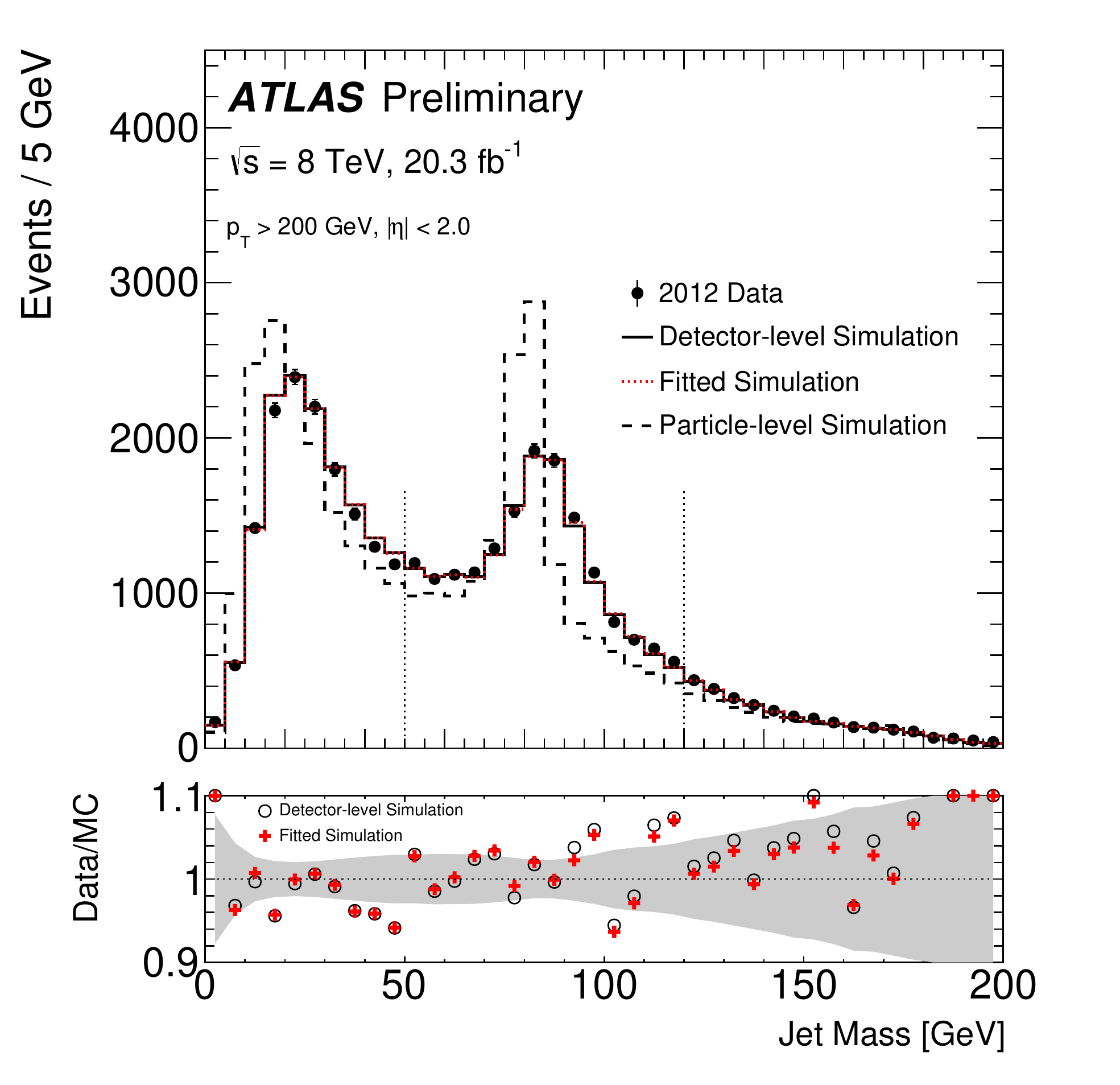}
\caption{The trimmed jet mass before (detector-level) and after (fitted) determining $s$ and $r$. The particle-level distribution is shown for comparison. Jets are required to have $\pt>200$ GeV. Adapted from Ref.~\cite{ATLAS-CONF-2016-008}.}
\label{fig:atlas:JMSJMR1}
\end{figure}

ATLAS has recently developed a data-driven approach to extract the JMS and JMR from an enriched sample of boosted \ttbar events, however the method can also be extended to other final states. This \textit{forward-folding} approach folds the particle-level mass spectra by a modified response function such that the JMS in a given bin of particle-level jet mass and reconstructed jet \pt is scaled by the scale parameter $s$ and the JMR is scaled by the resolution parameter $r$: 
\begin{align}
\label{eq:atlas:forwardfold}
M_\text{fold} = s \times M_\text{reco} + ( M_\text{reco} - \langle M^{m,\pt}_\text{reco} \rangle)(r-s)).
\end{align}
The values of $r$ and $s$ for which the $M_\text{fold}$ distribution best matches the data are extracted from a 2 dimensional $\chi^2$ fit as shown in figure ~\ref{fig:atlas:JMSJMR1} and detailed in Ref.~\cite{ATLAS-CONF-2016-035,ATLAS-CONF-2016-008}. 

\begin{figure}[tb]
\includegraphics[width=0.45\textwidth]{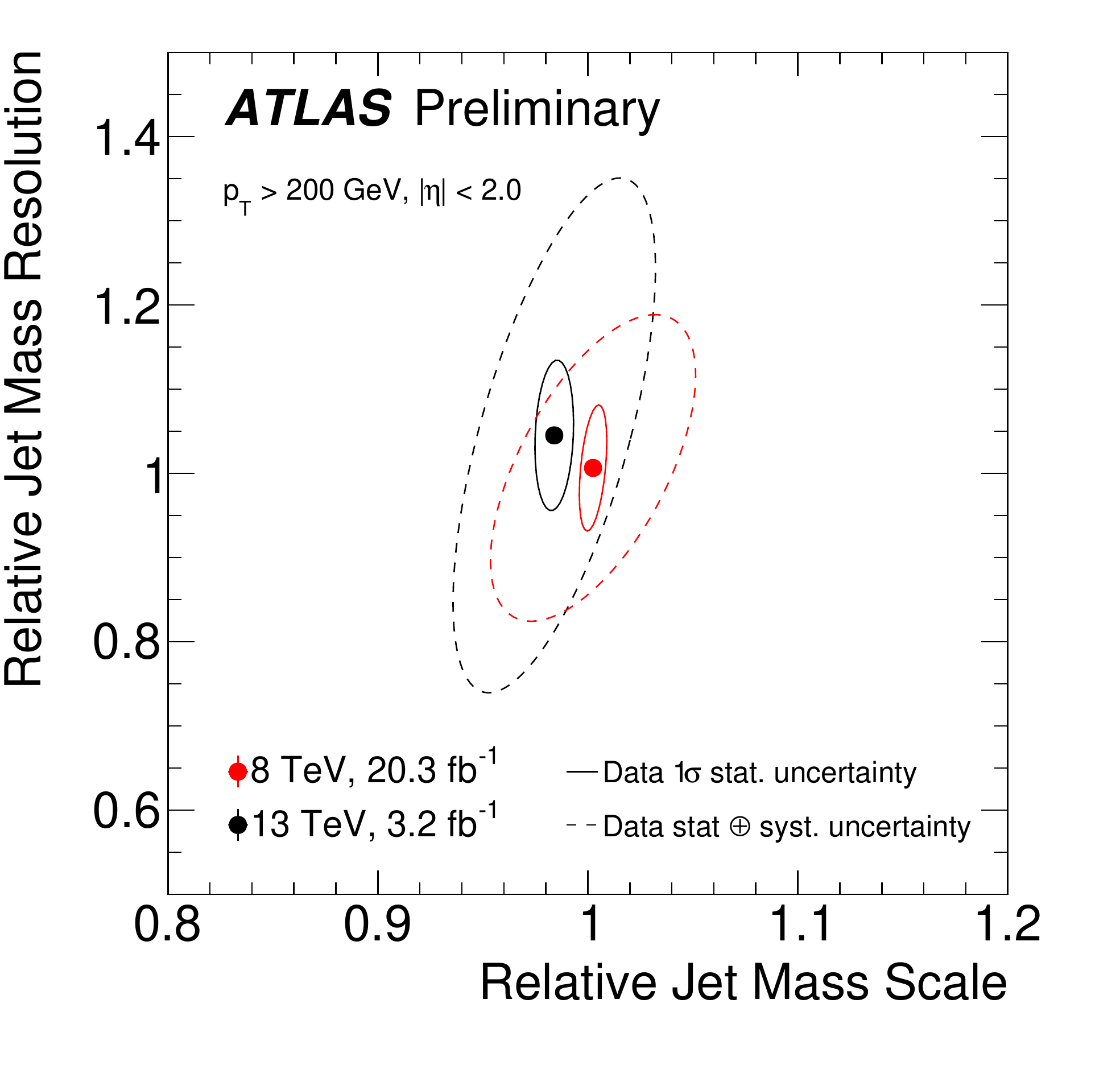}
\caption{The fitted values of the relative jet mass scale ($s$) and resolution ($r$) for trimmed \antiktten calorimeter jets from the 2012 and 2015 ATLAS datasets and the $1\sigma$ statistical and total uncertainty ellipses. The $\sqrt{s}=8$ and $13$ TeV selections are similar, although the trimming definition slightly changed between Runs (\rsub = 0.3 to \rsub = 0.2). Adapted from Ref.~\cite{ATLAS-CONF-2016-035}.}
\label{fig:atlas:JMSJMR2}
\end{figure}

With the forward-folding approach, the JMS and JMR for hadronically decaying boosted $W$ bosons with $\pt \gtrsim 200$ GeV are determined with 2--3\% and 20\% systematic uncertainties, respectively (see figure~\ref{fig:atlas:JMSJMR2}). As the jet mass and its detector-response depend on kinematics and jet substructure, the measurement was repeated differentially with an increased luminosity for boosted $W$ and top quarks in Ref.~\cite{ATLAS-CONF-2017-063}. It will be important to extend the technique to other final states in the future.  This may require hybrid data/simulation methods.  A detailed study of the various contributions to the JMS and JMR has been performed in context of the soft drop mass measurement~\cite{Aaboud:2017qwh}, described in section~\ref{sec:jetmassmeas}, by propagating experimental uncertainties on the inputs to the jet reconstruction to the jet mass. The dominating uncertainties are due to the theoretical modeling of jet fragmentation and the cluster energy scale.

As the forward-folding method is currently restricted to jets with $\pt < 350$ and 500\GeV for boosted $W$ bosons and top quarks, respectively, the results are combined with the so-called $R_{\rm trk}$ method which constrains the mass scale by comparing the calorimeter jet mass to the mass calculated from track jets and extends up to $\pt = 3000$\GeV~\cite{ATLAS-CONF-2017-063}. The $R_{\rm trk}$ method can also be generalized to other variables and is used in ATLAS to constrain the \pt scale of large-$R$ jets as well as to derive systematic uncertainties on jet substructure variables. 

\begin{figure}[tb]
\centering
\includegraphics[width=0.49\textwidth]{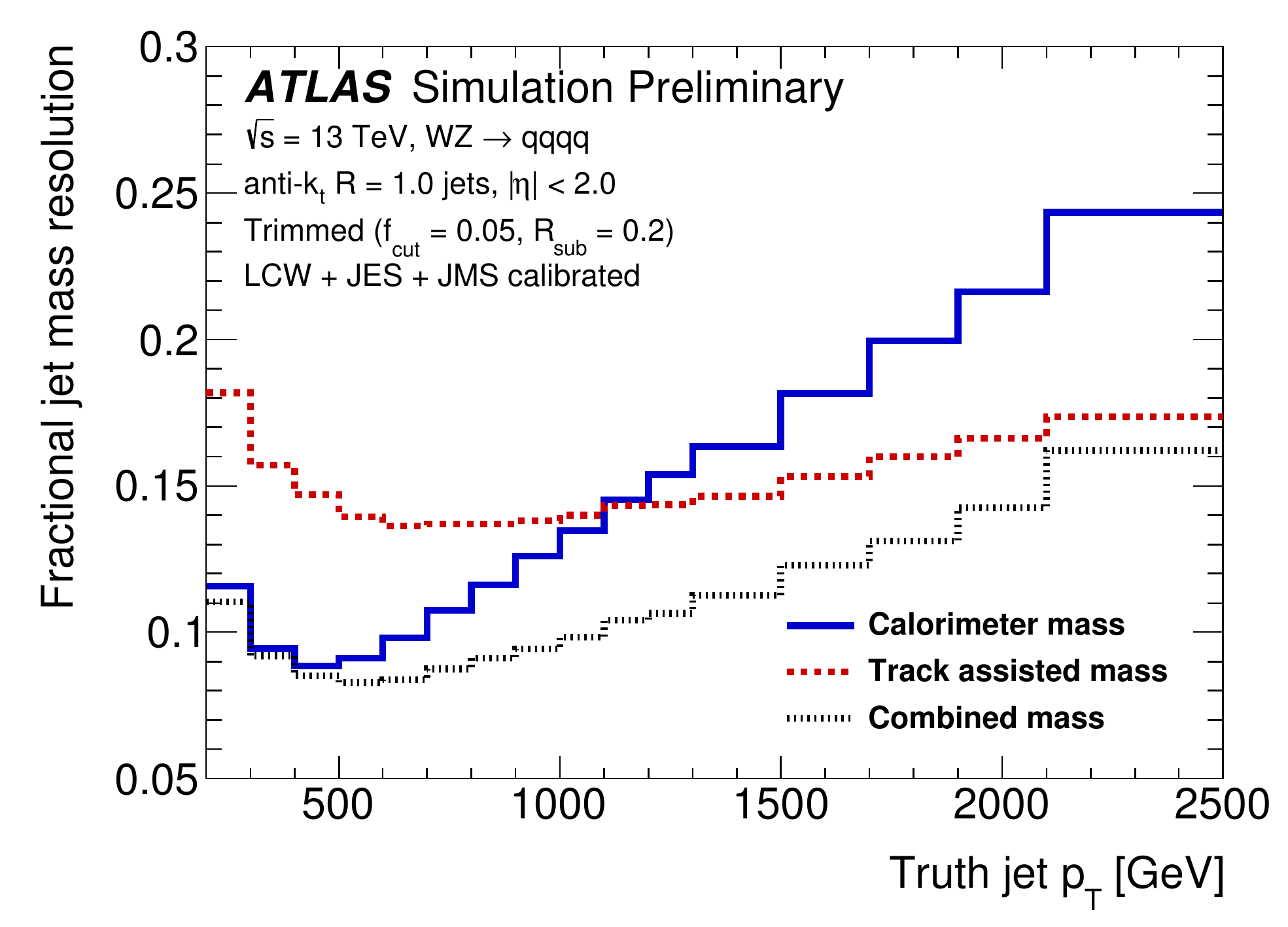}
\caption[ATLAS combined mass]{The ATLAS combined jet mass resolution. Adapted from~\cite{ATLAS-CONF-2016-035}.}
\label{fig:atlas-jes-massres}
\end{figure}

The concept of a \textit{Track-Assisted Mass}  for trimmed, large-$R$ jets has 
been studied in ATLAS \cite{ATLAS-CONF-2016-035} to maintain performance for highly boosted particles due to the limited granularity of the calorimeter. The track-assisted mass is defined as:
\begin{align}
m^{\mathrm{TA}} = \frac{\pt^{\mathrm{calo}}}{\pt^{\mathrm{track}}} \times m^{\mathrm{track}},
\end{align}
where $\pt^{\mathrm{calo}}$ is the transverse momentum of the calorimeter jet, $\pt^{\mathrm{track}}$ is the transverse momentum of the four-vector sum of tracks associated to the calorimeter jet, and
$m^{\mathrm{track}}$ is the invariant mass of this four-vector sum, where the track mass is set to the pion mass $m_{\pi}$. The track-assisted mass exploits the excellent angular resolution of the tracking detector and the ratio $\pt^{\mathrm{calo}}$ to $\pt^{\mathrm{track}}$ corrects for charged-to-neutral fluctuations. The \textit{Combined Mass} is defined as:
\begin{align}
m^{\mathrm{comb}} = \left( \frac{ \sigma_{\mathrm{calo}}^{-2} }{ \sigma_{\mathrm{calo}}^{-2}+\sigma_{\mathrm{TA}}^{-2} } \right)m^{\mathrm{calo}} + \left( \frac{ \sigma_{\mathrm{TA}}^{-2} }{ \sigma_{\mathrm{TA}}^{-2}+\sigma_{\mathrm{calo}}^{-2} }  \right)m^{\mathrm{TA}},
\end{align}
where $\sigma_{\mathrm{calo}}$ and $\sigma_{\mathrm{TA}}$ are the calorimeter-based jet mass resolution and the track-assisted mass resolution, respectively. The jet mass resolution for the calorimeter mass, track-assisted mass and combined mass are shown in figure~\ref{fig:atlas-jes-massres} for $W/Z$ boson jets as a function of jet \pt. Similar techniques that take advantage of the excellent angular resolution of the tracking detector at high \pt have been developed to correct topoclusters to improve the resolution of jet substructure variables~\cite{ATL-PHYS-PUB-2017-015}.

It is important to point out that in ATLAS unlike in CMS, the jet energy scale directly impacts the jet mass scale. As opposed to the description of the JES calibration for small-$R$ jets in section~\ref{sec:jetrec-jes}, the area subtraction, residual correction and Global Sequential Calibration (GSC) (see section~\ref{sec:quarkgluon}) are not applied to large-$R$ jets. 

In CMS, the jet mass is by default reconstructed as a combination of track and calorimeter measurements via the virtues of the particle flow algorithm. Thus the strategy for calibrating the jet mass in CMS differs from the one in ATLAS. In CMS, the individual PF objects are input to the jet reconstruction, and are locally calibrated to account for the detector's single particle response (see section~\ref{sec:jetrec-rec}). After correcting the individual inputs, the jet four-vector is corrected using JES corrections and small residual differences in the jet mass between data and simulation are corrected using dedicated samples.

\begin{figure}[tb]
\centering
\includegraphics[width=0.45\textwidth]{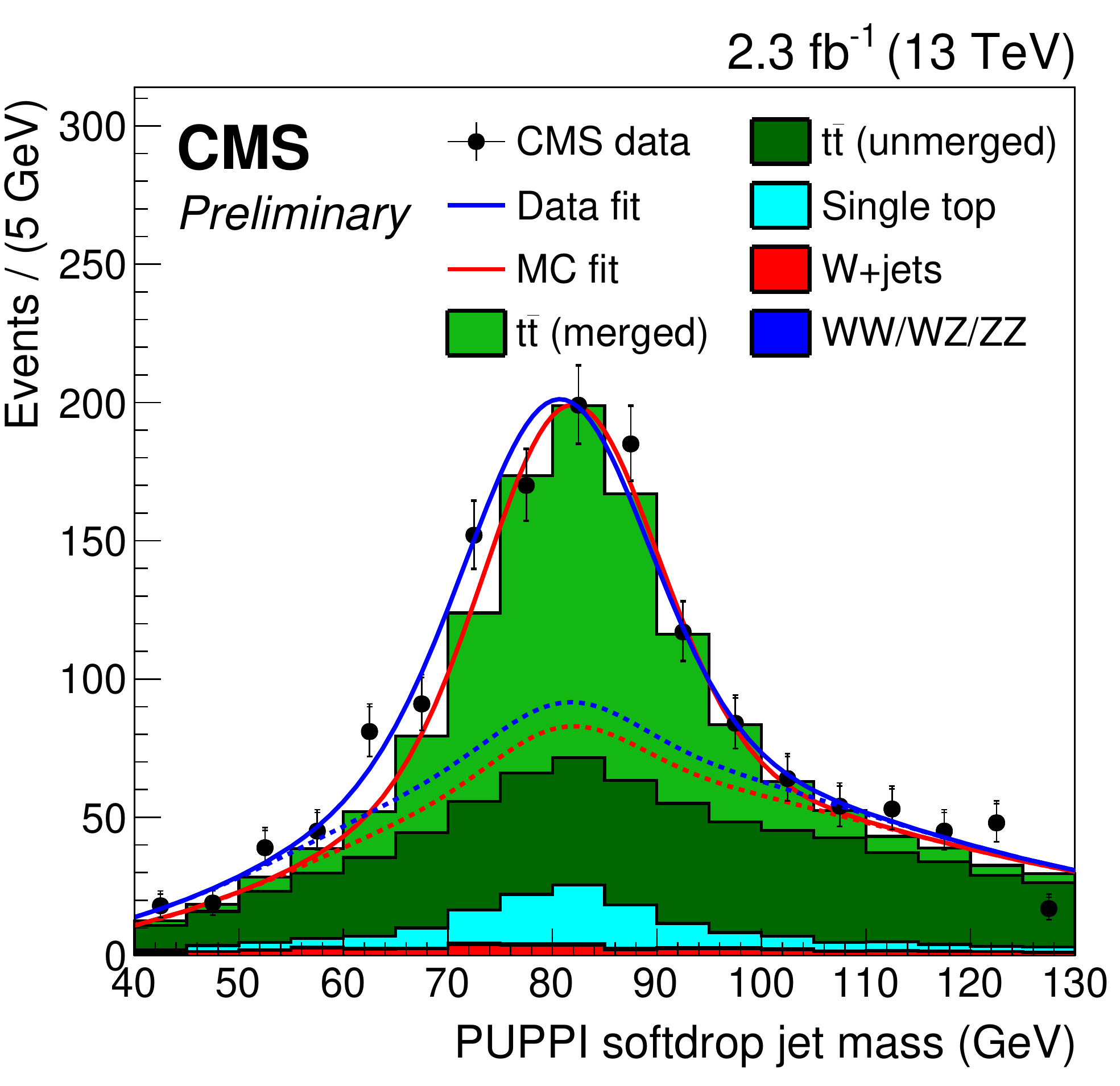}
\caption{\label{ttCMS} Jet mass distribution in a sample enriched with lepton+jets \ttbar events, where the hadronic $W$ jet with $\pt>200\GeV$ is selected, taken from Ref.~\cite{CMS-PAS-JME-16-003}.}
\end{figure}
The residual \textit{in-situ} jet energy corrections are not applied when reconstructing jet masses. Therefore, dedicated corrections are derived from simulation and data. Firstly, the jet mass response is corrected as a function of $\pt$ and $\eta$ using simulation of $W$ jets from boson pair production. Secondly, residual corrections are obtained from a data sample enriched in lepton+jets $t\bar{t}$ production where the hadronic $W$ jet can be studied in data~\cite{Khachatryan:2014vla, CMS-PAS-JME-16-003}. 
The selection is optimized for fully-merged hadronic $W$ decays. Large-$R$ jets in this sample show a peak at the $W$ mass in the jet mass distribution, as shown in figure~\ref{ttCMS} for the soft drop grooming case. The excellent performance of the PF algorithm results in a JMR of about 10\%. The absolute response and the resolution are well described by the simulation, within 1--2\% for the JMS and about 10\% for the JMR, which is about the same size as the statistical uncertainty of this measurement.
Residual differences in this distribution are used to calibrate the JMS and JMR in simulation, and can also be used for dedicated efficiency corrections on other jet substructure observables, such as the $N$-subjettiness ratio $\tau_{21} = \tau_2/\tau_1$.

Since these measurements are performed in samples of $W$ jets with $\pt \approx 200$~\GeV, additional systematic uncertainties apply at higher $\pt$~\cite{Sirunyan:2017acf}. A detailed study of the various contributions to the JMS has also been performed for fully merged top-jets in the context of an unfolded top-jet mass measurement~\cite{Sirunyan:2017yar}.
To summarize the impact of the various sources of systematic uncertainty to the measurement of residual corrections for jet substructure observables, we quote here the dominant uncertainties related to the scale factor measurement of an $N$-subjettiness ratio $\tau_{21}<0.4$ selection~\cite{CMS-PAS-JME-16-003}. The statistical uncertainty of 6\% (with 2.3/fb of data) is comparable to the systematic uncertainties related to the simulation of the $t\bar{t}$ topology (nearby jets, \pt spectrum) contributing 4\%, the choice of method to derive the scale factors contributing 6\% and the modeling of the $\pt$ dependence that rises from 5\% at $\pt = 500$~\GeV to 13\% at $\pt = 2000$~\GeV.

\begin{figure}
\centering
\includegraphics[width=0.49\textwidth]{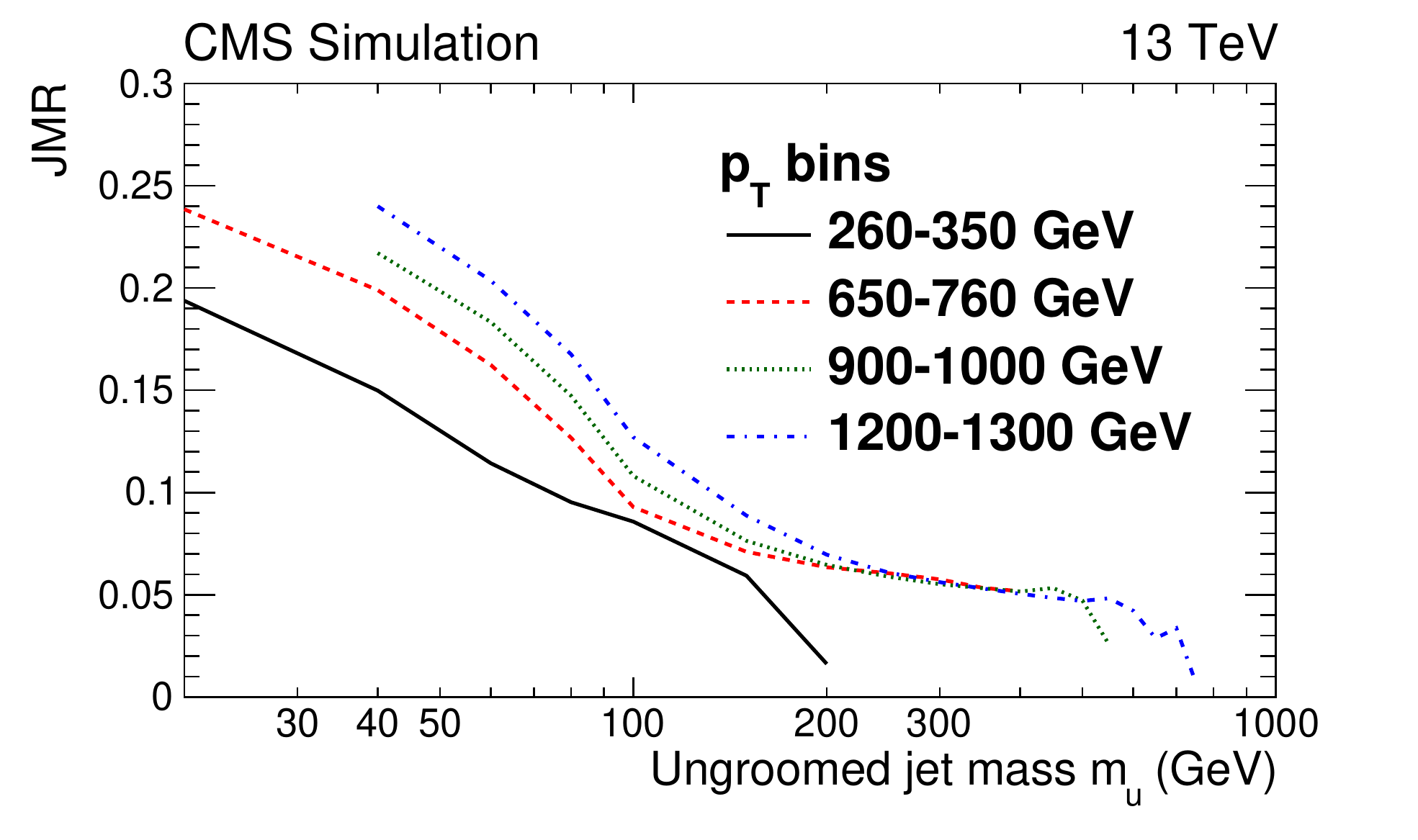}
\caption[CMS ungroomed mass]{The CMS jet mass resolution as a function of the 
ungroomed jet mass $m_\text{u}$ in different generated \pt bins. Adapted from~\cite{Sirunyan:2018xdh}.}
\label{fig:cms-massres}
\end{figure}
The relative JMR in CMS is shown in figure~\ref{fig:cms-massres} as a function of the ungroomed jet mass $m_\text{u}$ for \antikteight jets. The JMR is obtained from a sample of jets initiated by quarks and gluons. The resolution improves with increasing $m_\text{u}$ and is around 9--13\% for the most probable value of $m_\text{u} \approx 100$--$150\GeV$. For a given value of $m_\text{u}<200\GeV$, the resolution worsens with increasing jet \pt due to a higher degree of collimation. Remarkably, the resolution obtained in CMS is comparable to the one for the combined mass in ATLAS (figure~\ref{fig:atlas-jes-massres}), even though quark/gluon jets are compared with $W$/$Z$-jets and very different technologies are used to reconstruct the jet mass.

\subsection{Other Jet Substructure Observables}

Additional jet substructure observables are used for a variety of purposes, often to complement the jet mass.  Most uses of these observables are within the context of a dedicated tagger, described in the next section.  These observables can generally be classified into two categories: prong-taggers and haze-taggers.  The most widely used prong-taggers are the $N$-subjettiness ratios $\tau_{ij}^\beta$~\cite{Thaler:2010tr, Thaler:2011gf}, $C_2^\beta$~\cite{Larkoski:2013eya}, $D_2^\beta$~\cite{Larkoski:2014gra, Larkoski:2015kga}, and $N_2^\beta$~\cite{Moult:2016cvt}.  The latter three are ratios of energy correlation functions, which are sums over constituents inside jets weighted by the momentum fractions and pairwise opening angles to the power $\beta$.  For example, 
\begin{align}
N_2=\frac{{}_2e_3^{(\beta)}}{({}_1e_2^{\beta})^2},
\end{align}
where 
\begin{align}
{}_1e_2^{(\beta)}&=\sum_{1\leq i < j < k \leq n_J}z_iz_jz_k\min\left\{\Delta R_{ij}^\beta,\Delta R_{ik}^\beta,\Delta R_{jk}^\beta\right\}\\\nonumber
{}_2e_3^{(\beta)}&=\sum_{1\leq i < j < k \leq n_J}z_iz_jz_k\\
&\vspace{15mm}\times\min\left\{\Delta R_{ij}^\beta\Delta R_{ik}^\beta,\Delta R_{ij}^\beta\Delta R_{jk}^\beta,\Delta R_{ik}^\beta\Delta R_{jk}^\beta\right\},
\end{align}
where the sums run over the $n_J$ jet constituents with momentum fractions $z_i$ and opening angles $\Delta R_{ij}$.

The goal of haze-taggers is to generally characterize the radiation pattern within a jet without explicitly identifying the number of prongs.  The prong-taggers also are sensitive to the distribution of radiation around the subjet axes and so the distinction is not strict.  Popular haze-taggers include jet width, $n_\text{constituents}$ (or $n_\text{tracks}$), and $p_T^D$.

In applications of jet substructure taggers based on these variables the description of data by simulation is a crucial aspect. Differences in the distributions lead to differences in efficiencies and misidentification rates, which need to be quantified in dedicated measurements. Measurements of jet substructure observables, their calibration, and improving their description by adjusting free parameters in event generators is an important step in every analysis. 

\begin{figure}[tb]
\includegraphics[width=0.98\linewidth]{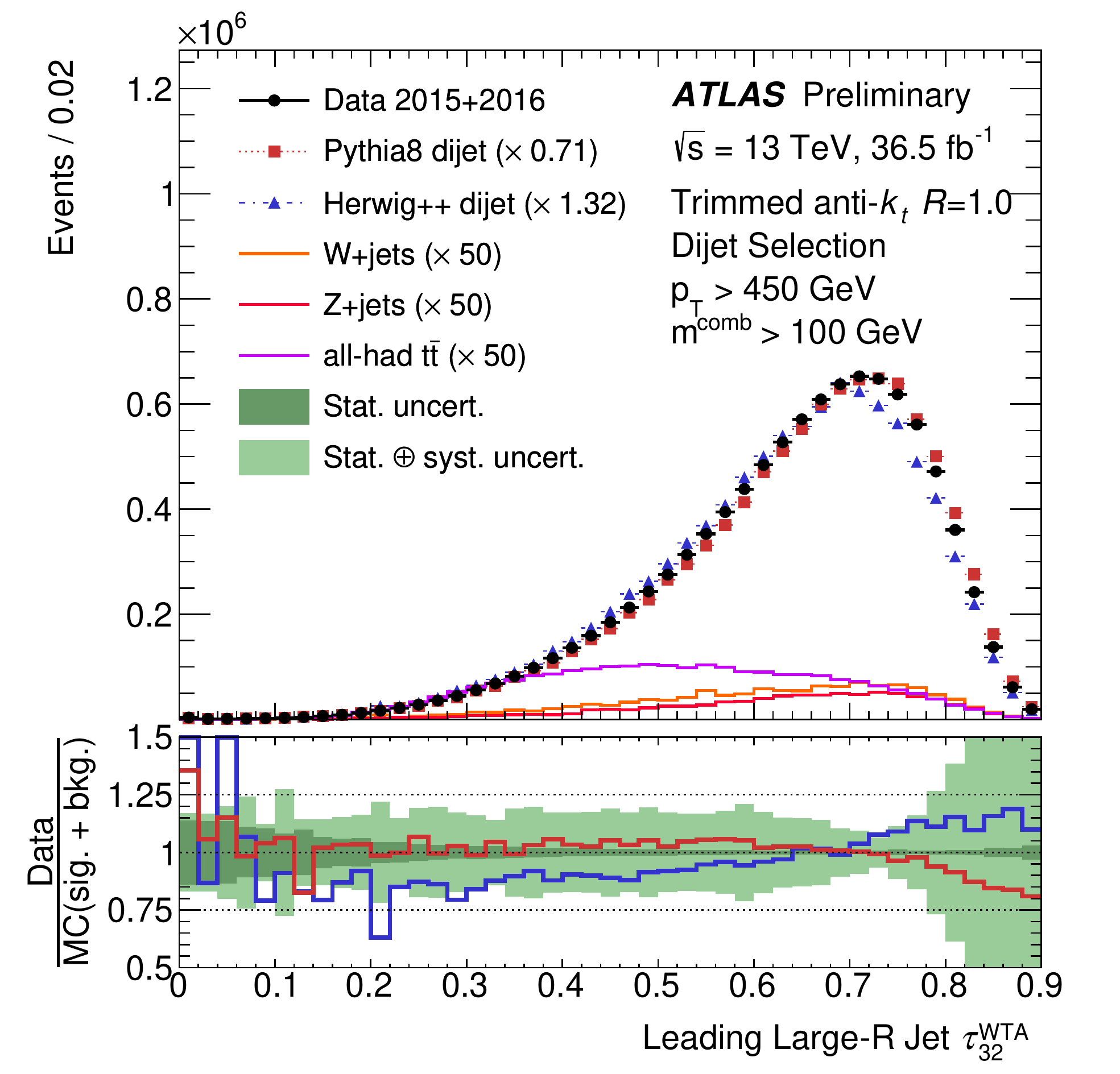}
\caption{Measured distribution of the $N$-subjettiness ratio $\toptau$ calculated on trimmed \antiktten jets for a dijet selection with $\pt>450\GeV$ and $\pt>200\GeV$ for the leading and sub-leading jet, respectively. The data are compared to simulated events, where the dijet samples have been normalized to the signal-subtracted data. Taken from Ref.~\cite{ATL-JETM-2017-005}. }
\label{fig:atlas_tau32}
\end{figure}

\begin{figure}[tb]
\includegraphics[width=0.98\linewidth]{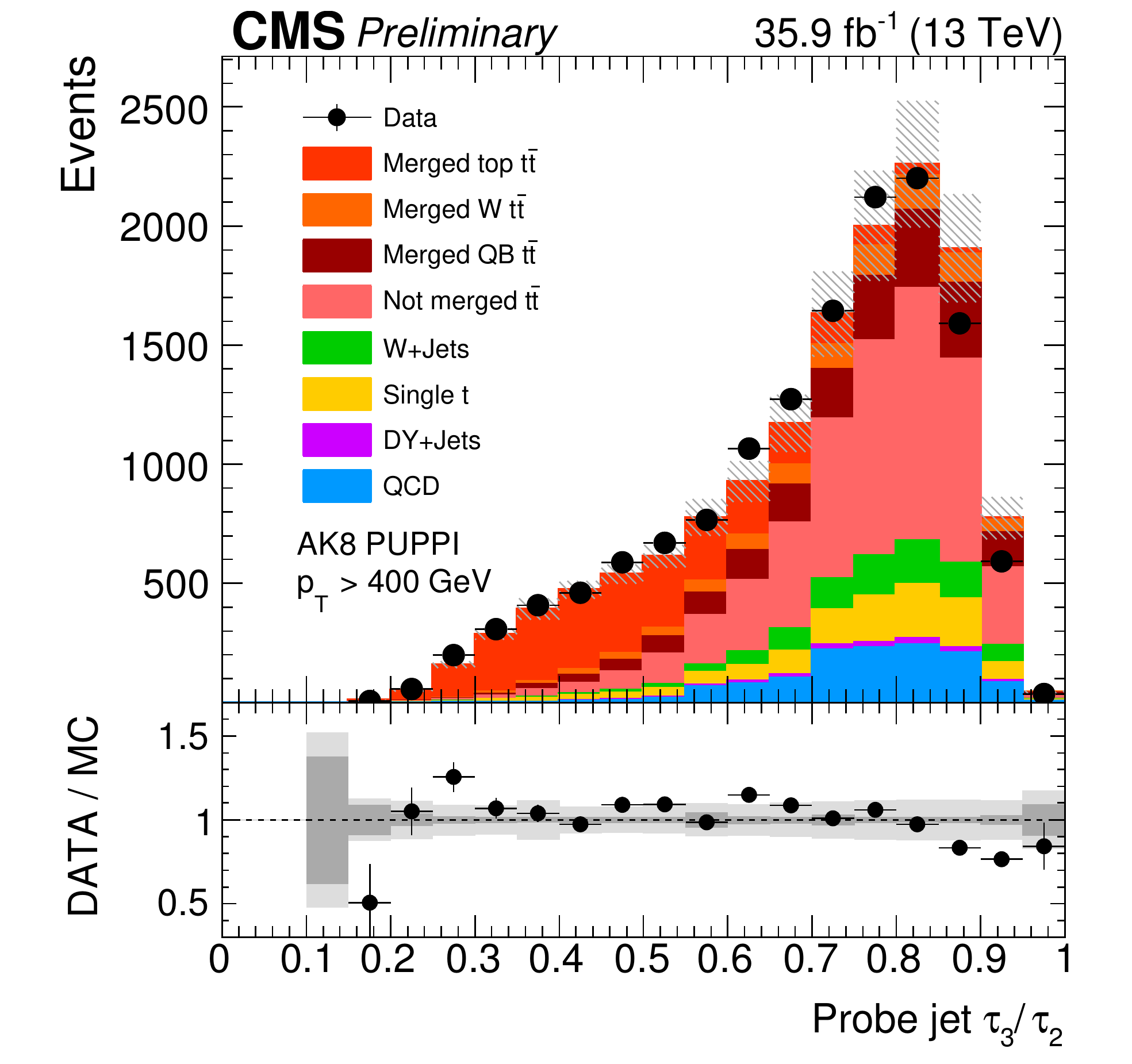}
\caption{Measured distribution of the $N$-subjettiness ratio $\toptau$ calculated on \antikteight jets with $\pt>400\GeV$ corrected with PUPPI in a \ttbar sample. The data are compared to simulated events, where the ``Merged QB'' \ttbar contribution consists of events in which the $b$ quark from the top quark decay and just one of the quarks from the $W$ boson decay are clustered into the jet. Taken from Ref.~\cite{CMS-DP-2017-026}. }
\label{fig:cms_tau32}
\end{figure}

As an example for three-prong taggers, the $N$-subjettiness ratio $\toptau = \tau_3 / \tau_2$ for $\beta=1$ is shown here. It is used in ATLAS and CMS for top tagging and studied in light quark and gluon jets from dijet production, as well as in fully-merged top-quark jets from dedicated \ttbar samples. The distribution of $\toptau$ with \runtwo data is shown in figure~\ref{fig:atlas_tau32} for a dijet selection and in figure~\ref{fig:cms_tau32} for a \ttbar selection. Overall good agreement between data and simulation is observed, which leads to data-to-simulation scale factors for top-tagging compatible with unity~\cite{CMS-DP-2017-026}.

\begin{figure}[tb]
\includegraphics[width=0.98\linewidth]{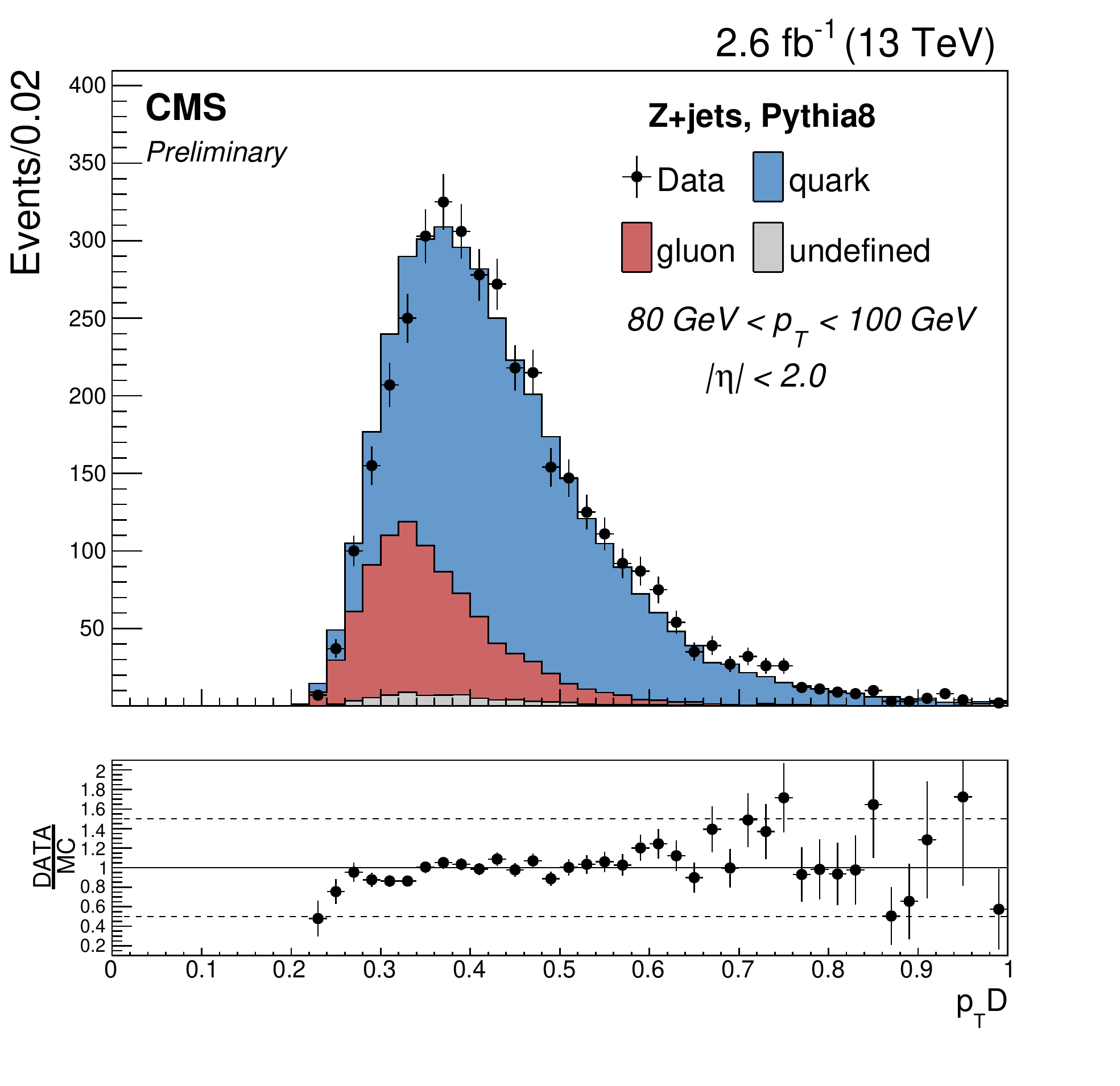}
\caption{Distribution of $p_T^D$ calculated on \antiktfour jets with $80 < \pt < 100\GeV$ a $Z$+jets sample. Taken from Ref.~\cite{CMS-PAS-JME-16-003}. }
\label{fig:cms_ptD}
\end{figure}

As an example for an haze-tagger distribution, the $p_T^D$ distribution is shown in Fig.~\ref{fig:cms_ptD}. 
The distribution from $Z$+jets production is well described by simulation, but a significant discrepancy is observed when selecting dijet events. This has important consequences for quark/gluon tagging, 
where dedicated template fits to data are performed to extract weights to correct the simulation (see section~\ref{sec:quarkgluon}). 
Similar conclusions are found for the jet width and constituent multiplicity distributions~\cite{CMS-PAS-JME-16-003}.

\section{Jet Tagging}\label{sec:tagging}

Particle identification is an experimental challenge that is traditionally met using custom-designed charged-particle detectors, muon chambers and calorimeters with granularity fine enough to allow shower shape measurements. Particle identification played an important 
role in the design considerations for the ATLAS and CMS detectors. 
Jet substructure techniques used for the identification of the particle origin of jets are a recent development, though. 
Several substructure variables have been developed by the theoretical community that can be used along with the jet mass for jet classification. The term `tagger' indicates the use of one or more of these variables (sometimes after grooming has been applied) to discriminate between jets coming from different types of particles. 

A rule of thumb for the decay of a
massive object such as a $W/Z/H$ boson is that the decay
products lie within a cone of radius $\Delta R = 2M / \pt$ in the laboratory rest frame,
where $M$ and \pt are the mass and transverse momentum of the
object\footnote{
Note that this rule of thumb gives only a lower bound on $\Delta R$, and it strictly holds only for 
two-body decays with massless decay products and $\pt \gg M$.}.
Using this for the example of a $W$ boson decay,
a $W$ boson with $\pt = 200\GeV$ will have its decay products captured
by a jet with a distance parameter of at least $0.8$, and the higher the \pt of the $W$ boson, the
more collimated the decay products. 
For top quarks, the value of $\pt$ for which all decay products are captured 
by a jet with $R=0.8$ is at least $400\GeV$.

\subsection{Quark/Gluon Discrimination}\label{sec:quarkgluon}

Since the first algorithmic definitions of jets, jet substructure observables have been widely used for quark-initiated (quark) versus gluon-initiated (gluon) jet tagging.  Most measurements and searches at the LHC target a final state with a particular partonic structure and the dominant backgrounds may have a different flavor composition.  Therefore, tagging jets as quark or gluon could increase the analysis sensitivity.  For example, jets produced in vector-boson scattering/fusion (VBF/VBS) are quark jets, while many of the background jets are gluon jets.  There are many other applications, ranging from high multiplicity supersymmetry searches, initial state jet tagging, etc.

The probability for a gluon to radiate a gluon is enhanced by a factor of $C_A/C_F=9/4\sim 2$ over the probability for a quark to radiate a gluon of the same energy fraction and opening angle~\cite{Altarelli:1977zs}.   As a result, gluon jets tend to have more constituents and a broader radiation pattern than quark jets.  There are also more subtle differences due to quark and gluon electric charges and spins.

There are three key challenges of quark versus gluon jet ($q/g$) tagging: (1) quark and gluon labeling schemes are not unique; (2) for a given labeling scheme, quark and gluon jets are not that different; (3) the differences that do exist are sensitive to both perturbative and non-perturbative modeling choices. Since quarks and gluons carry color charge and only colorless hadrons are observed, there is not a unique way to label a jet in simulation as originating from a quark or a gluon.  Many labeling conventions exist, ranging in simplicity and model-dependence from matching to out-going matrix element partons to parsing an entire jet clustering history~\cite{Banfi:2006hf,Buckley:2015gua} to using entirely observable phase-space regions~\cite{Metodiev:2018ftz,Komiske:2018vkc}; however, no treatment escapes the problem that the notion of a quark and gluon jet is not universal\footnote{This can be mitigated by jet grooming; see e.g. Ref.~\cite{Frye:2016aiz}.  Also, the non-universality may be `small' in practice~\cite{Bright-Thonney:2018mxq}.}: quark and gluon jet radiation depends on the production mechanism.  This means that the calibration and application of $q/g$ taggers must be treated with additional care compared with more universal classification tasks such as $b$ tagging.

\begin{figure}[t]
\centering
\includegraphics[width=0.45\textwidth]{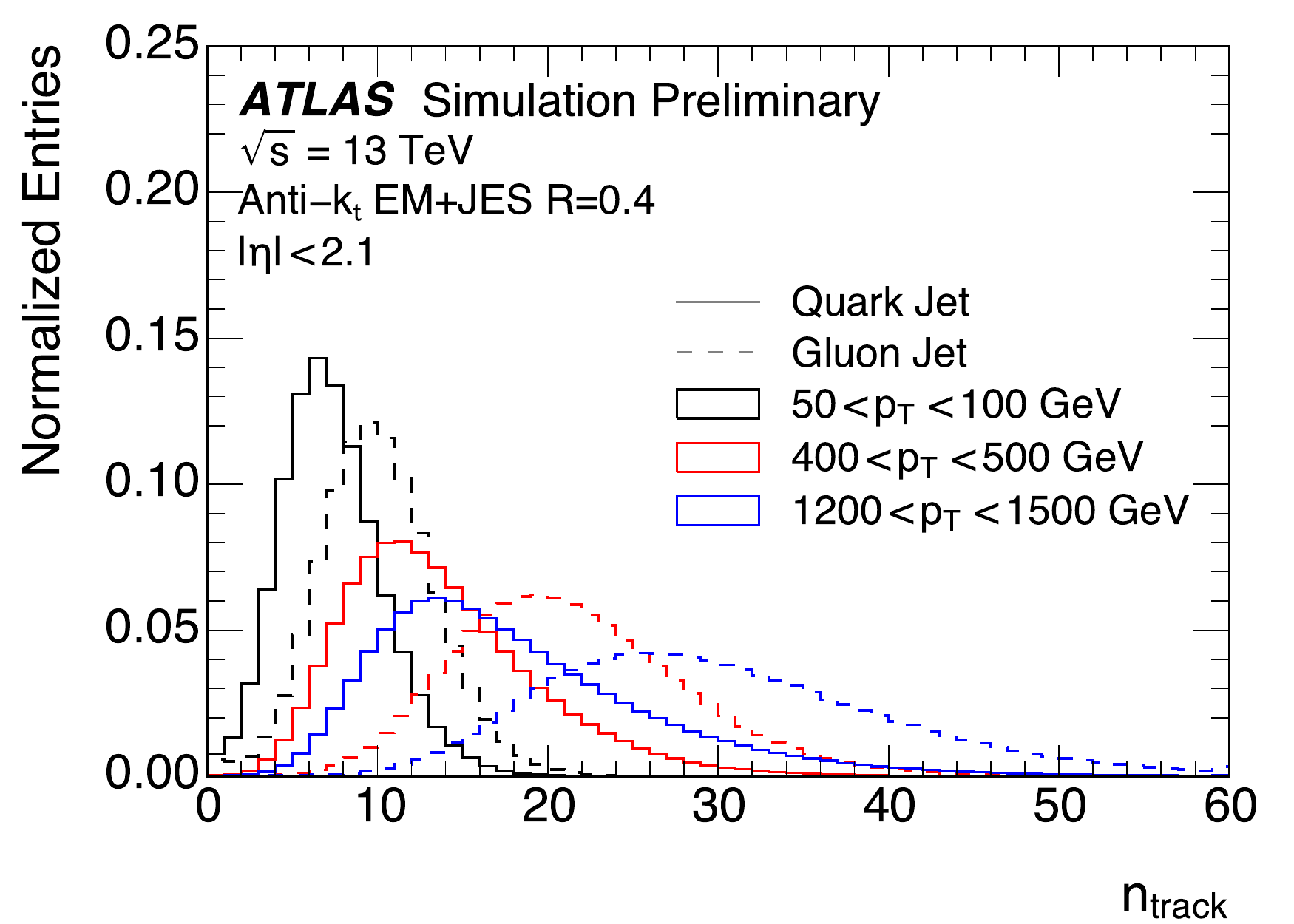}
\caption{The distribution of the number of tracks inside jets for quark and gluon jets in multiple jet $p_T$ ranges.  Reproduced from Ref.~\cite{ATL-PHYS-PUB-2017-009}.\label{fig:qg:1}}
\end{figure}

There is a plethora of jet substructure observables that can be used for $q/g$ tagging; see e.g. Ref.~\cite{Gallicchio:2011xq} for a large survey.  Many of these observables exhibit \textit{Casimir scaling} which results in nearly the same, limited discrimination power for all the observables~\cite{Larkoski:2013eya,Larkoski:2014pca}.  The most powerful single $q/g$ observable is the particle multiplicity inside a jet (shown in Fig.~\ref{fig:qg:1}), which does not exhibit Casimir scaling and recent theoretical advances~\cite{Frye:2017yrw} have shown that its discrimination power can be largely understood from perturbative theory. There is further $q/g$ separation possible when using the full radiation pattern inside a jet, though the combination of multiplicity and a Casimir scaling observable carries a significant fraction of the total discrimination power~\cite{Komiske:2016rsd}.  The modeling of $q/g$ tagging observables has a long history - see Ref.~\cite{Gras:2017jty} for a recent and detailed study.

Despite the challenges listed above, both ATLAS and CMS extensively use explicit or implicit quark versus gluon tagging. Explicit taggers are algorithms designed to directly isolate quark and gluon jets while implicit techniques are designed for another purpose that also happens to perform some quark versus gluon jet tagging. The explicit taggers developed by ATLAS~\cite{Aad:2014gea,ATLAS-CONF-2016-034,ATL-PHYS-PUB-2017-009,ATL-PHYS-PUB-2017-017} and CMS~\cite{CMS-PAS-JME-13-002,CMS-DP-2016-070,CMS-PAS-JME-16-003,CMS-DP-2017-027} include a variety of observables and data-driven calibration and validation techniques. These and related techniques have been successfully deployed in a variety of physics analyses (see e.g.~\cite{Chatrchyan:2012sn, Chatrchyan:2013jya, Khachatryan:2015bnx, CMS-PAS-JME-14-002, Sirunyan:2017wif, Aaboud:2016cns, Aad:2015owa}). Additionally, it has been shown that an improved $W$ tagger can be constructed by utilizing $q/g$ discrimination on subjets~\cite{CMS-PAS-JME-14-002}. 

Both ATLAS and CMS have developed likelihood-based discriminants for explicit $q/g$ tagging. The discriminants are constructed from variables sensitive to the radiation pattern of quark and gluon jets, also taking into account differences between light ($uds$) and heavy flavor ($cb$) quark jets, where the latter are more similar to gluon jets. ATLAS uses the number of tracks $n_{\mathrm{trk}}$ as an approximation for the number of jet constituents and the jet width~\cite{ATL-PHYS-PUB-2017-017} while CMS utilizes the number of particle-flow constituents $n_{\mathrm{const}}$, the jet axes and fragmentation functions~\cite{CMS-PAS-JME-16-003}. Since the distributions of these variables depend on $\eta$, $\pt$, and $\rho$, the likelihood discriminators are constructed differentially with respect to these variables. In \runtwo, ATLAS also introduced a simple and robust tagger using solely $n_{\mathrm{track}}$~\cite{ATL-PHYS-PUB-2017-009}, which has the advantage of a much-simplified uncertainty derivation.

\begin{figure}[tb]
\centering
\includegraphics[width=0.45\textwidth]{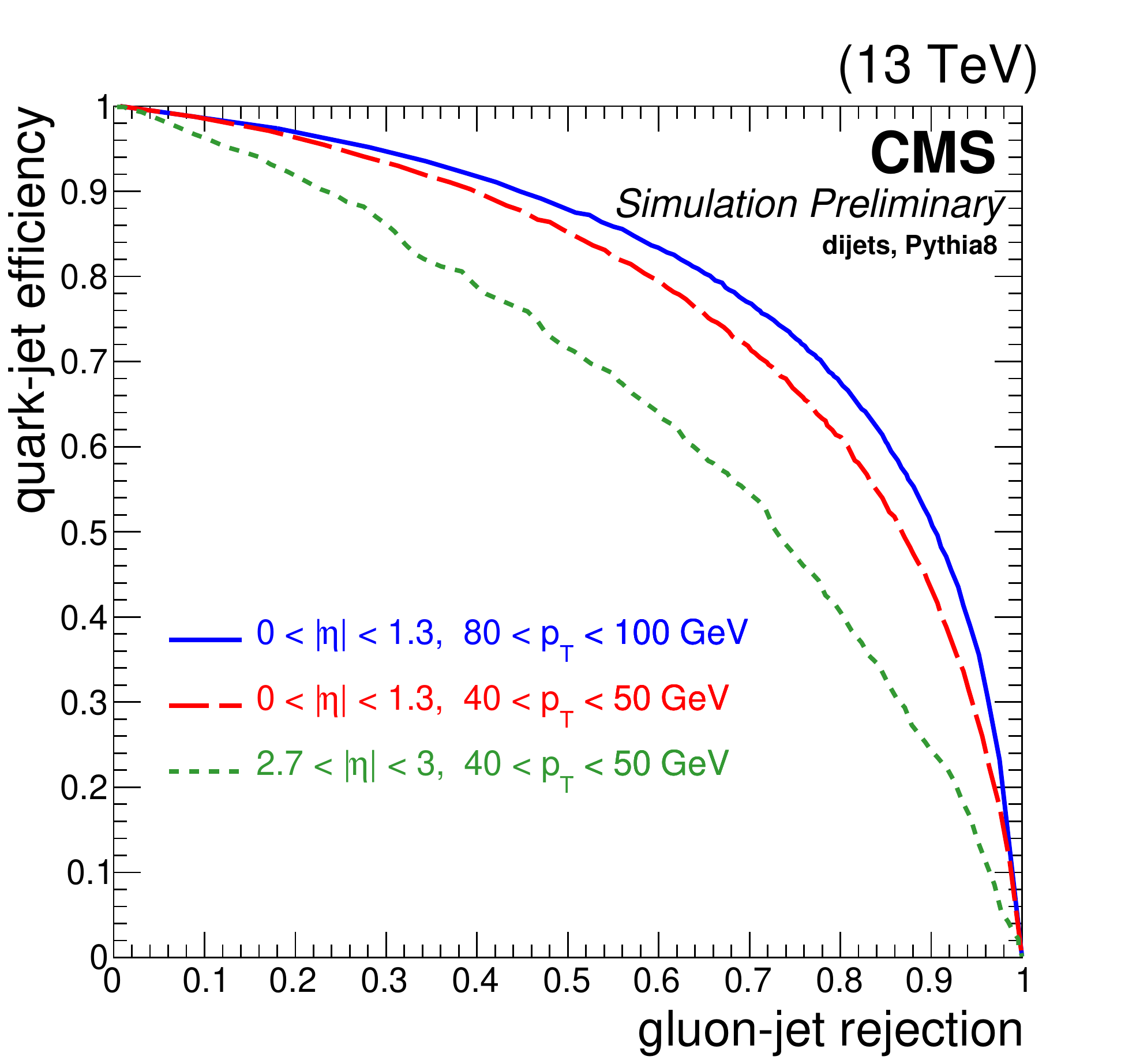}
\caption{The CMS $q/g$ tagging performance in simulation for two bins in jet $\pt$ and two bins in jet $|\eta|$.  Reproduced from Ref.~\cite{CMS-PAS-JME-16-003}.\label{fig:qg:2}}
\end{figure}

Figure~\ref{fig:qg:2} shows the CMS $q/g$ tagging performance in simulation. The $q/g$ label is obtained through a matching of jets on the detector level to outgoing partons from the matrix-element calculation. For a 50\% gluon or quark efficiency, the misidentification rate (quark or gluon) is about 10\%. This performance depends slightly on the jet $\pt$, in part because the particle multiplicity increases with $\pt$ (and therefore the performance improves). Outside the tracking acceptance ($|\eta|\gtrsim 2.5$), $q/g$ tagging significantly degrades due to the coarse calorimeter granularity and increased pile-up sensitivity.

ATLAS~\cite{ATL-PHYS-PUB-2017-017} and CMS~\cite{CMS-DP-2017-027} are also actively studying sophisticated approaches based on modern machine learning. While these methods hold great promise for their power and flexibility, simple combinations of a small number of features often achieves a similar performance. Machine learning architecture design and input optimization are still an active area of research and development.

\begin{figure}[tb]
\centering
\includegraphics[width=0.45\textwidth]{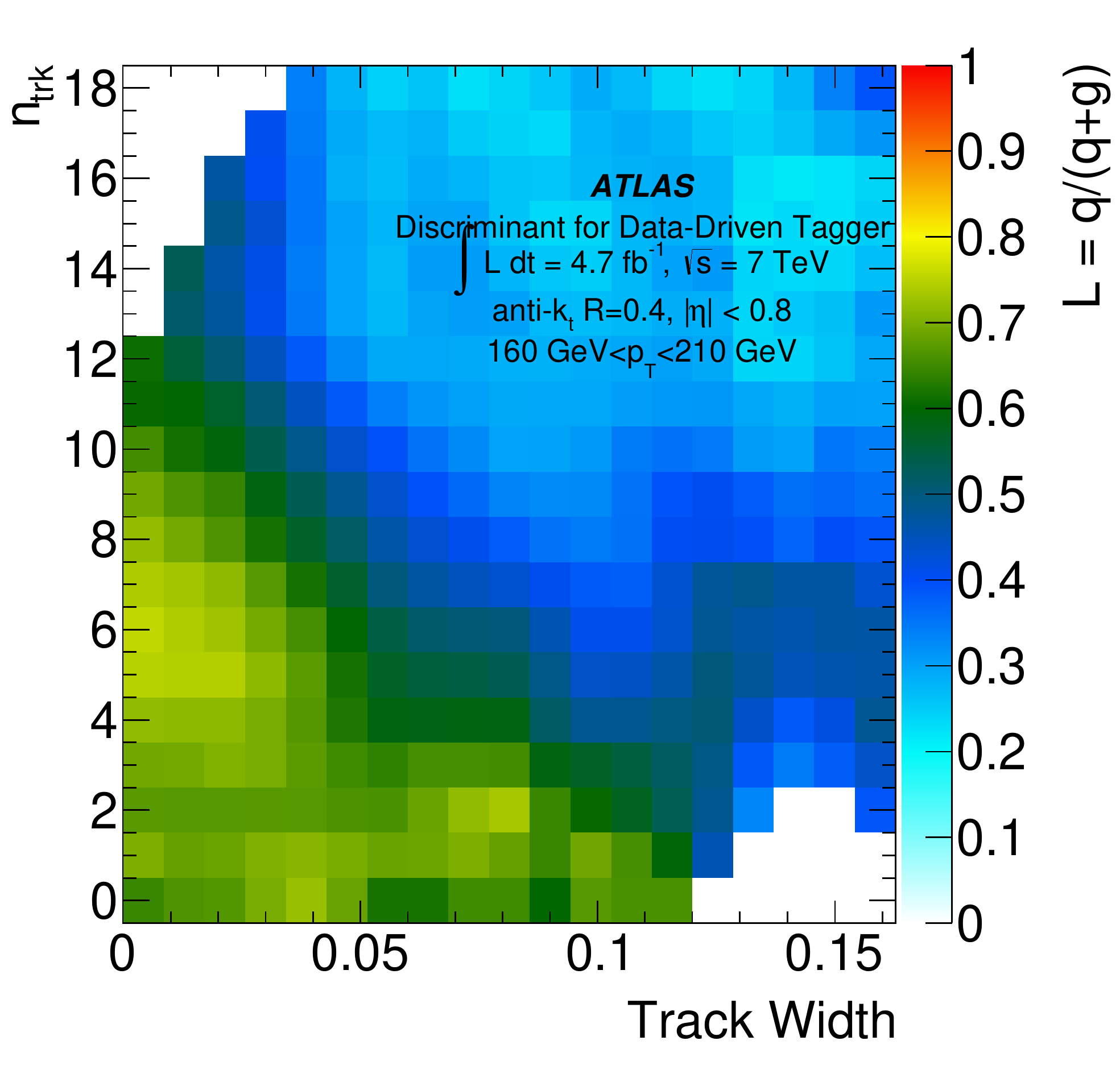}\\
\includegraphics[width=0.45\textwidth]{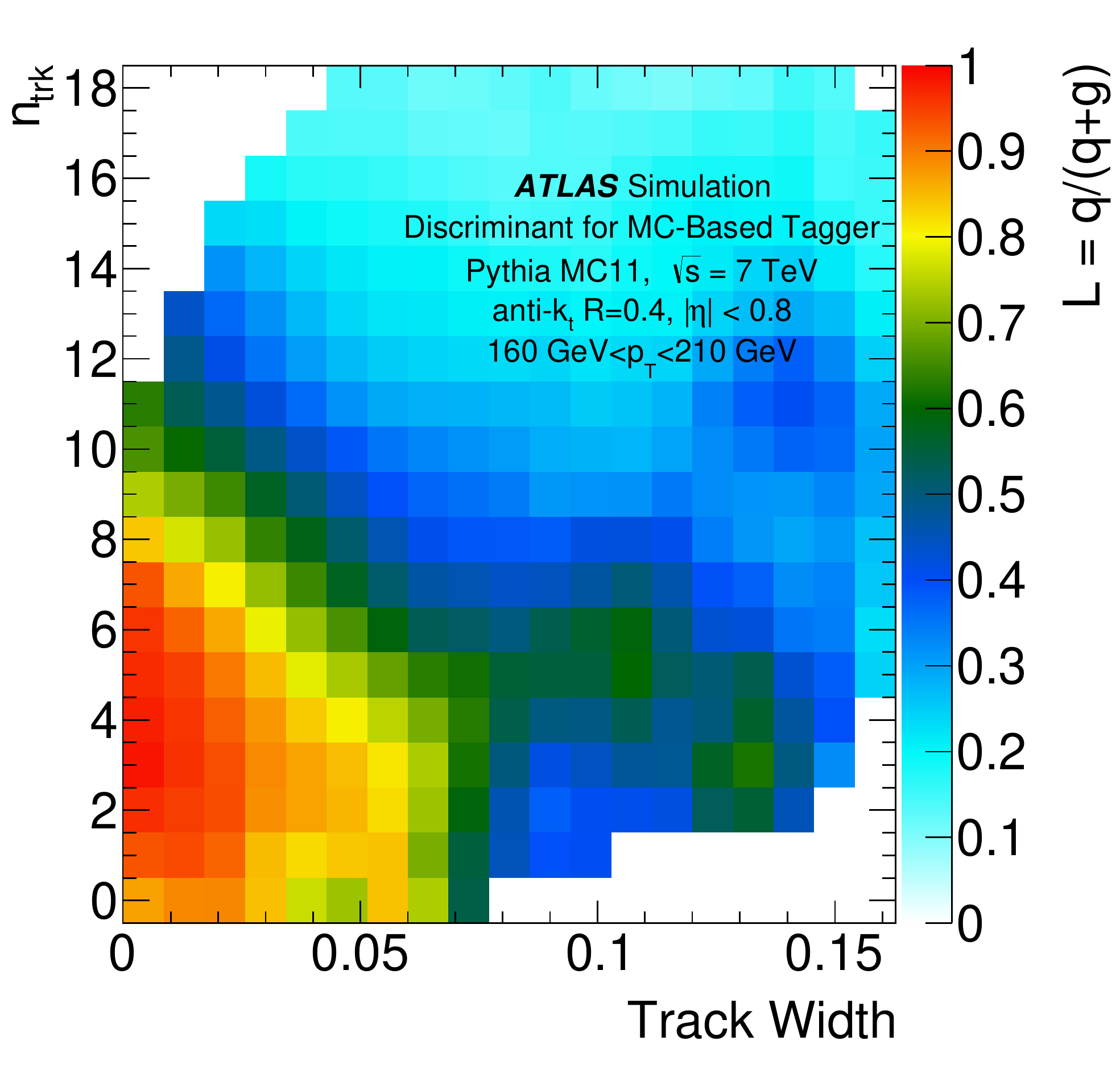}
\caption{The two-dimensional $q/g$ likelihood with ATLAS data (top) and simulation (bottom). Reproduced from Ref.~\cite{Aad:2014gea}.}
\label{fig:qg:4b}
\end{figure}
The modeling of $q/g$ discriminating observables is a key concern for tagging applications. Typically, Pythia~\cite{Sjostrand:2007gs,Sjostrand:2006za} tends to describe quarks better than Herwig~\cite{Bellm:2015jjp,Bahr:2008pv}, whereas the opposite is observed for gluons. Pythia tends to overestimate the $q/g$ tagging performance with respect to data, as illustrated quite strikingly in figure~\ref{fig:qg:4b}.  This figure shows that gluon jets tend to have more tracks and have a broader radiation pattern relative to quark jets\footnote{
The jet flavor is obtained as the type of the highest energy parton from the event record inside the jet cone.  This gives nearly the same result as the CMS definition discussed above for the two leading jets in a $2\rightarrow 2$ calculation, but also works well for additional jets in the event. }.  
The fact that the hot spot in the bottom left of figure~\ref{fig:qg:4b} is much more pronounced for MC than for data indicates that the simulation over-predicts the difference between quark and gluon jets. In contrast, Herwig (not shown) tends to underestimate the performance observed in data. 

\begin{figure}[tb]
\centering
\includegraphics[width=0.45\textwidth]{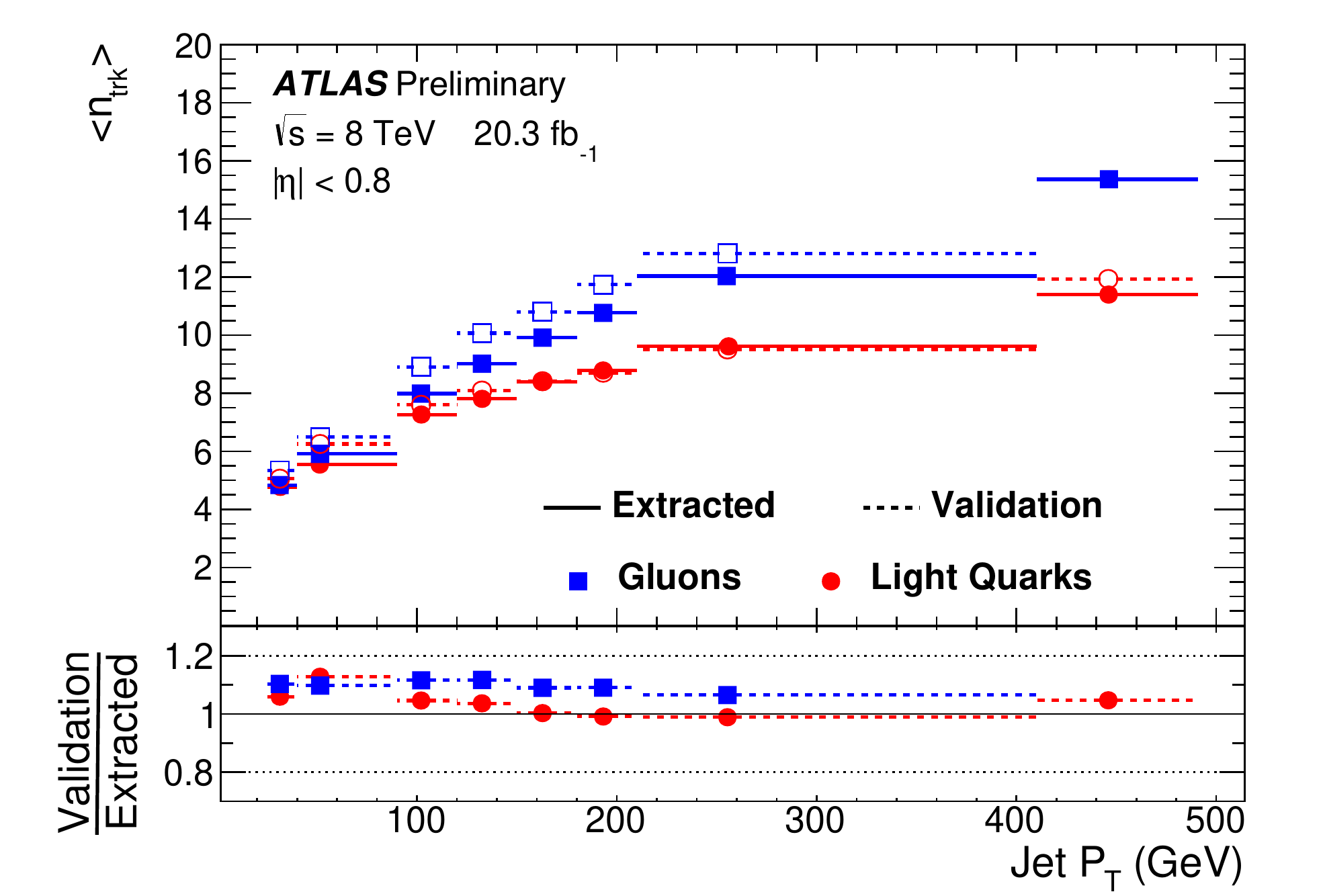}
\caption[ATLAS Ntrk for gluon and quark, template method.]{The average track multiplicity in ATLAS for $Z/\gamma$+jets (quark-enriched) and dijets (gluon-enriched). The dashed lines indicate the measurement on the validation samples : $Z/\gamma$+2-jets (quark-enriched) and trijets (gluon-enriched). Reproduced from Ref.~\cite{ATLAS-CONF-2016-034}.}
\label{fig:qg:5}
\end{figure}

Multiple samples with a different (but known) $q/g$ composition can be used to extract the distribution of $q/g$ tagging observables. ATLAS and CMS have both used $Z/\gamma$+jets and dijet samples, which are enriched in quark and gluon jets, respectively.  The extracted average $n_\text{track}$ from data is shown using this method in figure~\ref{fig:qg:5}.  As expected, gluon jets have more particles on average than quark jets and the multiplicity distribution increases with jet $\pt$.  

The \runtwo ATLAS tagger is based entirely on dijets, exploiting the rapidity dependence of the $q/g$ fraction to extract the track multiplicity separately for quarks and gluons. A \runone measurement is used to constrain the particle-level modeling, and dedicated track reconstruction uncertainties are used to complement the particle-level uncertainty with a \runtwo detector-level uncertainty. The uncertainties on $q/g$ tagging are $2$-$5$\% over a wide range of $200\GeV \lesssim \pt \lesssim 1\TeV$ at a working point of 60$\%$ quark jet efficiency~\cite{ATL-PHYS-PUB-2017-009}. The template-based calibration can also be used to directly construct the $q/g$ tagger in data; however, when more than two observables are used to construct the tagger, it becomes impractical to extract the high-dimensional templates.  

The likelihood-based discriminant used for $q/g$ tagging in CMS in \runtwo is calibrated with a template-based fit using two discriminant distributions obtained from a $Z$+jets and a dijet sample. The different quark and gluon fractions in each bin of the discriminant distributions are determined simultaneously and fitted by polynomial functions in order to obtain smooth interpolations~\cite{CMS-PAS-JME-16-003}. 

Despite its power, the template technique has some residual non-closure because the resulting calibrated tagger applied to another final state may not have the same performance. This is illustrated in figure~\ref{fig:qg:5}, which shows how the average track multiplicities extracted for quark and gluon jets (using high-purity $Z/\gamma$+jets and dijets data respectively) differ from the values obtained in the $\gamma$+2-jet and trijet samples used for validation.

Explicit tagging is often the focus of modern $q/g$ discrimination, but
there is a broad program of implicit tagging as well. One ubiquitous
example of this is the ATLAS jet calibration procedure (see section \ref{sec:jetrec-jes}). Since the calorimeter response is non-linear, a jet with a higher particle multiplicity will have a lower response for the same energy. After applying a simulation-based correction to eliminate this inclusive bias in the JES, a residual calibration is applied to correct for the dependence of the bias on the number of tracks associated to the jet and the jet width~\cite{ATLAS-CONF-2015-002}. After applying this residual GSC, the difference in response between quark and gluon jets is reduced. Implicit $q/g$ tagging also appears in pile-up jet identification~\cite{CMS-PAS-JME-13-005,Aaboud:2017pou}, boson and top tagging~\cite{CMS-PAS-JME-14-002,Sirunyan:2017wif,Aad:2015owa}, and elsewhere.

Despite its long history, quark versus gluon jet tagging is still a very active topic of research. Since most analyses at the LHC target processes with a known and asymmetric $q/g$ jet composition, $q/g$ tagging holds great promise for improving searches and measurements in the future.  Further studies are required to understand the limits of $q/g$ tagging performance and to mitigate the sample dependence for universal definitions and calibrations. Interestingly, recent studies have shown how modern machine learning classifiers can be directly trained on data even though there are no per-jet labels~\cite{Dery:2017fap,Metodiev:2017vrx}.

\subsection{Vector Boson Tagging }\label{sec:BosonTagging}

The hadronic, two-prong decays of weak vector bosons $V$ have a
distinct radiation pattern compared to individual high-$\pt$ quarks or
gluons.  In particular, boosted bosons tend to have two distinct subjets with relatively equal momentum sharing.  In contrast, most generic quark and gluon jets will have one prong and if they have two, the second one tends to be soft.  Furthermore, the mass of quark and gluon jets scales with their $p_T$ and is lower than the electroweak boson masses for low jet $p_T$ and higher for ultra-high $p_T$ jets.  For jets around 200 GeV, the decay products of a boosted $W$ and $Z$ boson are typically only captured by a jet of radius $R\sim 1$, while smaller radii can be used at higher jet $p_T$.  Good separation power between $W$ and $Z$
bosons is also desirable in a number of analyses, most notably
searches for diboson resonances (see section~\ref{sec:dibosonsearch}).  

ATLAS and CMS performed a broad range of studies during \runone and the beginning of \runtwo, systematically identifying the influence of pile-up reduction and grooming techniques on jet substructure observables used for $V$ tagging~\cite{Khachatryan:2014vla, Aad:2015rpa}. Simulated samples containing $W$ jets (rather than $Z$ jets) are primarily used for these studies, as $W$ jets are abundant in data thanks to the large quantity of $\ttbar$ events produced at the LHC. 

The optimization of the $V$ tagging algorithm is generally based on various factors concerning the tagged jet mass: $(i)$ a sensible JMS (i.e., tagged jet mass close to the $W$ mass), $(ii)$ a narrow jet mass response with an approximate Gaussian lineshape, $(iii)$ stability with respect to pile-up and jet $\pt$, and $(iv)$ good background rejection at a given signal efficiency. Considering all of these factors, ATLAS decided on using the trimming algorithm~\cite{Krohn:2009th} with \fcut = 0.05 and \rsub = 0.2 on \antiktten jets in \runtwo, while CMS opted for using \antikteight jets, treating the pile-up first with PUPPI and then applying soft drop grooming with \zsoftdrop = 0.1 and $\beta=0$.  

In addition to the comprehensive studies of grooming options~\cite{Aad:2015rpa,ATL-PHYS-PUB-2015-033,ATL-PHYS-PUB-2017-020,CMS-PAS-JME-14-001,CMS-PAS-JME-14-002}, ATLAS and CMS both investigated the discrimination powers for a plethora of jet substructure variables, including $N$-subjettiness~\cite{Thaler:2010tr, Thaler:2011gf}, Qjet volatility~\cite{Ellis:2012sn}, ratios of energy correlation functions $C_2^\beta$~\cite{Larkoski:2013eya}, $D_2^\beta$~\cite{Larkoski:2014gra, Larkoski:2015kga} and $N_2^\beta$~\cite{Moult:2016cvt}, angularities and planar flow~\cite{Almeida:2008yp}, splitting scales~\cite{Butterworth:2002tt, Thaler:2008ju}, the jet and subjet quark/gluon likelihood, and the jet pull angle~\cite{Gallicchio:2010sw}.

Both ATLAS and CMS developed simple taggers that rely on the combination of the jet mass with one other variable that improves the discriminating power between the signal and background. The standard ATLAS $V$ tagger for \runtwo was chosen  to be the trimmed jet mass and $D_2^{\beta=1}$~\cite{Aad:2015rpa}, known as `R2D2', while CMS decided to use the soft drop jet mass and the $N$-subjettiness ratio $\tau_{21} = \tau_2/\tau_1$. Despite the different choices of tagging observables and detector design, ATLAS and CMS reach a very similar background rejection at a given tagging efficiency. An active field of developments is the usage of multivariate techniques for boosted $V$ identification which have shown to be able to significantly improve the background rejection~\cite{Khachatryan:2014vla,ATL-PHYS-PUB-2017-004}.

\begin{figure}[tb]
\centering
\includegraphics[width=0.45\textwidth]{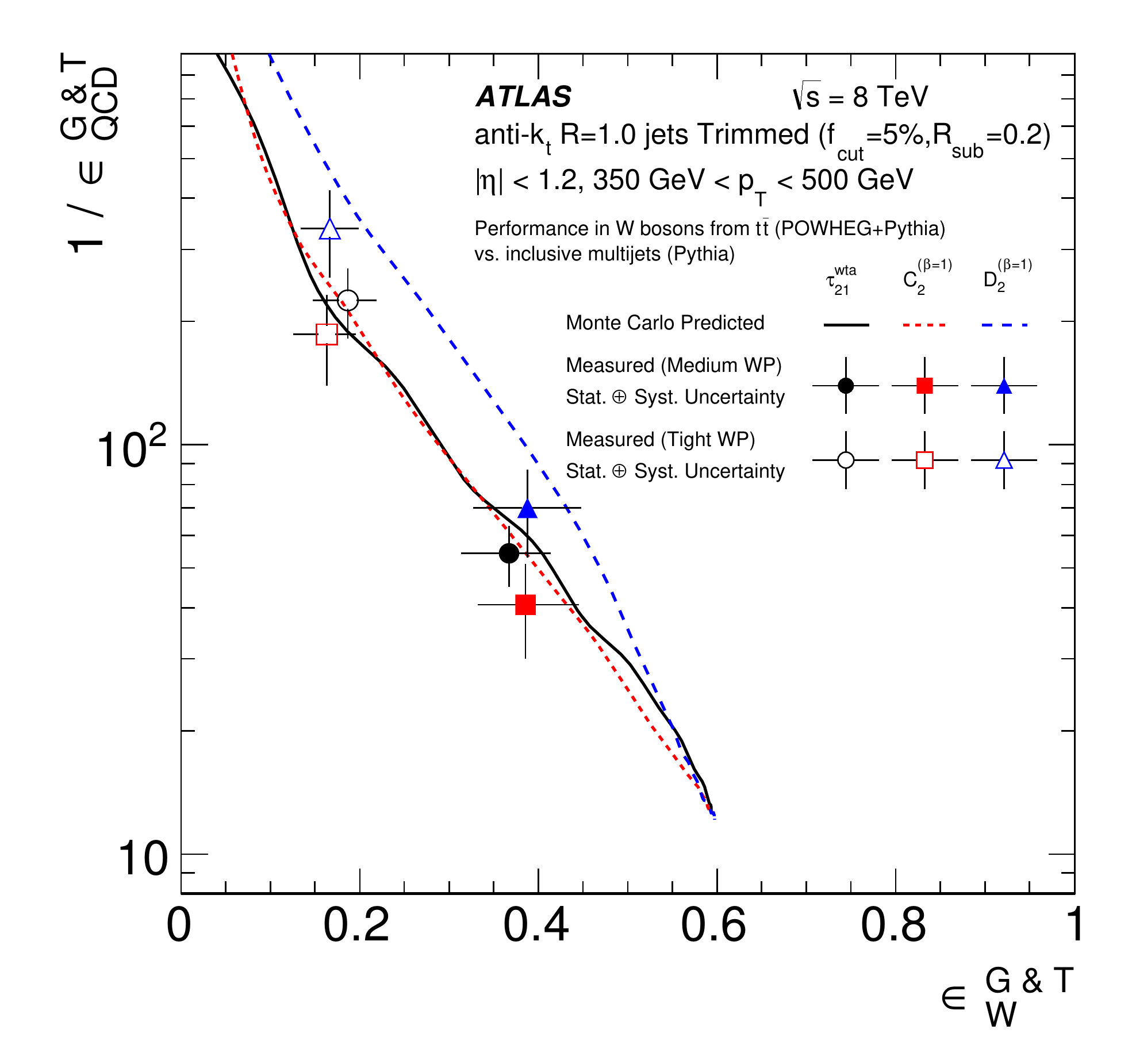}
\caption{\label{fig:wROC_ATLAS} Signal efficiency versus background rejection power compared with measurements from ATLAS for $350<\pt<500\GeV$. Taken from Ref.~\cite{Aad:2015rpa}.}
\end{figure}

In the ATLAS studies the variable $C_2^{\beta=1}$ in combination with the trimmed jet mass has been shown to be as good a discriminator as $\tau_{21}$\footnote{
	A different axis definition for the subjet axes is used in ATLAS when calculating $\tau_N$, known as {\it the-winner-takes-all} axis~\cite{Larkoski:2014uqa}, which is consistently found to perform slightly better than the standard subjet axis definition in tagging bosons.} as shown in figure~\ref{fig:wROC_ATLAS}. This is in contradiction to the study by CMS, where $C_2^{\beta}$ is one of the weaker observables; however, a direct comparison is difficult, since in ATLAS \textit{groomed} substructure variables are used, calculated for trimmed jets, while in CMS ungroomed variables are used. Also, the particulars of particle reconstruction have a large impact on the performance of individual observables. While a study of the performance of $D_2^{\beta}$ at CMS is still pending, the soft drop $N_2^\beta$ observable was found to give similar performance to $\tau_{21}$ in CMS~\cite{Sirunyan:2017nvi}.

CMS studied the quark/gluon likelihood (QGL) discriminator for its potential in $V$ tagging applications in \runone~\cite{CMS-PAS-JME-14-002}, finding that a combination of the groomed jet mass and the QGL achieved a similar discrimination power as the groomed jet mass and $\tau_{21}$. When adding the QGL to the \runone $V$ tagger (pruned jet mass and $\tau_{21}$), the misidentification rate was reduced slightly from 2.6\% to 2.3\% at a constant signal efficiency of 50\%. A similar reduction of the misidentification rate was observed when adding $C_2^{\beta=2}$, showing that $C_2^{\beta}$ carries additional information with respect to the groomed jet mass and $\tau_{21}$. However, the QGL and $C_2^{\beta}$ exhibit a considerable pile-up dependence, resulting in a degradation of their discrimination power with increasing activity. This pile-up dependence is expected to be reduced when using PUPPI in place of particle flow + CHS.

In figure~\ref{fig:wROC_ATLAS} the ATLAS measurements of signal efficiencies versus background rejection power are shown for $\tau_{21}$, $C_2^{\beta}$ and $D_2^{\beta}$, together with a selection on the trimmed jet mass (in this \pt range, the smallest mass window that captured 68\% of the signal jets was found to be 71-91\GeV -- see Ref.~\cite{Aad:2015rpa}, table 7). The measurements are shown with statistical and systematic uncertainties. It is reassuring that the points for all three observables lie on the predicted performance curves for the two different working points studied. 

In the ATLAS study, the most important systematic uncertainty is the jet substructure scale, which has been derived by comparing calorimeter-jets with track-jets. Once again, the distributions in data lie between the ones derived with Pythia and Herwig, leading to large modeling uncertainties~\cite{Aad:2015rpa, ATL-PHYS-PUB-2015-033}. A similar observation is made by CMS~\cite{Khachatryan:2014vla, CMS-PAS-JME-16-003}. Improving the modeling of jet properties and thereby reducing the differences between different event generators is a major task, but crucial for future precision studies using jet substructure.  

\begin{figure}[tb]
\centering
\includegraphics[width=0.45\textwidth]{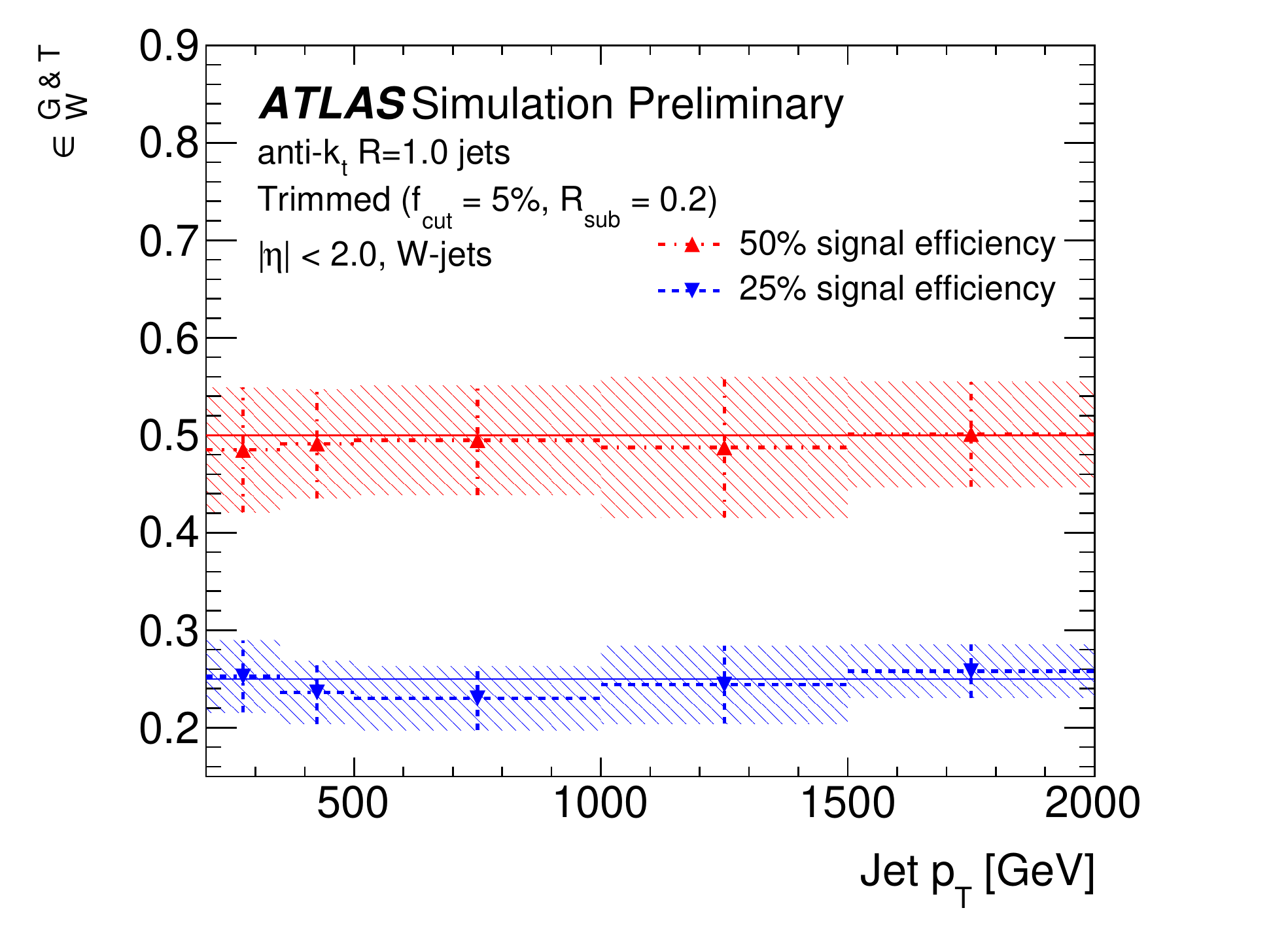} \\
\includegraphics[width=0.45\textwidth]{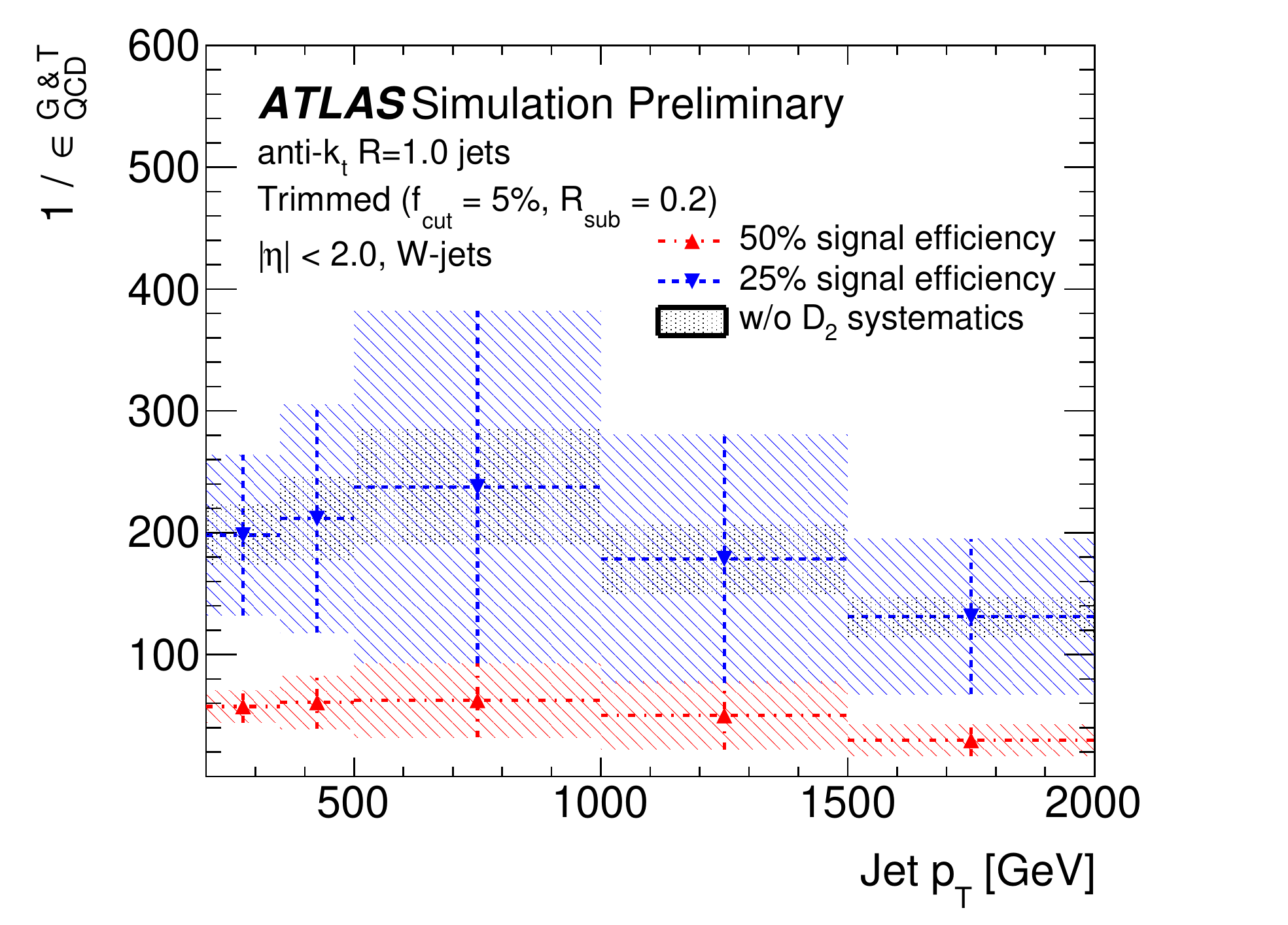} 
\caption{\label{fig:atl_es_eb} Efficiency (top) and misidentification rate (bottom) for tagging boosted $W$ bosons in ATLAS. Adapted from Ref.~\cite{ATL-PHYS-PUB-2015-033}.}
\end{figure}

A crucial aspect of $V$ tagging is the derivation of background rates from multijet production in real collision data when performing measurements. A commonly used method is the extrapolation from one or more control regions, which are defined orthogonally to the signal region. Usually, these control regions are defined by inverting the jet mass window selection, see e.g.~\cite{Khachatryan:2014gha, Khachatryan:2016yji, Sirunyan:2016cao, Aaboud:2017ahz, Aaboud:2017eta, ATLAS-CONF-2017-051}.
Transfer functions are derived from simulation, extrapolating the rates and shapes from the control to the signal regions. Even though these transfer functions are ratios of distributions, which results in a reduction of the impact of modeling uncertainties, a residual dependence on the simulation can not be eliminated. However, additional uncertainties in the high-$\pt$ tails of the transfer functions can be eliminated by ensuring a constant behavior as a function of $\pt$. The requirement is thus a flat signal or background efficiency (depending on the needs of the analysis). In order to achieve a flat signal efficiency, ATLAS developed a $\pt$-dependent selection on the value of $D_2^{\beta=1}$, as this distribution shows a strong dependence on $\pt$~\cite{ATL-PHYS-PUB-2015-033}. In contrast to the \runone studies described above, no $\pt$-dependent selection is made on the trimmed jet mass, as the calibrated jet mass is used to define the $V$ tagging working point. While the jet mass resolution still increases with $\pt$, a constant window of $\pm 15\GeV$ around the mean reconstructed $W$ or $Z$ boson mass is used. This results in a $\pt$-dependent signal and background efficiency, which can also be countered with the $\pt$-dependent cut on $D_2^{\beta=1}$. This leads to a constant signal efficiency, while the background efficiency shows a residual $\pt$ dependence, as shown in figure~\ref{fig:atl_es_eb}.

\begin{figure}[tb]
\centering
\includegraphics[width=0.36\textwidth]{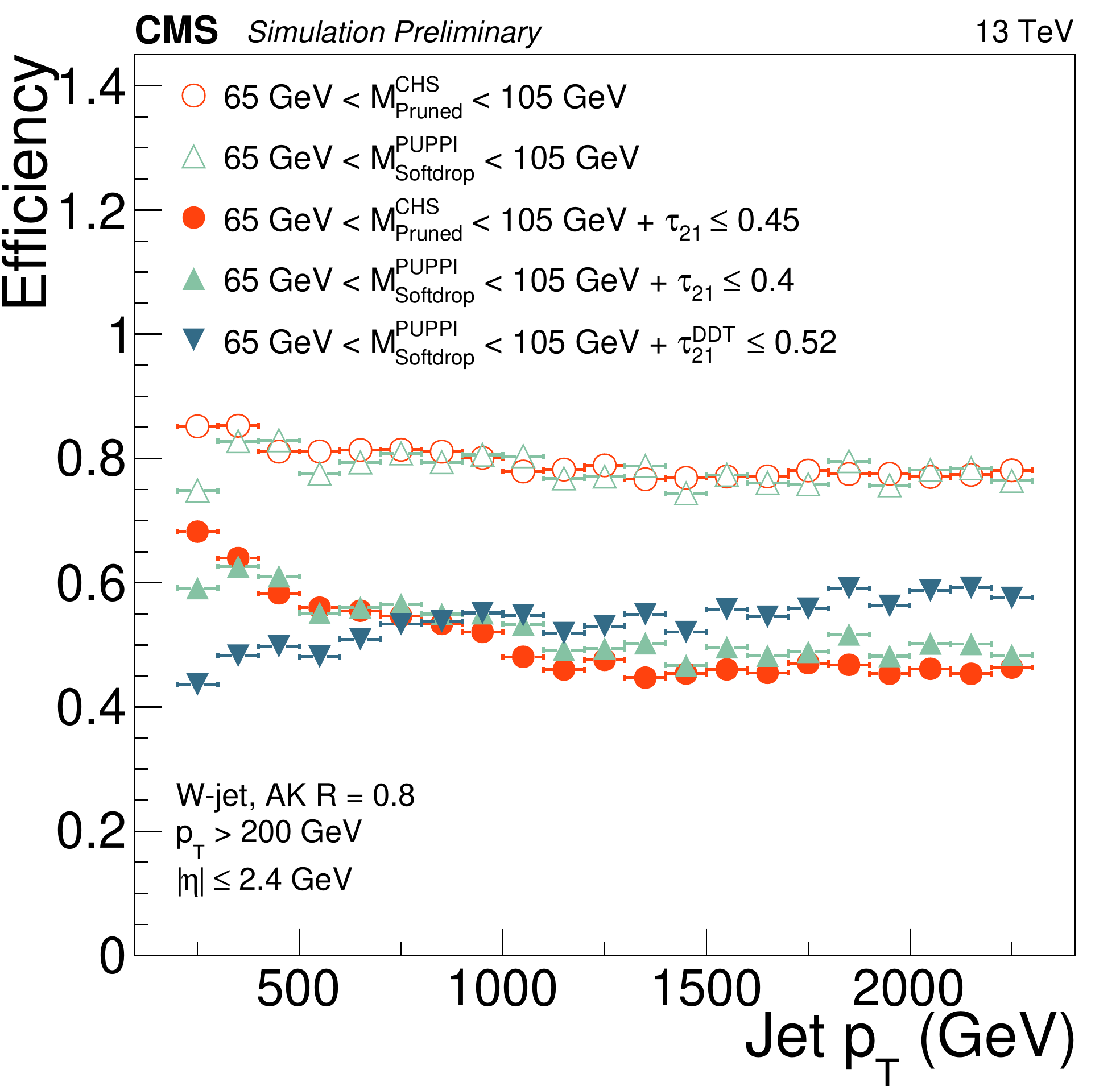} \\
\includegraphics[width=0.36\textwidth]{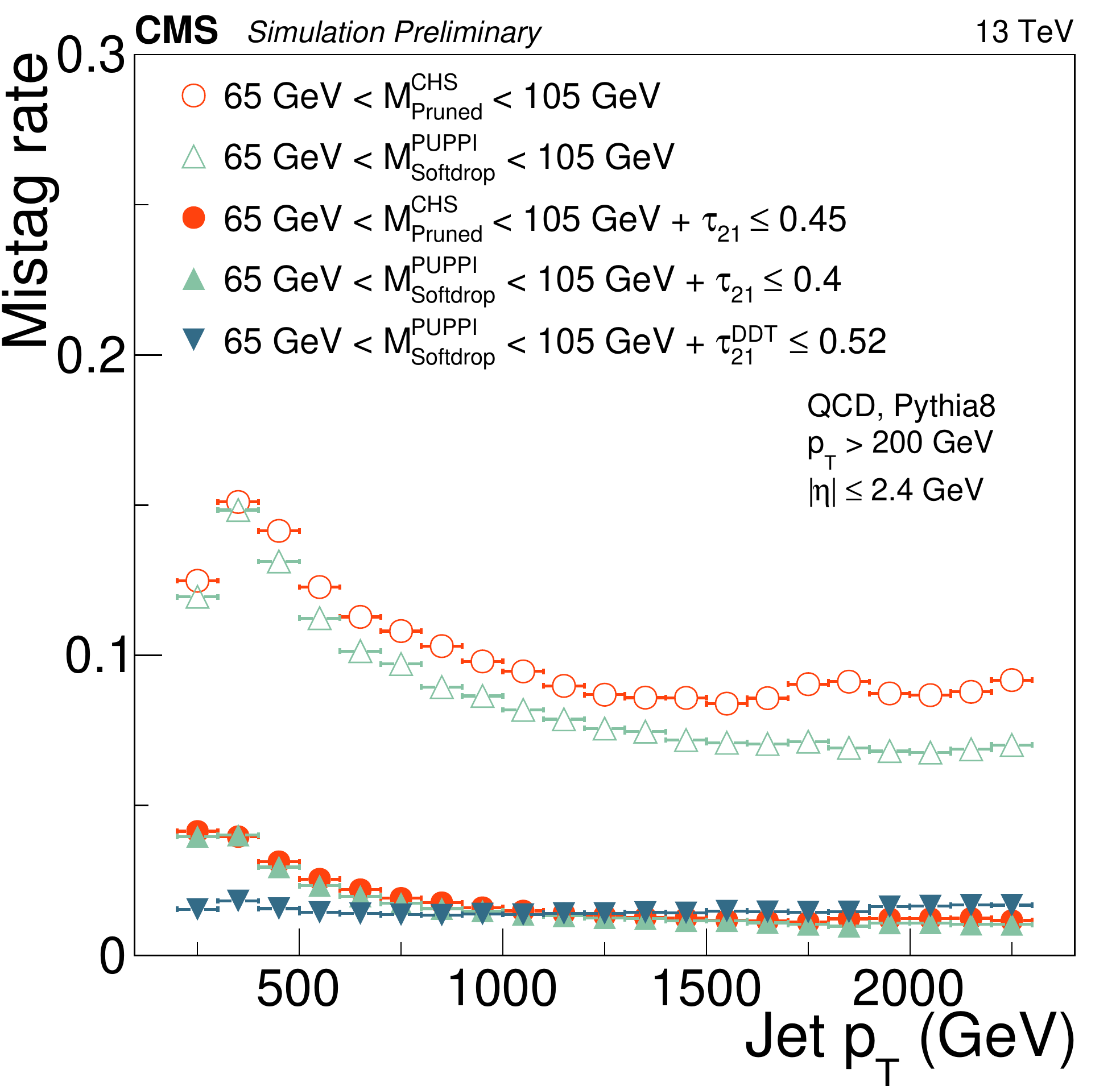}
\caption{\label{fig:vCMS} Efficiency and misidentification rate of various identification techniques for boosted $W$ tagging. Taken from Ref.~\cite{CMS-PAS-JME-16-003}.}
\end{figure}

Another possibility has been explored by CMS. Instead of introducing $\pt$-dependent selection criteria, a linear transformation of the ratio $\tau_{21}$ has been studied~\cite{CMS-PAS-JME-16-003}, given by 
$\tau_{21}^{\mathrm{DDT}} = \tau_{21} - M \cdot \log(m^2 / \pt / 1\GeV) $~\cite{Dolen:2016kst}, where $M$ is a constant determined from simulation. 
The replacement of $\tau_{21}$ with the designed decorrelated tagger (DDT) $\tau_{21}^{\mathrm{DDT}}$ does
not affect the overall performance of the tagger, but results in an
approximately flat misidentification rate as a function of $\pt$, as
shown in figure~\ref{fig:vCMS} (bottom). 
The effect of the DDT method on the $V$ tagging efficiency is shown in 
figure~\ref{fig:vCMS} (top). The efficiency increases as function of \pt with a slope 
somewhat smaller than the slope for the decreasing efficiency obtained with plain $\tau_{21}$. 
The development of decorrelated jet substructure taggers is an active field with new techniques e.g. described in Refs.~\cite{Shimmin:2017mfk,Aguilar-Saavedra:2017rzt,Moult:2017okx}.

A less-studied possibility to lift the $\pt$-dependence of substructure observables is the application of variable-$R$ jets~\cite{Krohn:2009zg}. By shifting the $\pt$-dependence to the jet-clustering level with a distance parameter proportional to $\pt^{-1}$, a stable position of the jet mass and jet substructure variables with respect to changes in $\pt$ can be achieved~\cite{ATL-PHYS-PUB-2016-013}. This can lead to a stable tagging performance without the necessity of $\pt$-dependent optimization steps, but further experimental studies are needed to commission this strategy for use in analyses.

For some analyses the requirement of $\pt \gtrsim 200\GeV$ is too restrictive, and hadronically decaying $V$ bosons with lower $\pt$ need to be selected. This poses a particular challenge due to the abundance of light flavor jets at the LHC and their indistinguishability from jets from $W$/$Z$ decays. An attempt was made by CMS to discriminate `resolved' (non-merged) hadronic $W$ decays from multijet background using the QGL, the sum of the jet charges of the dijet pair and the jet pull angle. Combining these variables into a Boosted Decision Tree, a misidentification rate of about 25\% is achieved for a signal efficiency of 50\%~\cite{CMS-PAS-JME-14-002}. While this is a first success, the performance is about an order of magnitude worse than $V$ tagging for fully merged decays, showing the power of substructure techniques in this field. 

\begin{figure}[tb]
\includegraphics[width=0.43\textwidth]{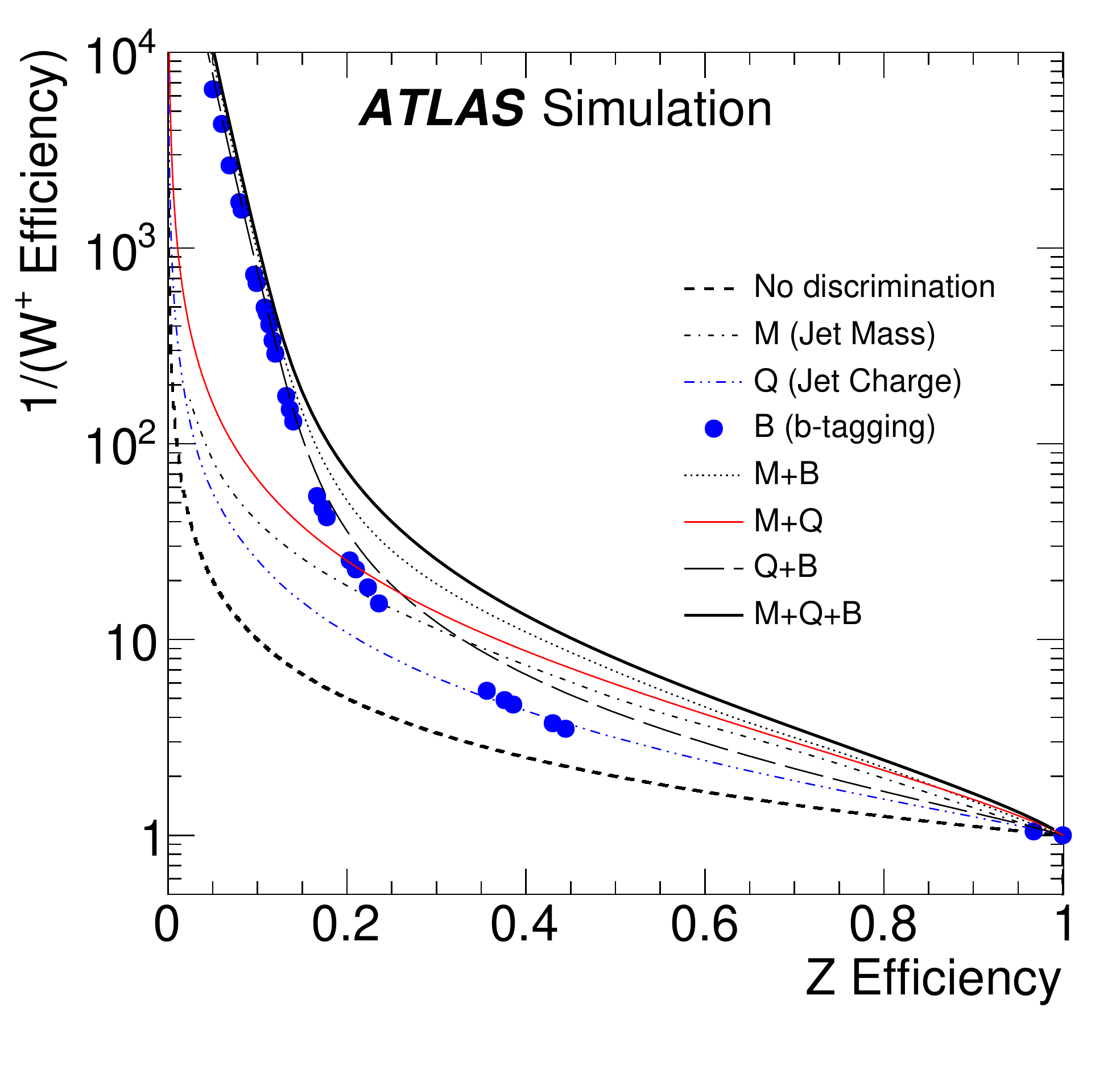}
\caption{Background rejection versus efficiency for discriminating $Z$ boson jets from $W$ boson jets for various jet observables and their combinations. Reproduced from Ref.~\cite{Aad:2015eax}.}
\label{fig:WZ:ROC}
\end{figure}

In addition to developing tools for distinguishing boosted hadronically decaying $W$ and $Z$ bosons from generic quark and gluon jets, ATLAS has also built a tagger to further classify a boson jet as either originating from a $W$ boson or a $Z$ boson~\cite{Aad:2015eax}.  While theoretically clean due to the color singlet nature of the $W$ and $Z$ boson, this task is particularly challenging because the jet mass resolution is comparable to the difference $m_Z-m_W$.  In order to improve the sensitivity of the tagger, jet charge and $b$ tagging information are combined with the jet mass. The jet mass distribution depends on the type of $W$ or $Z$ decay due to semi-leptonic $B$ and $D$ decays, so a full likelihood tagger is constructed by summing over the conditional likelihoods for each flavor type.  To maximize the discrimination power from $b$ tagging, multiple efficiency working points are used simultaneously in the tagger.  Figure~\ref{fig:WZ:ROC} illustrates the performance of the boson type-tagger in simulation.  A $W^+$ rejection near $8$ (corresponding to a misidentification rate of $12.5\%$) is achieved at a $Z$ boson efficiency of 50\%.  At this moderate $Z$ boson efficiency, all of the inputs offer useful discrimination information.  At low efficiencies, below the $b\bar{b}$ branching ratio for $Z$ bosons, 
$b$ tagging dominates over the jet mass and jet charge.

The boson type-tagger was optimized for a relatively low boson boost, $200\GeV < \pt <400\GeV$.  The discrimination power of all of the input variables degrades with $\pt$ due to the worsening jet mass resolution, tracking efficiency and momentum resolution, as well $b$ tagging efficiency.  However, there are recent developments to address each of these challenges, such as the track-assisted jet mass (section~\ref{sec:jetrec-mass}), pixel-cluster splitting~\cite{Aad:2014yva}, and track-jet $b$ tagging~\cite{ATL-PHYS-PUB-2014-013}.

\subsection{Top Tagging }\label{sec:TopTagging}

The three-prong decays of highly boosted top quarks in the fully hadronic decay channel offer richer phenomenology for their identification than the two-prong decays of $W$ and $Z$ bosons. This has been exploited in a number of algorithms, which usually aim at an optimal performance in a particular kinematic regime.   Flavor tagging also plays a key role for top tagging, which offers its own challenges because the $b$ jet from the $b$ quark may not be isolated from the radiation resulting from the associated $W$ boson decay. Due to the heavier mass of the top quark compared with the electroweak bosons, top tagging must also operate in a moderate boost regime where the decay products may not all be contained inside a single jet with $R\lesssim 1.0$.

The techniques for tagging boosted top quarks have evolved as fairly complex methods in comparison to the $V$ taggers; these techniques include:

\begin{enumerate}
\item[(a)]{The Johns Hopkins / CMS top tagger (CMSTT) \cite{Kaplan:2008ie}} was designed for tagging top quarks with \pt $>$ 1\TeV. The algorithm is based on a decomposition of the primary jet into up to four subjets by reversing the CA clustering sequence. It has been adapted by the CMS Collaboration~\cite{CMS-PAS-JME-09-001, CMS-PAS-JME-10-013}, and was adopted as the standard top-tagging algorithm in CMS in \runone, where it was typically used in the region of $\pt>400$\GeV, with an average identification efficiency of 38\% at 3\% misidentification rate~\cite{CMS-PAS-JME-13-007}. 
\item[(b)]{The HEPTopTagger (HTT)~\cite{Plehn:2009rk, Plehn:2010st}} was designed to target \ttH production in the $H\to\bbbar$ decay channel. In \ttH production the top quark \pt distribution peaks around 150\GeV and is steeply falling towards increasing \pt, where it is already an order of magnitude smaller at $\pt \sim 400\GeV$. This results in a requirement of non-zero signal efficiency already at $\pt \approx 200\GeV$, where the top quark decay is only moderately boosted. The HTT achieves this with a large jet distance parameter of $1.5$ and a sequence of declustering, filtering and re-clustering of the original CA jet. The performance of the HTT was studied by the ATLAS and CMS Collaborations on data with a center-of-mass energy $\sqrt{s}=7$ and $8\TeV$~\cite{Aad:2013gja, Aad:2016pux, CMS-PAS-JME-13-007}. Efficiencies of 10\% with misidentification rates of 0.5\% for jets with $200 < \pt < 250\GeV$ were observed. The efficiency increases with increasing jet \pt, where a plateau is reached for $\pt > 400\GeV$, with efficiencies of approximately 40\% at 3\% misidentification rate, very similar to the performance achieved with the CMSTT. 
\item[(c)]{Shower Deconstruction~\cite{Soper:2011cr, Soper:2012pb}} was designed to be analogous to running a parton shower Monte Carlo generator in reverse, 
where emission and decay probabilities at each vertex, color connections, and kinematic requirements are considered. 
Small-radius (generally $R=0.2$) subjets are reconstructed with the CA algorithm and all possible \textit{shower histories} that can lead to 
the observed leading final state \antiktten jet are calculated. Each shower history is assigned a probability weight factor based on the aforementioned considerations (to be signal-like or background-like), then a likelihood ratio  $\chi({p}_N)$ is constructed, and the $\log\chi({p}_N)$ is used as the discriminating substructure variable. 
For top quark tagging, efficiencies of 80\% with misidentification rates of 50\% for jets with $500 < \pt < 1000\GeV$ were observed. The efficiency increases with increasing jet \pt, where a plateau is reached for $\pt > 2000 \GeV$, with efficiencies of $\sim$ 80\% at 10\% misidentification rate. Recently, the Shower Deconstruction algorithm was optimized for top quarks with $\pt > 800$ GeV in context of the $W'$ to $tb$ hadronic search~\cite{Aaboud:2018juj} by using exclusive \kt subjets. 
\end{enumerate}

\begin{figure}
\includegraphics[width=0.98\linewidth]{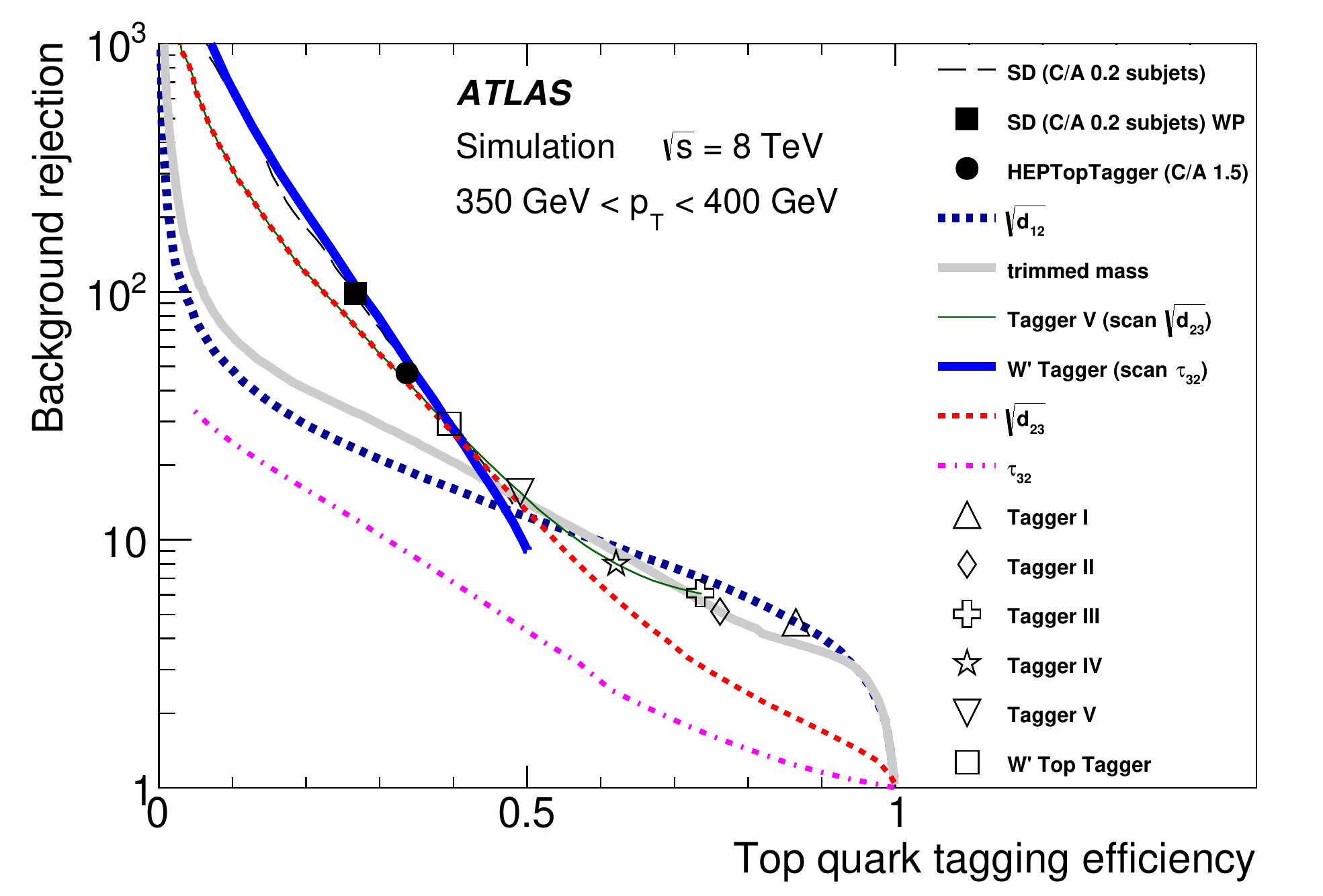}
\caption{Top quark tagging efficiency versus background rejection for various substructure variables and combinations in ATLAS. Taken from Ref.~\cite{Aad:2016pux}.}
\label{fig:atlas:ttbar_perf8_roc}
\end{figure}

In addition to the dedicated techniques described above, simpler algorithms using grooming and substructure similar to $V$ tagging methods have been investigated by ATLAS. 
A performance study at 7\TeV~\cite{Aad:2013gja} investigated a variety of performance metrics relating to the usage of groomed jets. Different grooming algorithms were investigated for their resilience to pile-up and mass resolution. It was concluded that trimmed \antikt jets with a distance parameter of $1.0$ and trimming parameters of $\rsub=0.3$ and $\fcut=0.05$ were a good candidate for a one-fits-all large-$R$ jet definition. This jet definition became standard in ATLAS for $W$/$Z$/$H$ and top quark tagging in \runone. 
A later ATLAS study~\cite{Aad:2016pux} investigated the various
methods available for tagging hadronic, highly boosted top quarks. The
so-called Tagger V has $\Mjet >100\GeV$, $\sqrt{d_{12}}> 40\GeV$ and
$\sqrt{d_{23}}>20\GeV$, where $\sqrt{d_{ij}}$ is the $\kt$-splitting
scale~\cite{Butterworth:2002tt}. The efficiency versus rejection is
shown for various taggers in figure~\ref{fig:atlas:ttbar_perf8_roc}. The difference between Taggers III and V is the additional requirement on $\sqrt{d_{23}}$ in Tagger V. At efficiencies smaller than 45\%, the $W^\prime$ tagger, based on $\sqrt{d_{12}}$ and the $N$-subjettiness ratios $\tau_{21}$ and $\toptau$, has better background rejection than Taggers III and V. ATLAS also tested the HTT and Shower Deconstruction~\cite{ATLAS-CONF-2017-064}, which have been found to have good background rejection (larger than 50) for efficiency values smaller than about 35\%. However, similar as for the CMS experiment, the background efficiencies of the two taggers show a significant rise with increasing \pt.

CMS has focused on enhancing the performance of CMSTT and HTT by identifying observables which carry discriminatory power, but have only small or moderate correlations with the observables used in the main algorithm. Typically, correlation coefficients of about 0.3 or less are required for noticeable improvement when augmenting an algorithm with additional variables. Examples for discriminating variables which fulfill this are $N$-subjettiness ratios, energy correlation functions and their ratios, and $b$ tagging. 
A study by the CMS Collaboration showed that at 20\% signal efficiency, the background rejection of the CMSTT can be improved by a factor of 5 when adding information from $\toptau$ and subjet $b$ tagging information~\cite{CMS-PAS-JME-13-007}. At higher efficiencies, the improvements become smaller. For the HTT, improvements of similar size are observed for $\pt>200\GeV$, becoming less significant at higher \pt.

The ATLAS choice of $R=1.0$ jets compared to CMS ($R=0.8$) results in an earlier rise of the tagging efficiency with increasing jet \pt. 

\begin{figure}
\begin{center}
\includegraphics[width=0.9\linewidth]{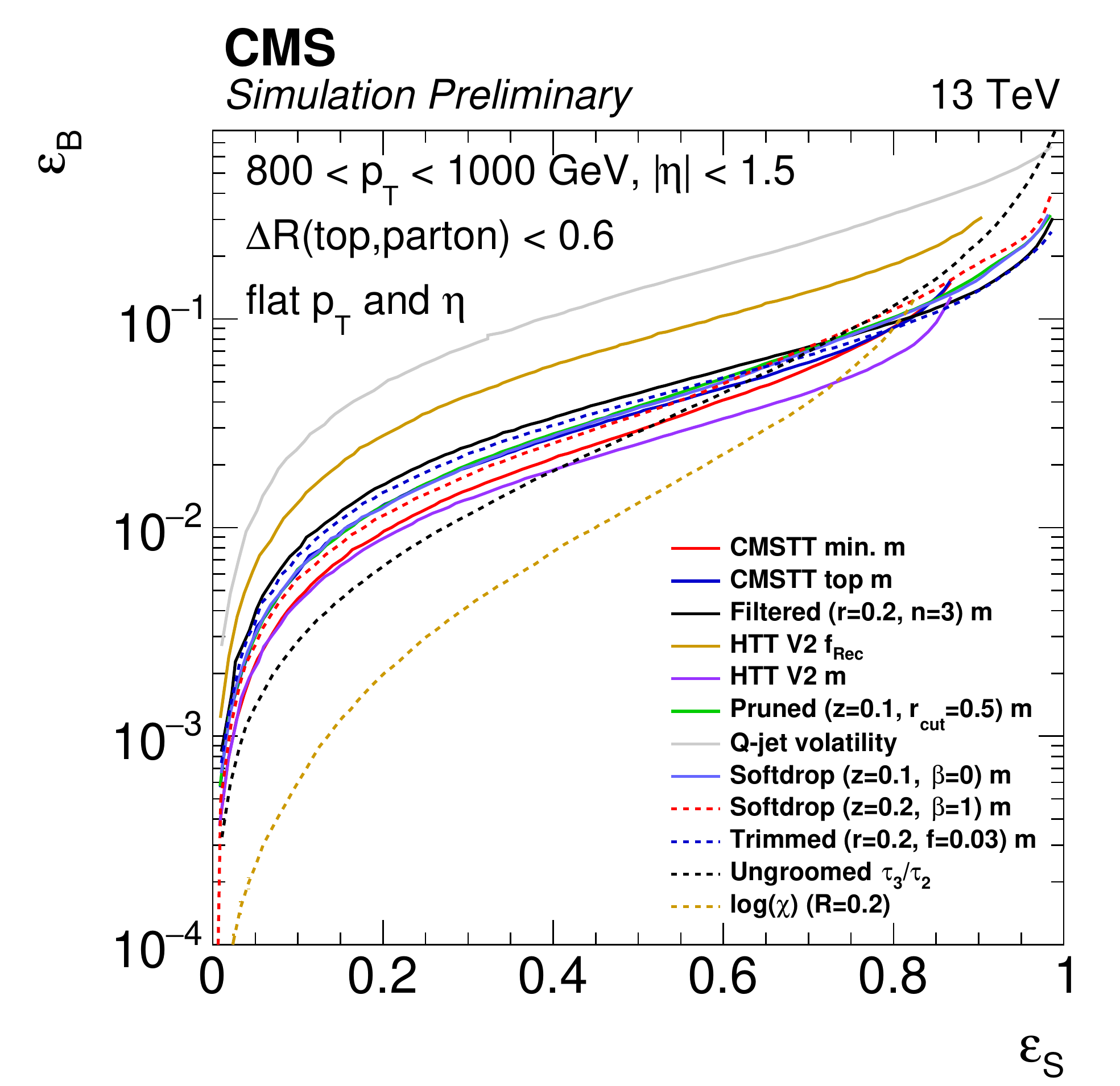}
\caption{Background versus signal efficiency for the single variables studied in the optimization of top tagging for CMS Run 2 analyses. 
Taken from Ref.~\cite{CMS-PAS-JME-15-002}.}
\label{CMSopt_single}
\end{center}
\end{figure}

The large difference in performance of the single variable $\toptau$
between ATLAS (figure~\ref{fig:atlas:ttbar_perf8_roc}) and CMS (figure~\ref{CMSopt_single}) is due to jet grooming. 
Although the CMS study shows only the ROC curves for $800<\pt<1000\GeV$, the overall picture does not change when studying top quarks in the region of $\pt\approx400\GeV$. Instead, in ATLAS $\toptau$ is calculated from trimmed jets, which results in less discrimination power when used as sole tagging variable compared to ungroomed $\toptau$. However, groomed $\toptau$ can still lead to considerable improvements when combined with other variables.

As with $V$ tagging discussed above, ATLAS and CMS took advantage of the LHC shutdown between \runone and \runtwo to perform broad studies of the different top-taggers available, with emphasis on their stability with respect to pile-up and other detector effects, instead of the utmost gain in performance~\cite{CMS-PAS-JME-15-002, ATL-PHYS-PUB-2015-053}. Single variables and their combinations are studied and compared with Shower Deconstruction, CMSTT, HTT, and an improved version of the {HTT} with shrinking cone size (HTTv2)~\cite{Kasieczka:2015jma}.

Figure~\ref{CMSopt_single} shows a comparison based on simulation of the single variable performance in CMS, where signal jets are generated through a heavy resonance decaying to $\ttbar$ and background jets are taken from QCD multijet production. Note that for this study reconstructed jets are matched to a generated parton, and the distance between the top quark and its decay products must be less than 0.6 (0.8) for a reconstructed $R = 0.8\ (1.5)$ jet, to ensure that the top quark decay products are fully merged and reconstructed in a single jet. The best single variable in terms of efficiency versus background rejection is the discriminator $\log\chi$, calculated with Shower Deconstruction. The second best variables are the $N$-subjettiness ratio $\toptau$ at low efficiency and the jet mass calculated with the HTTv2 at high efficiency values. The individual groomed jet masses show similar performance, and the CMS Collaboration moved to using the soft drop mass due to its beneficial theoretical properties~\cite{Larkoski:2017jix}. The default for CMS \runtwo analyses was chosen to be the soft drop jet mass combined with $\toptau$ for top tagging at high $\pt$. Generally, at high boost, the combination of a groomed mass with $\toptau$ leads to a large gain in background rejection.

\begin{figure}
\begin{center}
\includegraphics[width=0.9\linewidth]{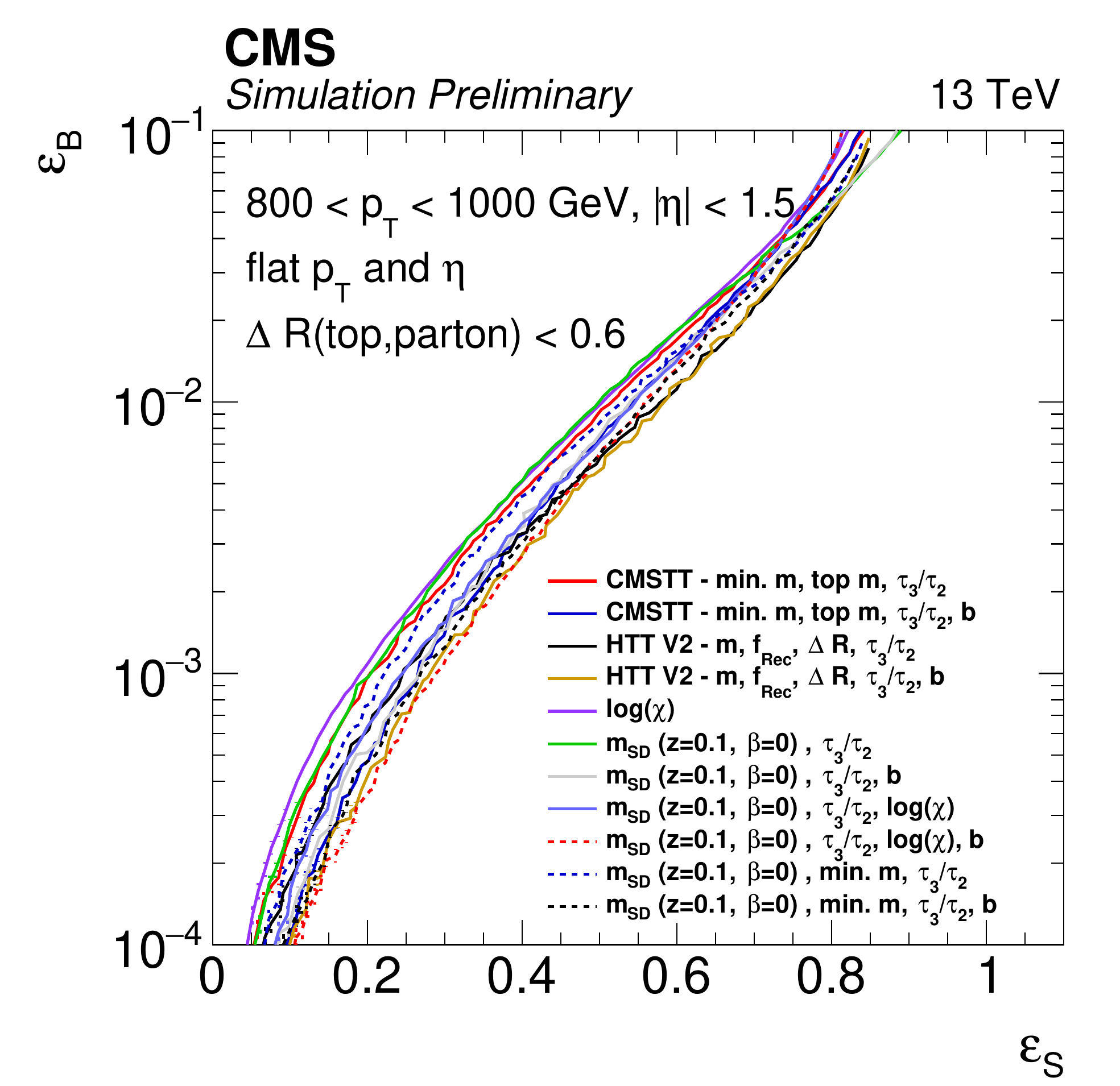}
\caption{Background efficiency versus signal efficiency for combined variables studied in the optimization of top tagging for CMS Run 2 analyses.
Taken from Ref.~\cite{CMS-PAS-JME-15-002}.}
\label{CMSopt_comb}
\end{center}
\end{figure}

The CMS study also investigated combining single variables with more
complex taggers. Combining Shower Deconstruction with the soft drop
mass, $\toptau$, and subjet $b$ tagging can lead to improvements, as
shown in figure~\ref{CMSopt_comb}; however, the efficiency and misidentification rate for this combination were found not to be stable as a function of jet $\pt$ (the combined algorithms were studied using working points corresponding to a background efficiency of 0.3).  At low boosts, the dedicated HTTv2 shows the best performance. In this kinematic region, using groomed $\toptau$, obtained by using the set of particles from the soft drop jet instead of the original jet, helps to improve the performance. 

In the shutdown between \runone and \runtwo, ATLAS commissioned a
single top tagger for use by physics analyses. The rationale behind
this approach was the potential benefit of having an efficient top
tagger with well-understood efficiency and associated systematic
uncertainties validated in the \runone dataset. Similarly as for
\runone, the supported top-tagger makes use of \antiktten trimmed
jets, but with a parameter of $\rsub=0.2$ instead of 0.3 as used in
\runone. Candidate top jets are required to satisfy a calibrated mass
window requirement $122.5 < \Mjet < 222.5\GeV$ and a $\pt$-dependent,
one-sided cut on $\toptau$~\cite{ATL-PHYS-PUB-2015-053}. 
The variable $\toptau$ has been chosen since it shows the best background 
rejection in combination with a small correlation with \Mjet,
a reduced \pt-dependence, and good performance 
across a large range in $\pt$. 

A common problem of top-tagging algorithms is the rise of the misidentification rate with increasing \pt, which is due to the peak of the mass distribution for quark- and gluon-initiated background jets shifting to higher values. For some taggers, for example the CMSTT, this shift also results in a decrease of the efficiency once a very high \pt threshold is crossed (larger than 1\TeV)~\cite{CMS-PAS-JME-10-013}. A possible solution to this is offered by the variable-$R$ (VR) algorithm, introduced in section~\ref{sec:BosonTagging}. The ATLAS Collaboration studied the performance of the VR algorithm for top-tagging and reported a stabilization of the position of the jet mass peak for a large range of \pt~\cite{ATL-PHYS-PUB-2016-013}. The VR jets are shown to improve the performance of the jet mass, $\sqrt{d_{12}}$ and $\toptau$ for top tagging, when compared to trimmed jets. An interesting development using VR jets is the Heavy Object Tagger with variable-$R$ (HOTVR)~\cite{Lapsien:2016zor}, which combines the VR algorithm with a clustering veto, resulting in a single jet clustering sequence producing groomed jets with subjets. 

Most top-taggers target either the region of low to intermediate boosts, or the highly boosted regime. However, in typical searches for new physics at the LHC non-vanishing efficiency for the full kinematic reach is crucial. Several attempts of combining different reconstruction and identification algorithms have been made. A search for resonances decaying to \ttbar by the ATLAS Collaboration uses a cascading selection from boosted to resolved~\cite{Aad:2015fna}, where the resolved topology is reconstructed and identified using a $\chi^2$-sorting algorithm. To efficiently identify top quarks over a broad \pt range in the search for top squark pair production, reclustered variable-$R$ jets are used with $R = 0.4$ jets as inputs to the jet reclustering algorithm~\cite{Aaboud:2017aeu,Aaboud:2016lwz}.

\begin{figure}
\includegraphics[width=0.95\linewidth]{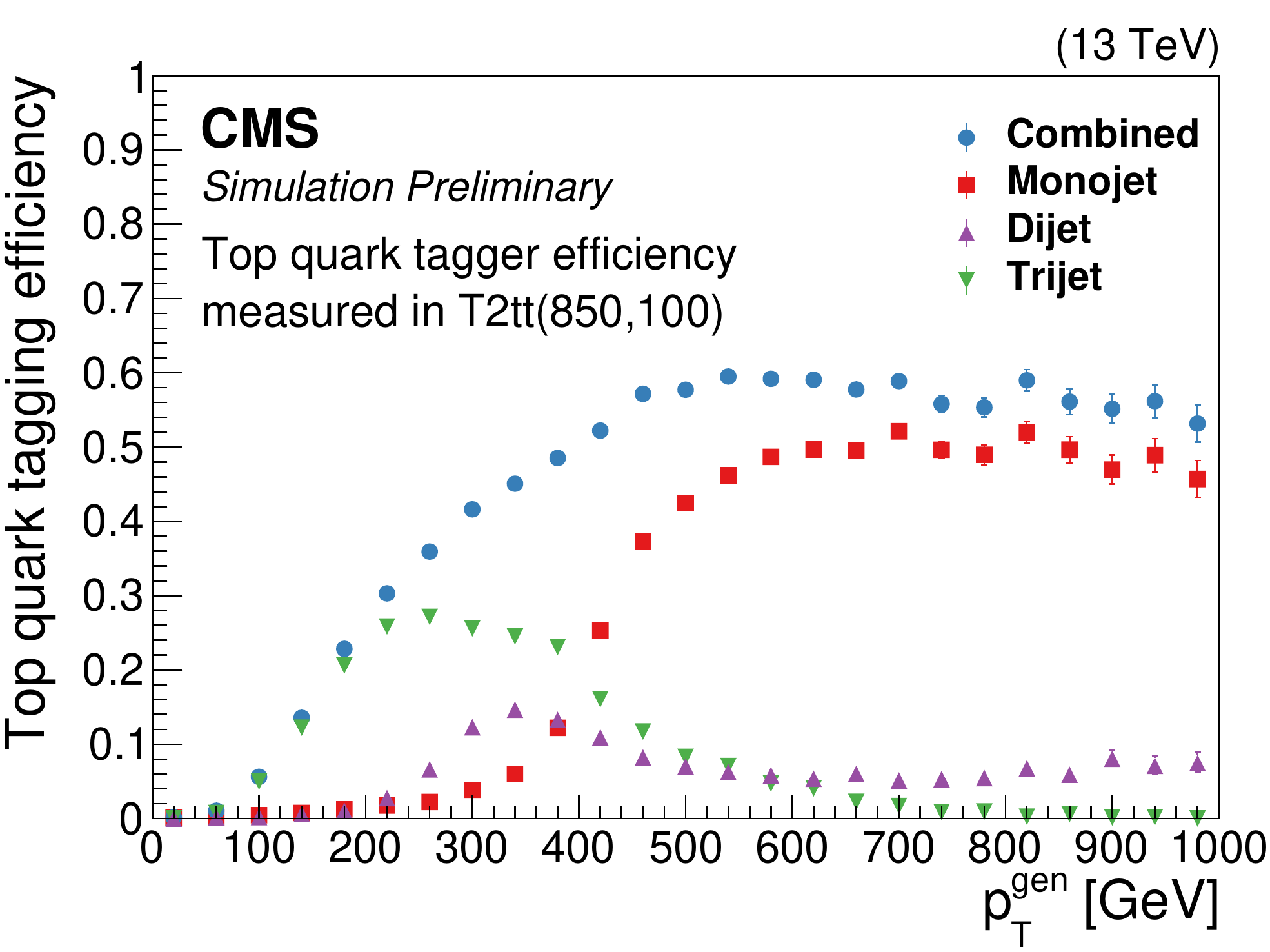}
\caption{Top tagging efficiency of three different top tagging methods and the combined efficiency, as a function of the generated top quark \pt. Taken from Ref.~\cite{CMS-PAS-SUS-16-050}. }
\label{fig:CMS_SUSY_RTT}
\end{figure}

A search for supersymmetry in CMS~\cite{CMS-PAS-SUS-16-050} uses three
distinct topologies: fully-merged top quark decays (Monojet), merged
$W$ boson decays (Dijet) and resolved decays (Trijet). The efficiency
of the three categories is shown in figure \ref{fig:CMS_SUSY_RTT}, where the turn-on of the combined efficiency starts at values as low as $\pt \approx 100\GeV$. The resolved trijet category is identified using three \antikt jets with a distance parameter of 0.4, where the large combinatorial background is suppressed through a multivariate analysis, which achieves a misidentification rate of approximately 20\%. 
There exist other approaches to cover the transition from low to high Lorentz boosts, using a single algorithm. In the HTTv2 algorithm, the jet size is reduced until an optimal size $R_{\rm opt}$ is found, defined by the fractional jet mass contained in the smaller jet. This results in better performance at high \pt, while keeping a low misidentification rate at low \pt.

An important step towards the commissioning of top taggers within an experiment are measurements of the efficiency and misidentification rate in real collision data. Generally, high-purity samples of top-jets in data are obtained using a tight signal selection (an electron or muon, well-separated from a high-\pt large-$R$ jet, and an additional $b$-tagged jet) to ensure that events contain a fully-merged top quark decay in a single large-$R$ jet. This can never be fully achieved, as no requirements on the substructure of the large-$R$ jet can be imposed without biasing the efficiency measurement. This results in an efficiency measurement that will be based on a sample also containing partially-merged or even non-merged top quark decays. These can be subtracted from the efficiency measurement by using simulated events, as done in a study by the ATLAS Collaboration~\cite{Aad:2016pux}, with the drawback of relying on a specific simulation and the ambiguous definition of a fully-merged top quark decay. By not correcting for non-merged top quark decays, efficiency values are obtained smaller than the ones suggested by ROC curve studies, see for example~\cite{CMS-PAS-JME-15-002}.  Instead of subtracting the top-backgrounds, the CMS collaboration performs a simultaneous extraction of the efficiencies for fully- and partially-merged categories~\cite{CMS-DP-2017-026}.  

Measurements of the misidentification rate can be carried out by selecting a dijet sample, which is dominated by light-flavor jets. The disadvantage of this approach is the high \pt threshold of unprescaled jet triggers, which results in measurements starting from $\pt>400\GeV$ or higher. A solution to this is the tag-and-probe method, in which the tagged jet can be required to fail top-tagging selection criteria, resulting in a sample with negligible contamination of \ttbar production, even after requiring the probe jet to be top-tagged~\cite{CMS-PAS-JME-15-002}. Another approach is to use a non-isolated electron trigger, where the electron fails offline identification criteria. This yields events mainly from light-flavor multijet production, where a jet is misidentified as an electron at the trigger level. While the top-tag misidentification rate can be measured starting from smaller values of \pt with this strategy, a non-negligible amount of \ttbar contamination has to be subtracted after requiring a top-tagged jet~\cite{Aad:2016pux}. 

\begin{figure}
\includegraphics[width=0.9\linewidth]{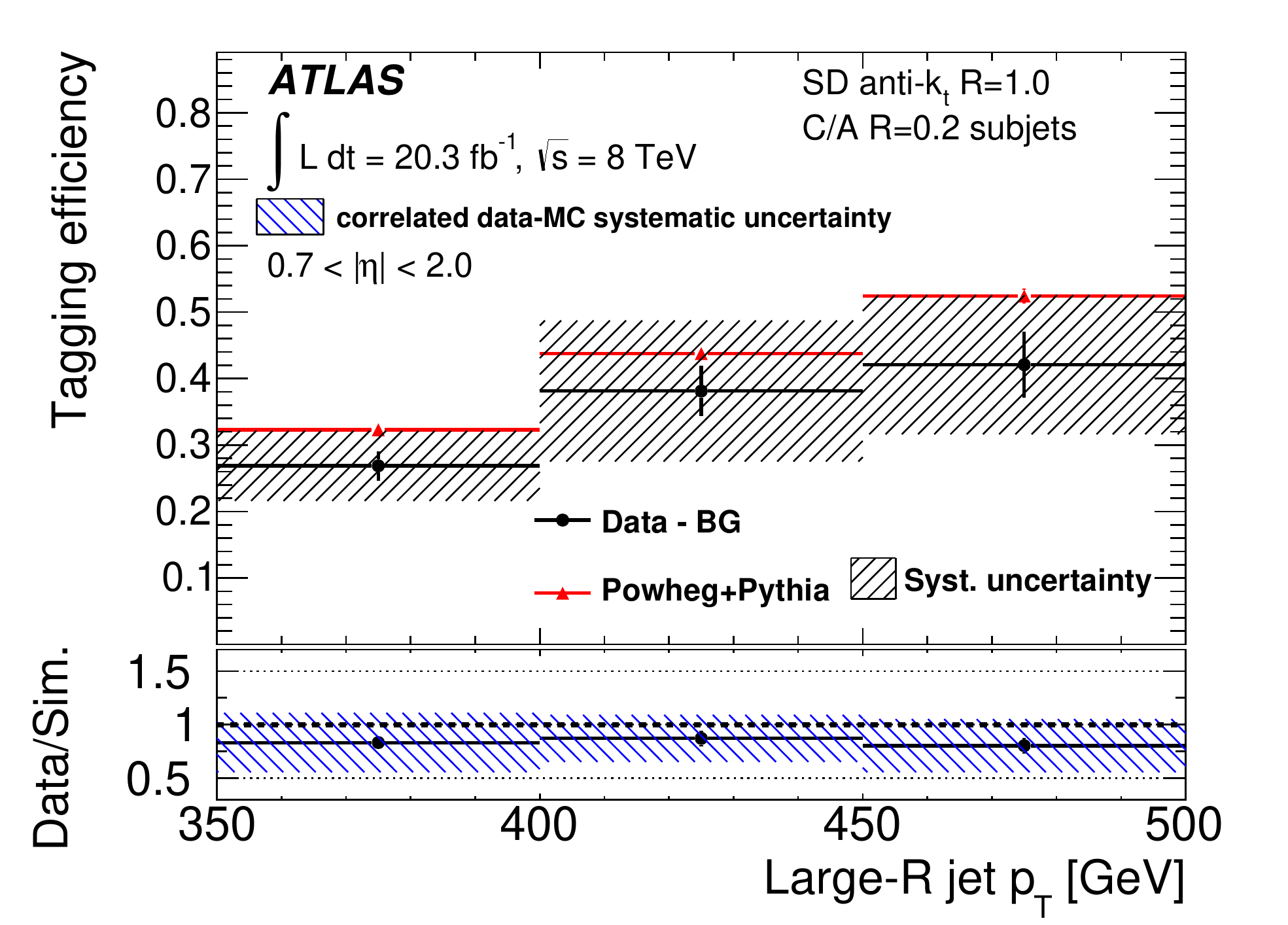}
\includegraphics[width=0.9\linewidth]{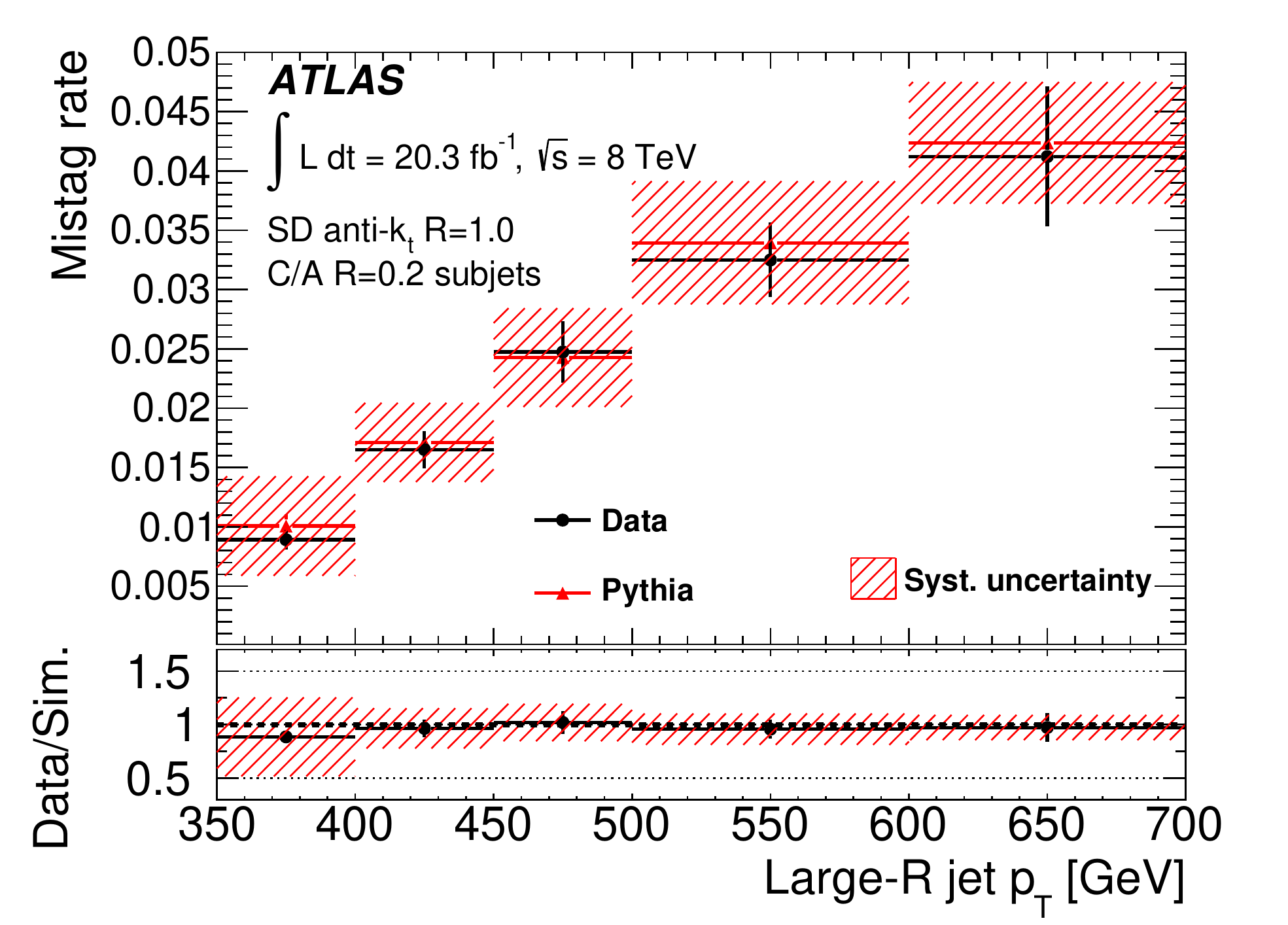}
\caption{ATLAS measurement of the efficiency (top) and misidentification rate (bottom) for trimmed jets with a distance parameter of $1.0$ tagged with Shower Deconstruction. Taken from Ref.~\cite{Aad:2016pux}. }
\label{fig:eff_SD_ATLAS}
\end{figure}

As an example, the efficiency and misidentification rate of Shower
Deconstruction with the requirement $\log(\chi) > 2.5$, as measured in
ATLAS, are shown in figure \ref{fig:eff_SD_ATLAS}. The efficiency of
30\% with a misidentification rate of 1\% for $350<\pt<400\GeV$ agrees
well with the values obtained from figure~\ref{fig:atlas:ttbar_perf8_roc}. Note that the largest uncertainty of the efficiency measurement stems from the choice of the Monte-Carlo (MC) event generator used to simulate \ttbar production. The uncertainty of the misidentification rate measurement is dominated by the energy scales and resolutions of the subjets and large-$R$ jets. 

\subsection{$H\to b\bar{b}$ Tagging}\label{sec:btagging}

The identification of jets originating from the fragmentation of $b$ quarks 
($b$ tagging) is a crucial task in many areas of particle physics. 
Algorithms used for $b$ tagging usually rely on the distinct signature 
of $B$ hadron decays, for example the presence of a secondary vertex 
due to the long $B$ hadron lifetime of about 1.5\,ps.  

ATLAS and CMS both use dedicated $b$ tagging algorithms that have 
been developed and optimized over more than a decade. 
Both experiments use multivariate techniques with various input parameters 
related to the secondary vertex or charged particle tracks originating 
from the $B$ hadron decay. For \runtwo analyses, CMS uses the CSVv2 algorithm~\cite{BTV-16-002} and ATLAS uses the MV2c10 algorithm~\cite{ATL-PHYS-PUB-2016-012}. Typically, efficiencies of around 
70\% with misidentification rates of 1\% for light quark and gluon jets and 
20\% for charm jets are achieved with these algorithms.

While $b$ tagging in busy hadronic environments plays an important 
role for top tagging, it is the key challenge for tagging boosted 
$H\to b\bar{b}$ signatures.  
Other jet substructure observables can improve performance, but are often 
less powerful once two $b$ tagged jets or subjets are required 
(as this necessarily forces the jet to have two-prongs).  The lighter 
mass of the Higgs boson compared with the top quark also means that 
the $b$-jets from the $H$ decay become merged at a lower parent particle boost.

The boosted $H\to b\bar{b}$ signature is present in many models of physics beyond the Standard Model: resonant $HH$ and $VH$ production, searches for boosted mono-$H$, or vector-like quark searches in the $tH$ and $bH$ final states. Because of the large predicted branching fraction for the $H \to b\bar{b}$ decay of about 58\%, its coupling to $b$ quarks is one of the most interesting to study. For a large fraction of Higgs bosons with \pt $>$ 300\GeV, the two $b$ quark jets merge into a single jet for a jet distance parameter of $R = 0.8$ or $1.0$, as used in CMS and ATLAS, respectively. Several phenomenological studies have explored $H \to b\bar{b}$ tagging algorithms using jet substructure, though ultimately the optimal performance comes from using a combination of substructure information and the track and vertex information related to the $B$ hadron lifetime.

The approaches to identify boosted $H \to b\bar{b}$ candidates that have been explored (and used) at CMS and ATLAS include:

\begin{enumerate}
\item[(a)] Subjet $b$ tagging~\cite{Chatrchyan:2012jua, CMS-PAS-BTV-13-001, ATLAS-CONF-2016-002, Aad:2015ydr, ATLAS:2016wlr, ATL-PHYS-PUB-2014-013}, where `standard' $b$ tagging is applied to each of the subjets (the standard for CMS is the CSVv2 algorithm~\cite{CMS-PAS-BTV-15-001}, and for ATLAS is MV2c20~\cite{ATL-PHYS-PUB-2016-012}). Tagging $b$-jets in dense environments is of particular importance here, and was studied by ATLAS in Ref.~\cite{ATL-PHYS-PUB-2014-014}. In CMS subjets with $R=0.4$ are clustered with the $\kt$ algorithm using the constituents of the large-$R$ jet, while for ATLAS track jets with a radius of $0.2$ are matched to the large-$R$ jet using the ghost-association technique. At high $\pt$ the subjets start to overlap causing the standard $b$ tagging techniques to break down due to double-counting of tracks and secondary vertices when computing the subjet $b$ tag discriminants. 
\item[(b)] Double-$b$
  tagging~\cite{ATLAS:2016wlr,CMS-PAS-BTV-15-002,BTV-16-002}, where in
  ATLAS, the term double-$b$ tagging means that the two leading $\pt$
  track jets must pass the same $b$ tagging requirement. In CMS, the
  double-$b$ tagger~\cite{CMS-PAS-BTV-15-002,BTV-16-002} uses the $N$-subjettiness axes and the pruned 
  \antikteight jet mass with a window of $50 < M < 200\GeV$ to reduce the multijet background.
\end{enumerate}

\begin{figure}[tb]
	\centering
    \includegraphics[width=0.45\textwidth]{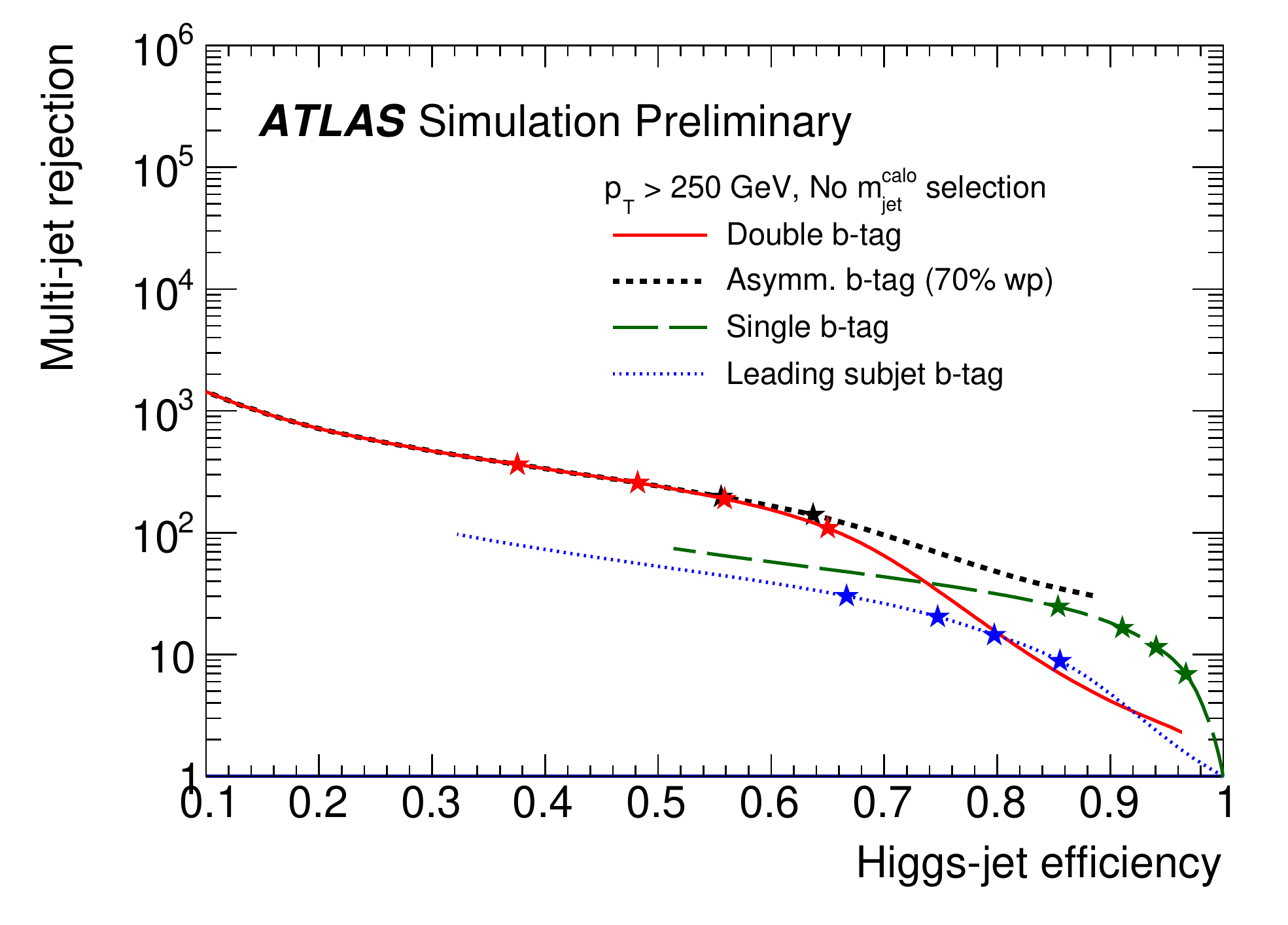}
    \caption{The rejection of inclusive multijets versus Higgs-jet efficiency using all large-$R$ jets with $\pt>250\GeV$ for single, double, asymmetric, and leading subjet $b$ tagging requirements. Taken from Ref.~\cite{ATLAS:2016wlr}. \label{fig:atlas}}
 \end{figure}

The Higgs-jet efficiency versus the inclusive multijet rejection are
shown in figure~\ref{fig:atlas} for ATLAS subjet $b$ tagging, where the
performance curves are shown for double-$b$ tagging, leading subjet $b$ tagging, and asymmetric $b$ tagging\footnote{
Asymmetric $b$ tagging means that among the two leading $\pt$ track jets, the track jet with the largest $b$ tagging weight must pass the fixed 
70\% $b$ tagging working point threshold, while the $b$ tagging requirement of the other jet is varied.} 
requirements. None of the curves reach a Higgs-jet efficiency of 100\% due to the imperfect efficiency to reconstruct the track jets
needed for $b$ tagging and, in the case of asymmetric $b$ tagging, also due to the 70\% $b$ tagging working
point requirement on one of the track jets.

The CMS double-$b$ tagging algorithm~\cite{CMS-PAS-BTV-15-002,BTV-16-002} attempts to fully exploit the strong correlations between the $b$ hadron flight directions and the energy flows of the two subjets, while adapting the variables used in the CSVv2 algorithm. The flexibility of the double-$b$ tagger is ensured by avoiding a strong performance dependence on the jet $\pt$ and mass. 

\begin{figure}[tb]
\begin{center}
\includegraphics[width=0.4\textwidth]{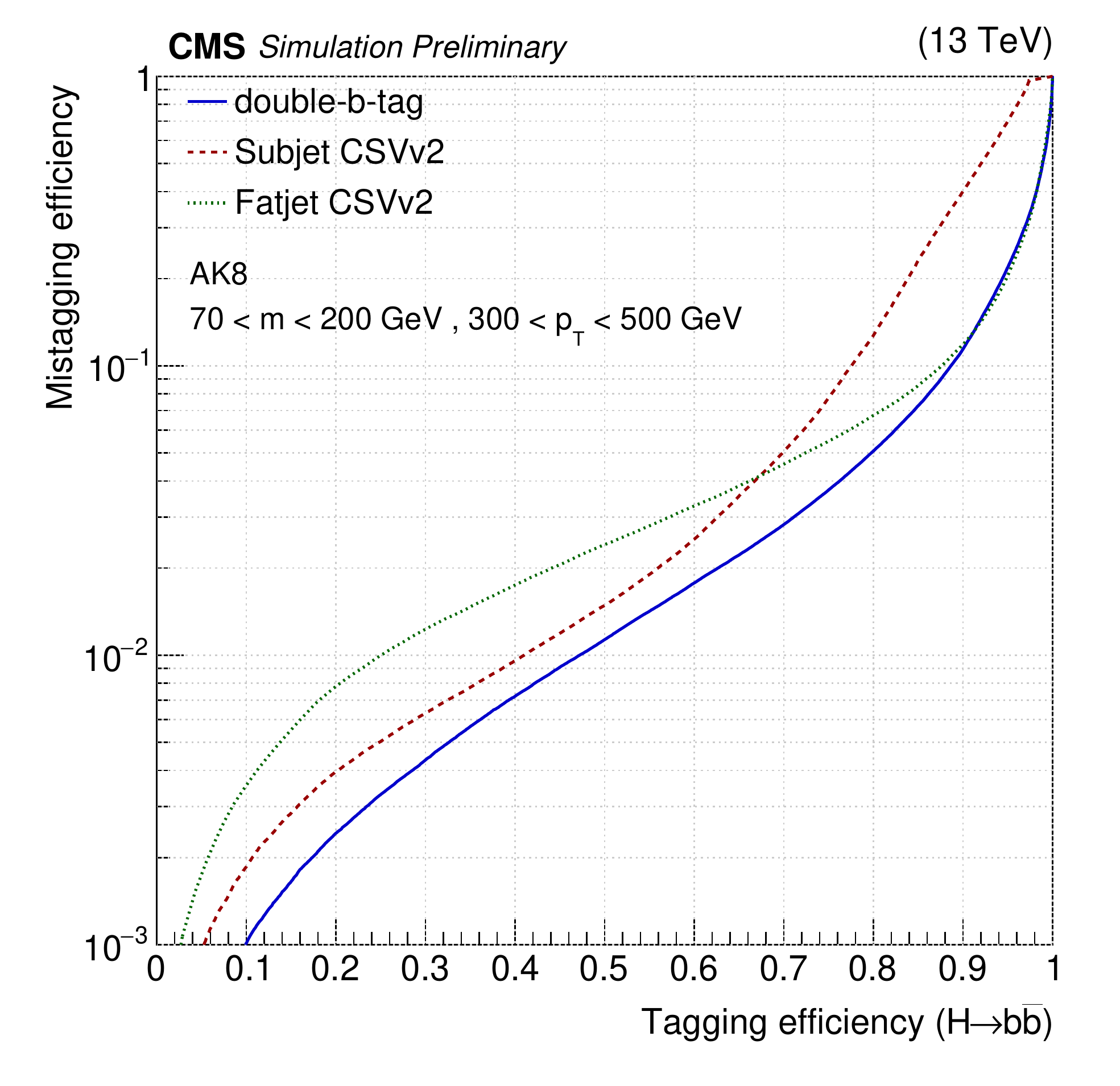}
 \caption{The misidentification rate for inclusive multijets versus Higgs-jet efficiency using jets with $300 < \pt < 500\GeV$ and pruned jet mass $70 < m < 200\GeV$ for three different $b$ tagging requirements. Taken from Ref.~\cite{CMS-PAS-BTV-15-002}.}\label{fig:cmsdoubleb}
 \end{center}
 \end{figure}

With the double-$b$ tagger, at the same signal efficiency, the misidentification rate is uniformly lower by about a factor of two compared to the subjet $b$ tagging approach. Given the different kinematic properties expected for a $b\bar{b}$ pair originating from the decay of a massive resonance compared to gluon splitting, the misidentification rate for the gluon splitting background reduces from 60\% to 50\% at 80\% signal efficiency and from 20\% to 10\% at 35\% signal efficiency. At high $\pt$, even larger performance improvements are observed, which is an important gain for searches for heavy resonances, where very high $\pt$ jets are expected. In figure~\ref{fig:cmsdoubleb} the signal efficiencies and misidentification rates for the double-$b$ tagger are shown as a function of jet $\pt$ for three operating points: loose, medium and tight, which correspond to 80\%, 70\% and 35\% signal efficiency, respectively, for a jet $\pt$ of about 1000\GeV. The misidentification rate is mostly flat across the $\pt$ range considered while the signal efficiency decreases with increasing $\pt$, as expected from the degradation of the tracking performance inside high $\pt$ jets.

Due to the small cross section of producing events with boosted $H \to b\bar{b}$ or $Z \to b\bar{b}$ jets, the efficiency of the ATLAS and CMS Higgs identification algorithms is measured using QCD multijet events enriched in jets from gluon splitting, $g \to b\bar{b}$ with a topology similar to that of boosted $H \to b\bar{b}$ jets.

CMS selects topologies as similar as possible to a signal jet by requiring the jet $\pt > 300\GeV$ and pruned mass $>$ 50\GeV~\cite{CMS-PAS-BTV-15-002,BTV-16-002}. Each jet has to contain at least two muons, each with $\pt > 7\GeV$ and $|\eta|<2.4$.  Each pruned subjet is required to have at least one muon among its constituents and within $\Delta R< 0.4$ from the subjet axis ("double-muon tagged"). The  double-muon tag enriches events with gluons splitting into $b\bar b$ where both $b$ quarks give rise to a semi-leptonic $B$ hadron decay. Such $g\to b\bar b$ events are proxies for the signal topology. An alternative selection that requires at least one muon is also examined as a cross-check for the measurement ("single-muon tagged"). While this single-muon selection allows for a larger dataset in which to perform the tagger efficiency measurement, the gluon splitting topology in this inclusive phase space is less signal-like relative to the double-muon selection. Thus, to maximize the similarity between the $g\to b\bar{b}$ and the $H\to b\bar{b}$ topologies, the measurement is performed requiring double-muon tagged jets. It is worth noting however that the jet mass depends on the number of muons and a large fraction of the signal will not contain two muons.

ATLAS performed a similar measurement selecting events with at least one \antiktten jet with $\pt > 250\GeV$ that has two ghost-associated $R = 0.2$ track jets~\cite{ATLAS:2016wlr}. As opposed to the measurement from CMS, only one of the subjets is required to have a muon associated to it. Kinematic and substructure variables are compared in data and MC after correcting for flavor composition differences of the large-$R$ jet observed between data and MC simulation and are found to be in good agreement.

One of the major backgrounds for analyses selecting boosted $H$ or $Z$ bosons decaying to $b\bar{b}$ is $t\bar{t}$ production. The misidentification rate for boosted top quark jets faking $H$ jets was measured in data by CMS~\cite{CMS-PAS-BTV-15-002,BTV-16-002} in enriched data samples of lepton+jets $t\bar{t}$ events.

As previously discussed, for high \pt of the Higgs boson, the two subjets from $b$ quarks start overlapping and the performance of identifying the subjets as fixed-radius track jets decreases significantly. To improve the performance of the ATLAS standard $H\rightarrow b\bar{b}$ identification algorithm for searches that require the presence of high \pt Higgs bosons, the ATLAS Collaboration studied alternative methods like the use of variable-radius track jets, exclusive \kt subjets, calorimeter subjets reconstructed in the center-of-mass frame of the Higgs jet candidate~\cite{ATL-PHYS-PUB-2017-010} and the combination of three jet shape and jet substructure variables into a multivariate discriminator~\cite{ATLAS-CONF-2012-100}. For highly boosted Higgs bosons, these reconstruction techniques significantly outperform the usage of fixed-radius track jets.

\section{Standard Model Cross Section Measurements\label{sec:crosssections}}

The measurement of jet properties is crucial to constrain the Standard Model in new energy regimes and constitutes an important test of perturbative calculations of jet structure over a wide region of phase space. Moreover jet cross section measurements provide constraints on the parton distribution functions and the strong coupling constant, $\alpha_s$.  The precise knowledge of jet properties also improves the precision of other measurements and searches by constraining the modeling of important background processes. Jet substructure observable measurements are challenging as they require a precise measurement of the radiation pattern within the jet and thus a detailed understanding of the jet constituent properties. Section~\ref{sec:meas} describes measurements of various jet substructure properties, starting from the most widely used and well-understood: the jet mass.

Jet substructure properties can also be used to extend measurements of SM cross sections to higher energy, where access to the hadronic branching ratios of $W/Z/H$ bosons and top quarks is important.   Section~\ref{sec:meas2} introduces cross section measurements for SM objects at high $p_\text{T}$.  The use of jet substructure in these cases is similar to the application for the searches described in the next section (Section~\ref{sec:searches}).

\subsection{Measurements of Jet Substructure\label{sec:meas}}

\subsubsection{Jet mass \label{sec:jetmassmeas}}

\begin{figure}[tbp]
\includegraphics[width=0.43\textwidth]{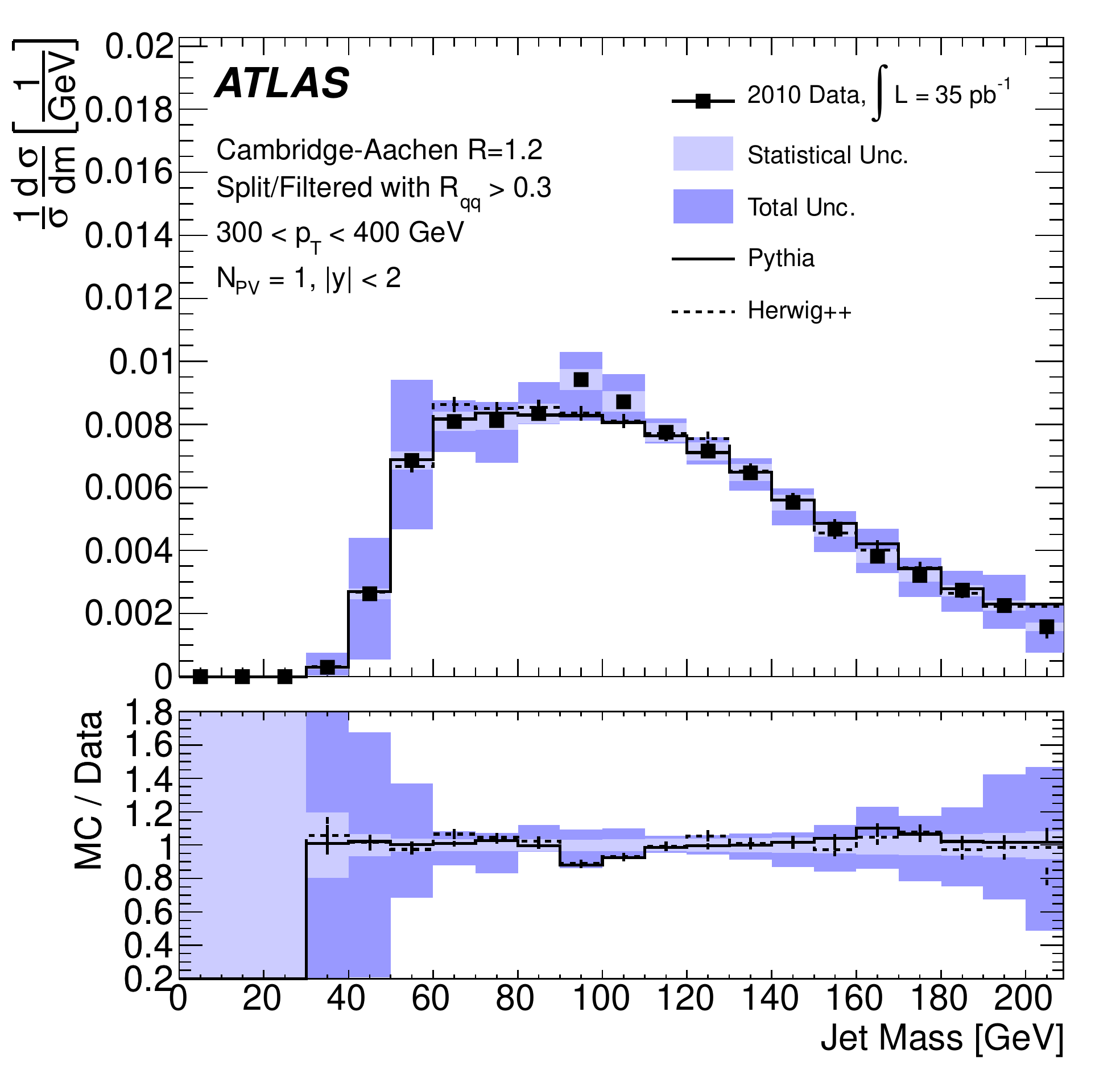}
\caption{Normalized differential cross section as a function of the jet mass for CA jets with $R = 1.2$ after splitting and filtering, taken from Ref.~\cite{ATLAS:2012am}.}
\label{fig:mass1}
\end{figure}

The first measurement of the normalized dijet differential cross section as a function of the jet mass was performed by the ATLAS Collaboration with a dataset corresponding to 35~pb$^{-1}$ of 7~\TeV $pp$ collisions~\cite{ATLAS:2012am}. Both the cross section for groomed and ungroomed CA $R = 1.2$ jets was measured separately to gain sensitivity to both the hard and soft jet physics and to gain a deeper understanding of the various effects involved in QCD radiation. For the ungroomed jet mass, large discrepancies were observed in the tails of the mass distribution between the predictions from the MC event generators Pythia and Herwig++, and the data, whereas the core of the mass distribution agreed within approximately 20\% over the considered \pt range. The largest discrepancies occur at low jet masses which is sensitive to the underlying event description, hadronization model and pile-up effects. The normalized cross section after applying the split filtering algorithm~\cite{Butterworth:2008iy} is shown in figure~\ref{fig:mass1} with the mass drop parameters $\mu_{\rm frac} = 0.67$ and $y_{\rm filt} = 0.09$, 
and a filtering parameter of $R_{\mathrm{filt}}=\min(0.3,\Delta R /2)$. After removing soft radiation from the jet which is difficult to model, the MC prediction is in excellent agreement with the data within statistical precision. The CMS Collaboration performed a similar measurement with \antikt $R = 0.7$ jets using various grooming techniques in selected dijet events using 5~fb$^{-1}$ of $\sqrt{s} = 7$~\TeV data and found as well that the agreement between data and the MC prediction improves significantly after grooming techniques are applied~\cite{Chatrchyan:2013vbb}. Furthermore a measurement of the cross section was performed in $V$+jet final states which overall show a slightly better data/MC agreement than that observed in dijet events suggesting that the simulation of quark jets is better than for gluon jets.

The CMS (ATLAS) Collaboration measured the double-differential jet cross section in balanced dijet events at $\sqrt{s} = 13$~\TeV for groomed \antikt $R = 0.8$ jets with the soft drop algorithm with $\zsoftdrop = 0.1$ and $\beta = 0$ ($\beta = 0,1,2$)~\cite{Sirunyan:2018xdh, Aaboud:2017qwh}. The soft drop algorithm was chosen as it allows to compare the unfolded measurement directly to theoretical calculations which exceed the precision of parton shower MC simulations. The jet energy of the ungroomed jets used in the ATLAS measurement are corrected for pile-up effects and calibrated to the generator-level while no explicit mass calibration is applied to the groomed jets as the unfolding procedure accounts for differences between the reconstructed and generator-level mass. The CMS Collaboration applied calibration factors derived from simulation and using \textit{in situ} techniques (from boosted $W$ bosons) to correct the jet energy and mass scale. Furthermore the jet energy and mass are smeared in MC simulation to match the resolution measurements in data. Various sources of systematic uncertainties, categorized as experimental and theoretical uncertainties, that impact the jet mass measurement are taken into account. While CMS evaluated the effect of the jet energy and mass scale uncertainties on the measurement by varying the energy and mass by their respective uncertainties, ATLAS evaluated the experimental uncertainties based on the accuracy of the modelling of the topological cluster energies and positions as well their reconstruction efficiency.  Theoretical uncertainties on the physics model are taken into account by comparing the response matrix for various MC generators. 

\begin{figure}[tb]
\includegraphics[width=0.49\textwidth]{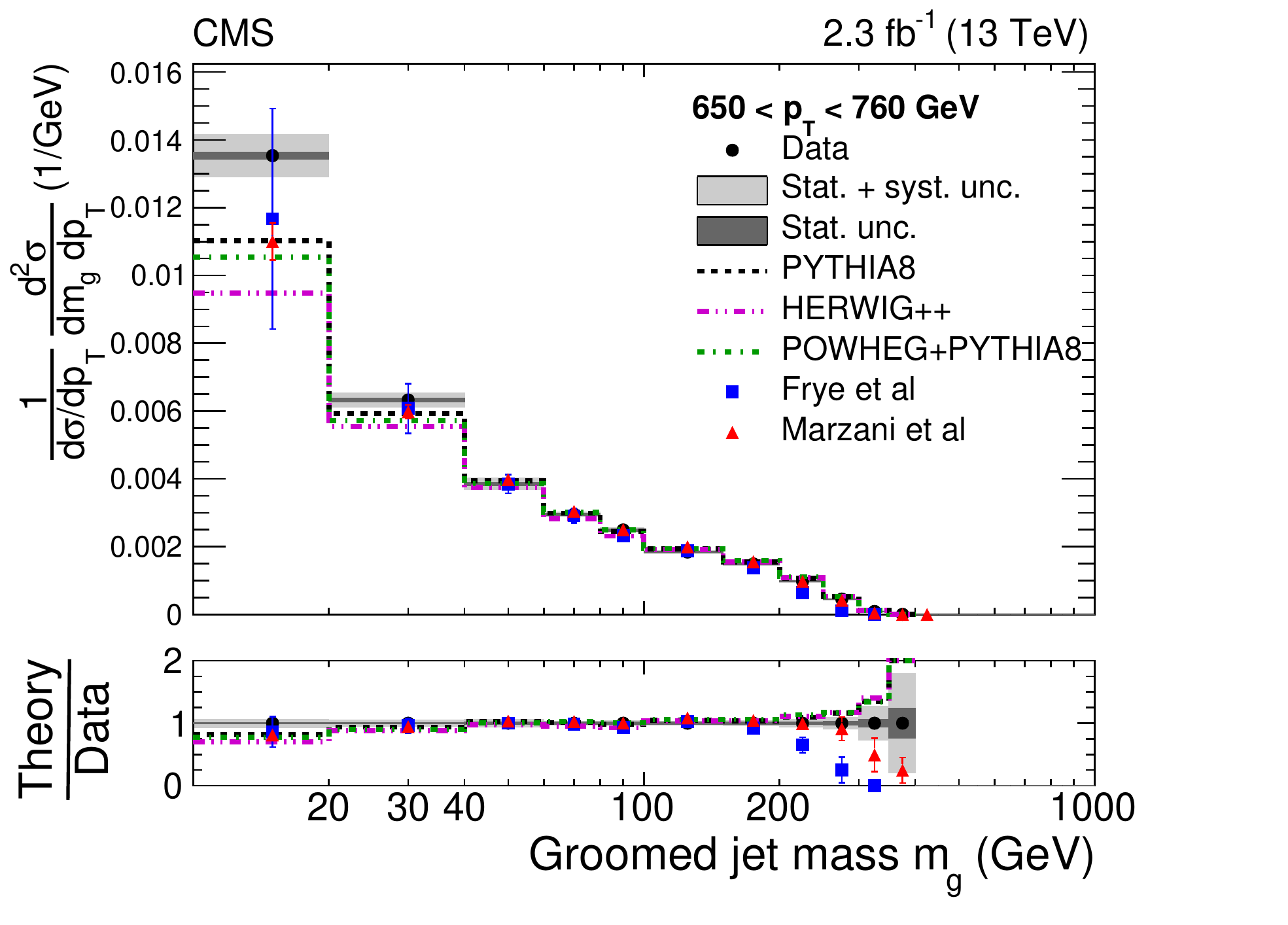}
\caption{Normalized differential cross section as a function of the mass for jets groomed with the soft drop algorithm in data and for two theoretical calculations. Taken from Ref.~\cite{Sirunyan:2018xdh}.}
\label{fig:mass2}
\end{figure}

\begin{figure}[tb]
\includegraphics[width=0.49\textwidth]{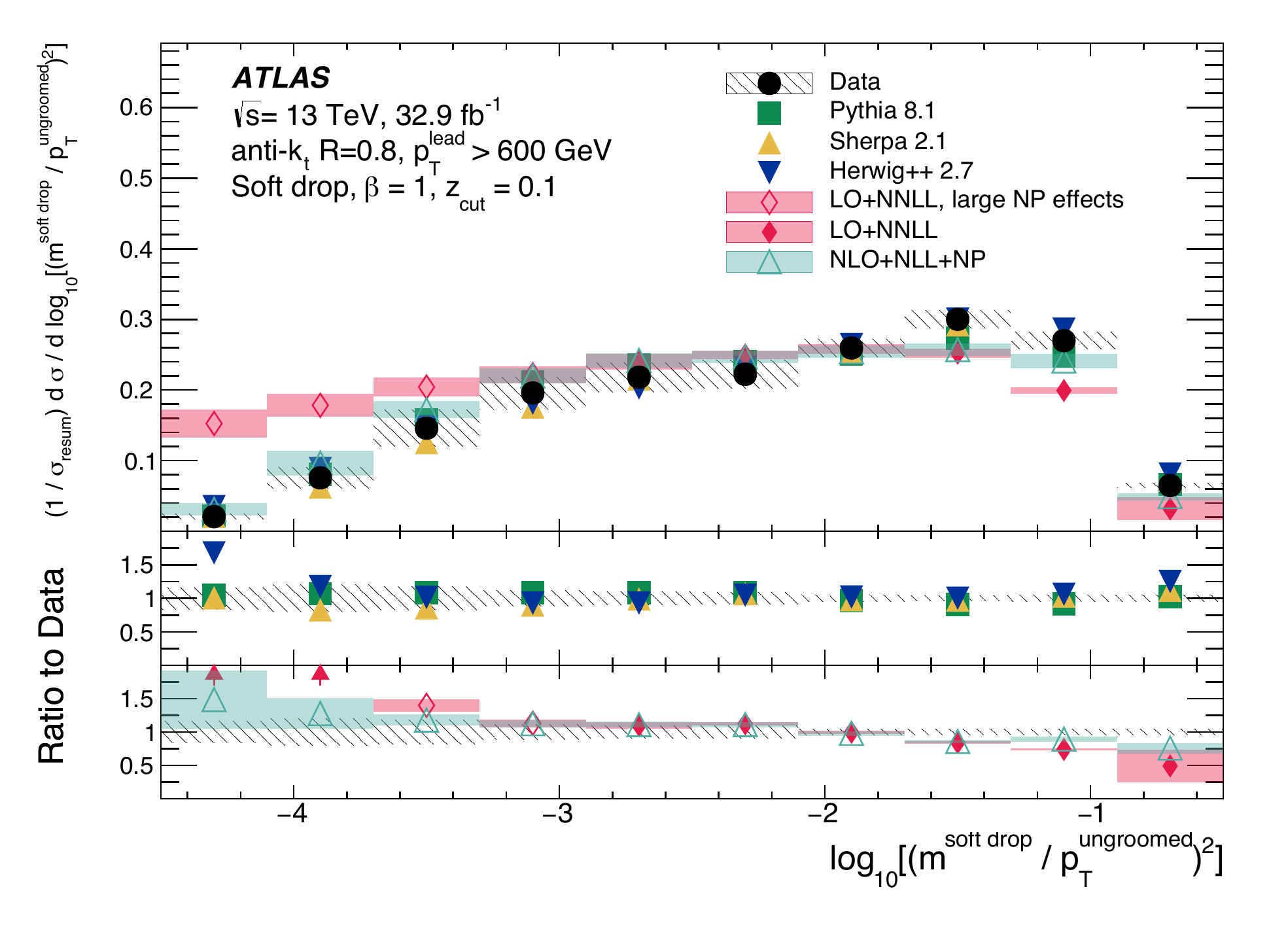}
\caption{Comparison of the unfolded $\log_{10}\rho^2$ distribution for $\zsoftdrop = 0.1,\ \beta = 1$ in data to various Monte Carlo particle-level predictions and theory predictions, normalized to the integrated cross section measured in the resummation regime $-3.7 < \log_{10}\rho^2 < -1.7$. Taken from Ref.~\cite{Aaboud:2017qwh}.}
\label{fig:ATLAS_SD_meas}
\end{figure}

The comparison of the normalized cross section with two analytical calculations as measured by CMS is shown in figure~\ref{fig:mass2}. ATLAS measured instead the $\log_{10}\rho^2$ distribution, shown in figure~\ref{fig:ATLAS_SD_meas}, where $\rho$ is the ratio of the soft drop jet mass to the ungroomed jet \pt. Both measurements are compared to calculations at next-to-leading order with next-to-leading-logarithm and leading order with next-to-next-to-leading-logarithm accuracy. Good agreement between the data and the predictions is observed in resummation regime $-3.7 < \log_{10}\rho^2 < -1.7$.  For higher jet masses, where fixed-order effects play an important role, the NLO+NLL calculation provides a better description than the LO+NNLL calculation. 

\begin{figure}[tb]
\centering
\includegraphics[width=0.43\textwidth]{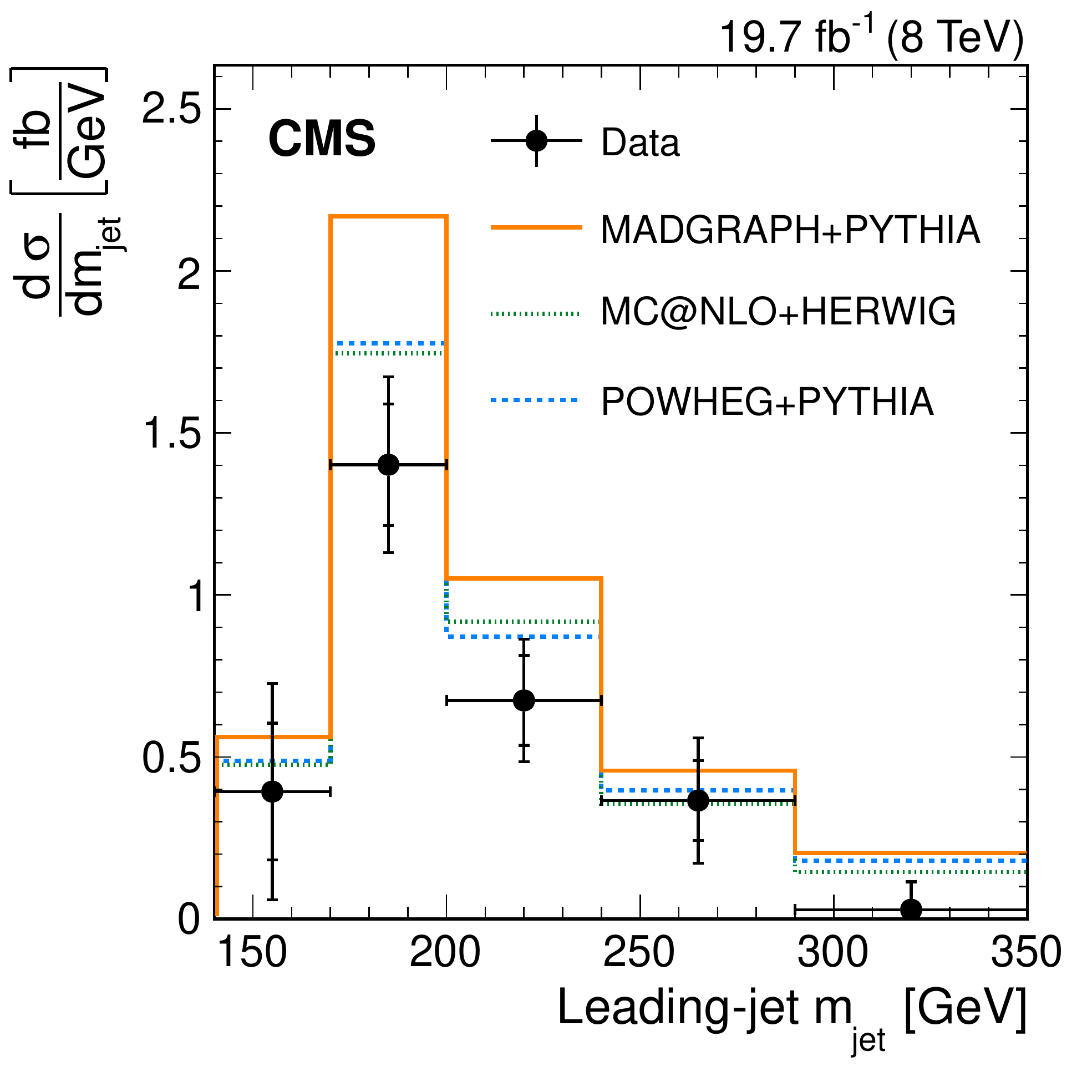}
\caption{Particle-level differential $t\bar{t}$ cross section measurement as a function of the leading jet mass compared to the predictions for three different Monte-Carlo event generators. Taken from Ref.~\cite{Sirunyan:2017yar}.\label{fig:cms:ttbar}}
\end{figure}

In addition to generic QCD jets, the jet mass has also been measured for boosted top quarks in lepton+jets $t\bar{t}$ events collected by the CMS Collaboration at 8 TeV~\cite{Sirunyan:2017yar}. This measurement is the first jet mass distribution  unfolded at the particle level probing three prong decays. Large-$R$ jets are reconstructed with the CA algorithm using a distance parameter of $1.2$. The larger value of $R$ in this measurement compared to the default $R=0.8$ applied for top tagging applications in CMS is due to an optimization of of statistical precision versus the width of the jet mass distribution at the particle level and the JMR. The number of fully-merged top quarks grows with increasing $R$, but so does the width of the jet mass distribution and the suceptibility to pile-up and the underlying event. 
The leading jet $\pt$ is required to be above 400\GeV to ensure the hadronic top quark decay to be fully captured within the large-$R$ jet. 
No substructure selection is applied on the high-\pt large-$R$ jet in order not to bias the jet mass measurement. 
A requirement of $\pt > 150\GeV$ is imposed on the subleading jet to select the $b$ quark from the leptonically decaying top quark. A veto on additional jets with $\pt > 150\GeV$ is applied, which results in a fraction of 65\% of fully-merged top quark decays within the large-$R$ jet. The particle-level differential \ttbar cross section as a function of the leading jet mass is shown in figure~\ref{fig:cms:ttbar}. The shown simulations predict a larger cross section than observed in the measurement, consistent with the \ttbar cross section measurements from the ATLAS and CMS Collaborations at high \pt. The shape of the jet mass distribution is well described by the simulations. 
The experimental systematic uncertainties are dominated by the uncertainties on the jet mass and energy scale, but are smaller than the uncertainties due to the signal modeling, coming from the choice of the top quark mass, the parton showering and the choice of the factorization and renormalization scales. 

\begin{figure}[tb]
\centering
\includegraphics[width=0.475\textwidth]{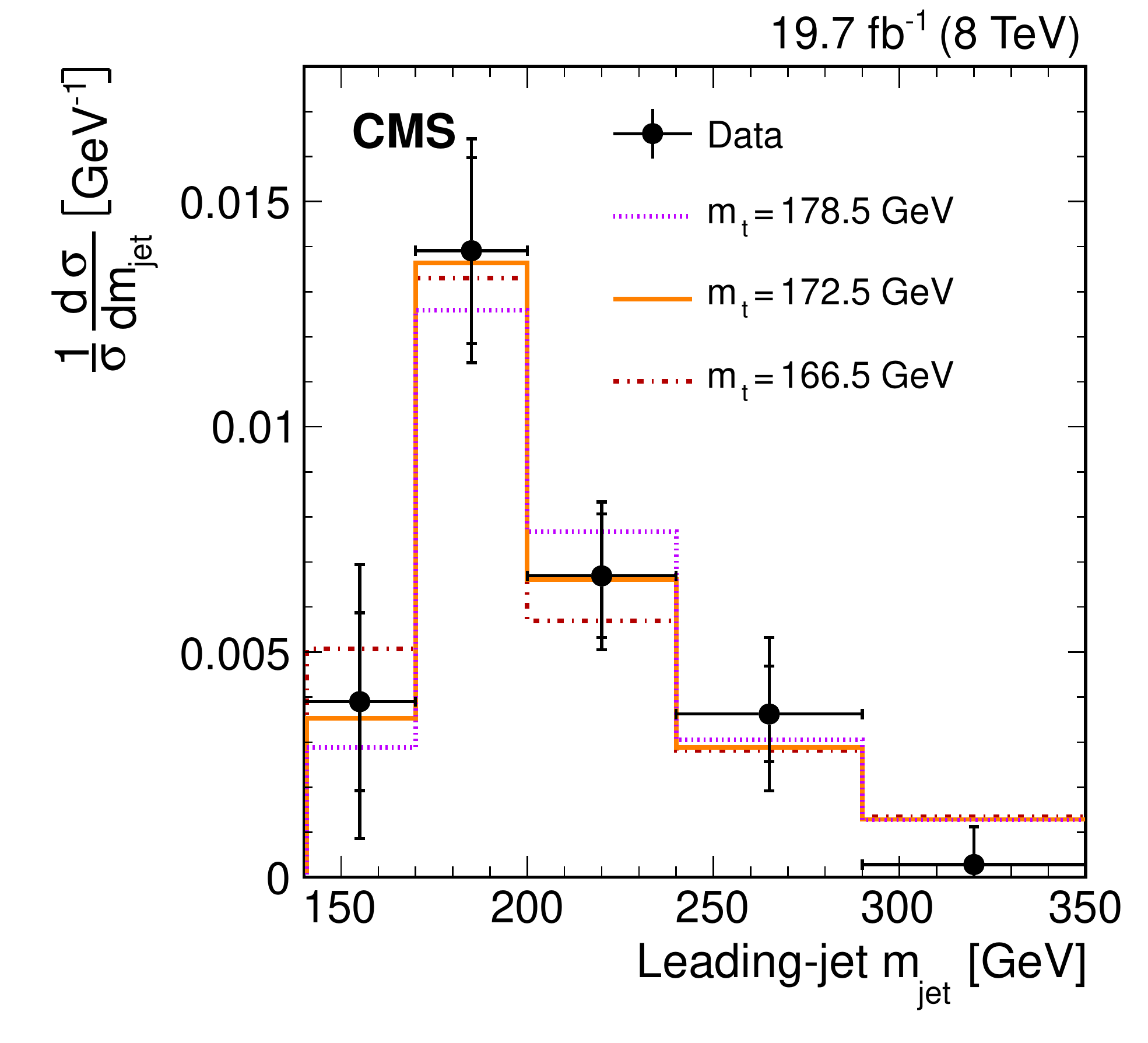}
\caption{Normalized particle-level differential $t\bar{t}$ cross section measurement as a function of the leading jet mass compared to predictions using three different top quark mass values. Taken from Ref.~\cite{Sirunyan:2017yar}.\label{fig:cms:ttbar2}}
\end{figure}

The normalized mass distribution from boosted top quarks, shown in figure~\ref{fig:cms:ttbar2}, can be used to extract the top quark mass. 
The normalized distribution is used since only the shape can be reliably calculated, and it has the additional benefit that systematic uncertainties partially cancel. 
The top quark mass is measured to be 
$m_{t} = 170.8 \pm 6.0\ (\rm stat) \pm 2.8\ (\rm sys)\
\pm 4.6\ (\rm model) \pm 4.0\ (\rm theo)\GeV$ 
in agreement with top quark mass measurement in resolved $t\bar{t}$ events (see
e.g.\ Refs.~\cite{,Tevatron:2014cka, Khachatryan:2015hba, Aad:2015nba, Aaboud:2016igd}), albeit with a much larger uncertainty. 
This constitutes a proof-of-principle, presenting the possibility to extract a fundamental SM parameter from a jet mass distribution. 
This is of particular interest, as ambiguities arise in the interpretation of  traditional $m_{t}$ measurements~\cite{Hoang:2018zrp}, which can be circumvented by measurements and analytical calculations in the highly-boosted regime~\cite{Butenschoen:2016lpz, Hoang:2017kmk}.
Future measurements at $\sqrt{s}=13\TeV$ will allow for a higher statistical precision and, in combination with jet grooming and pile-up mitigation techniques, lead to a large improvement in the total precision of the measurement. 
Measurements at higher jet \pt will facilitate comparisons with analytical calculations.

\subsubsection{Jet Charge}
\label{sec:jetcharge}

\begin{figure}[tb]
\centering
\includegraphics[width=0.43\textwidth]{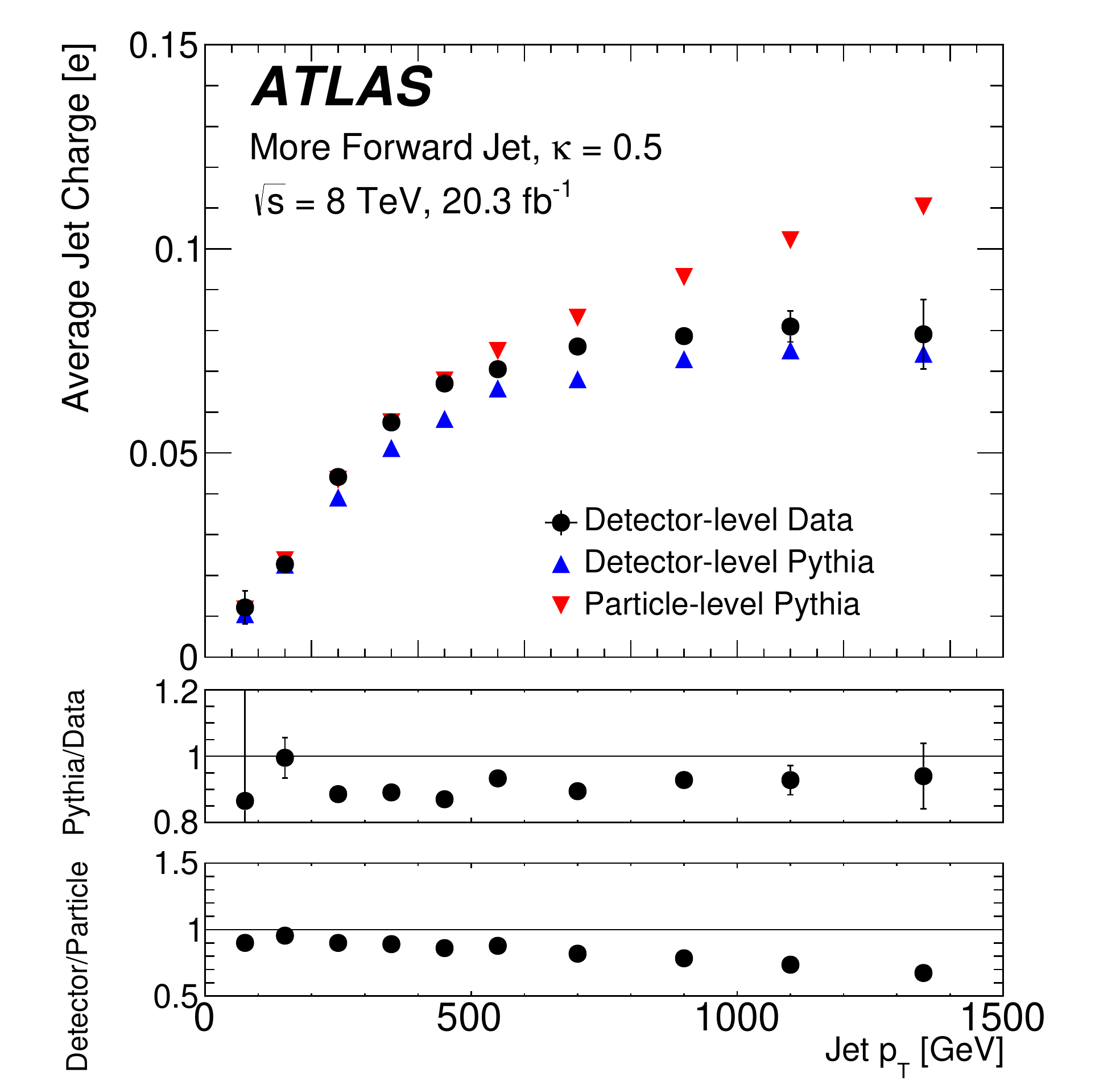}
\caption{The detector- and particle-level average jet charge as a function of jet $\pt$. Reproduced from Ref.~\cite{Aad:2015cua}.}
\label{fig:ATLASMeasurements:jetchargereco}
\end{figure}
\begin{figure}[tbp]
\centering
\includegraphics[width=0.43\textwidth]{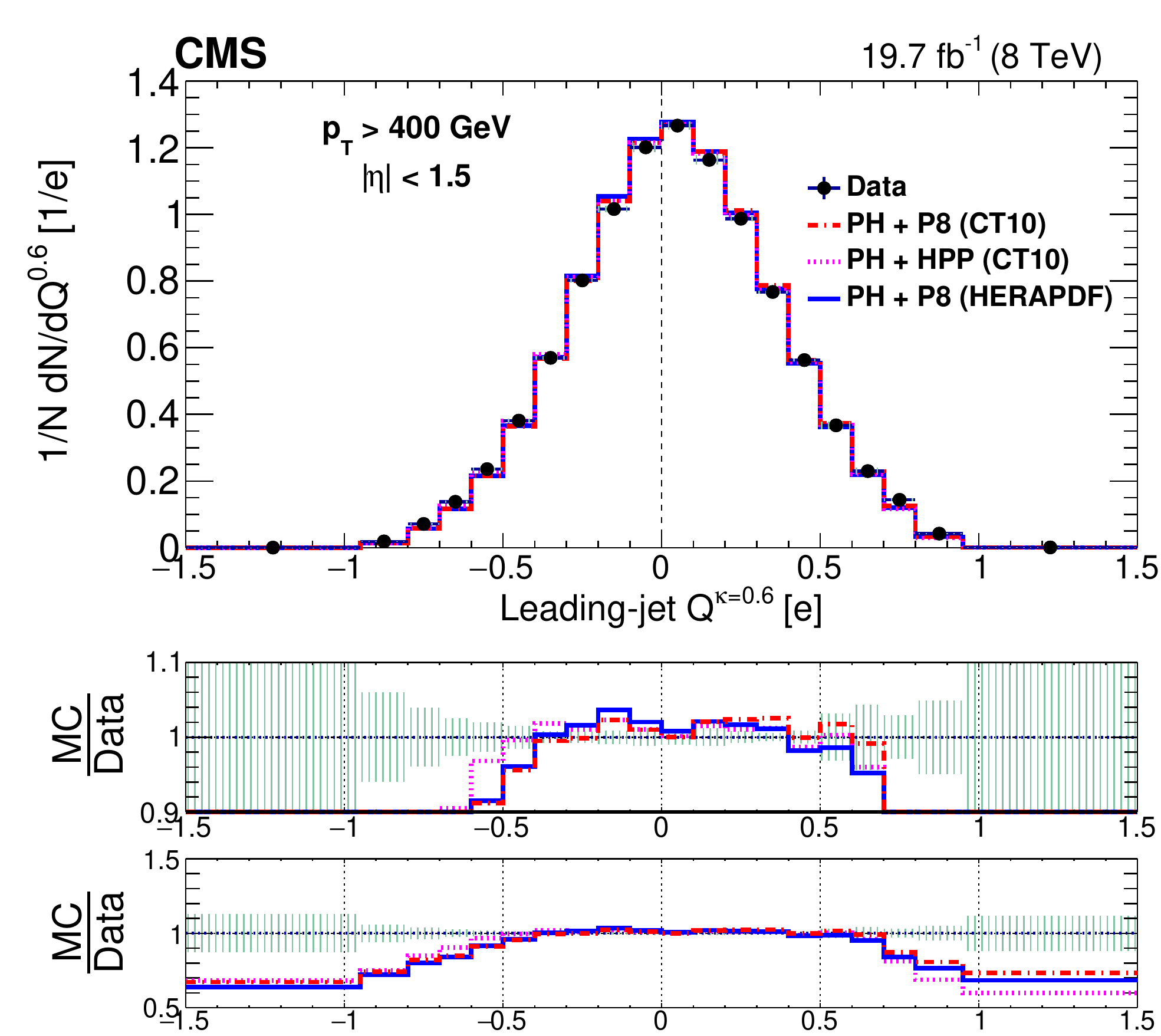}
\caption{Unfolded jet charge distribution for $\kappa = 0.6$ in data and MC prediction. Taken from Ref.~\cite{Sirunyan:2017tyr}}.
\label{fig:CMSMeasurements:jetchargeunfolded}
\end{figure}

The jet charge~\cite{Aad:2015cua,Sirunyan:2017tyr} is defined as the energy weighted sum of the electric charges of the jet constituents
\begin{equation}
Q_{\kappa} = \sum_{i\in J} \left(
\frac{p_{\mathrm{T},i}}{p_{\mathrm{T},J}}
\right)^{\kappa} q_i\,,
\end{equation}
where $q_i$ is the electric charge of particle $i$ and the free parameter $\kappa$ that controls the sensitivity to soft particles within the jet. 
The ATLAS (CMS) Collaboration measured the jet charge for different values of $\kappa$ using \antikt jets with a radius parameter of $R = 0.4$ ($R = 0.5$) in a sample of dijet events. The ATLAS Collaboration distinguishes between the two leading jets using the pseudorapidity instead of the $p_\text{T}$ to avoid cases where the leading particle-level jet is reconstructed as the sub-leading detector-level jet due to the jet energy resolution and to gain sensitivity to different jet flavors. The average jet charge at detector- and particle-level for the more forward of the leading jets and for $\kappa=0.5$ is shown in figure~\ref{fig:ATLASMeasurements:jetchargereco}.  Due to the increasing fraction of scattering valence up quark jets (up quark charge $>0$), the average jet charge increases with $p_\text{T}$. The difference of the average jet charge distribution at detector-level and particle-level in figure~\ref{fig:ATLASMeasurements:jetchargereco} shows that the unfolding corrections are large and growing at high $\pt$, due to the loss of charged-particle tracks inside jets as a result of track merging. The average jet charge as predicted by Pythia 8~\cite{Sjostrand:2007gs} using the Perugia tunes~\cite{Skands:2010ak} is smaller than that observed in data due to a well-known over-estimation of the multiplicity inside jets. The dominating systematic uncertainties are the track $\pt$ resolution and the choice of MC generator used to construct the response matrix (Pythia 6 versus Herwig++) for the CMS Collaboration whereas the uncertainties on the unfolding procedure, the jet energy resolution at low $\pt$ and uncertainties on the tracking at high $\pt$ dominate the measurement of the ATLAS Collaboration. The unfolded jet charge distribution ($\kappa = 0.6$) of the leading jet in data is compared to the prediction from Powheg+Pythia8 (PH+P8) and Powheg+Herwig++ (PH+HPP) in figure~\ref{fig:CMSMeasurements:jetchargeunfolded}. 
The different hadronization and fragmentation model used by Pythia8 and Herwig++ have the largest impact on the jet charge distribution. Variations of the jet charge can also be observed for different PDF sets however the effect of the relative flavor fraction in the dijet samples is significantly smaller than the choice of the showering and fragmentation model. It was further found that the predicted jet charge distribution has a significant dependence on the chosen value of $\alpha_s$ that describes final state radiation whereas it is insensitive to NLO QCD effect in the matrix element calculation, color-reconnection and multiple parton interactions. These findings are consistent between the ATLAS and CMS Collaboration. 

\begin{figure}[tbp]
\centering
\includegraphics[width=0.43\textwidth]{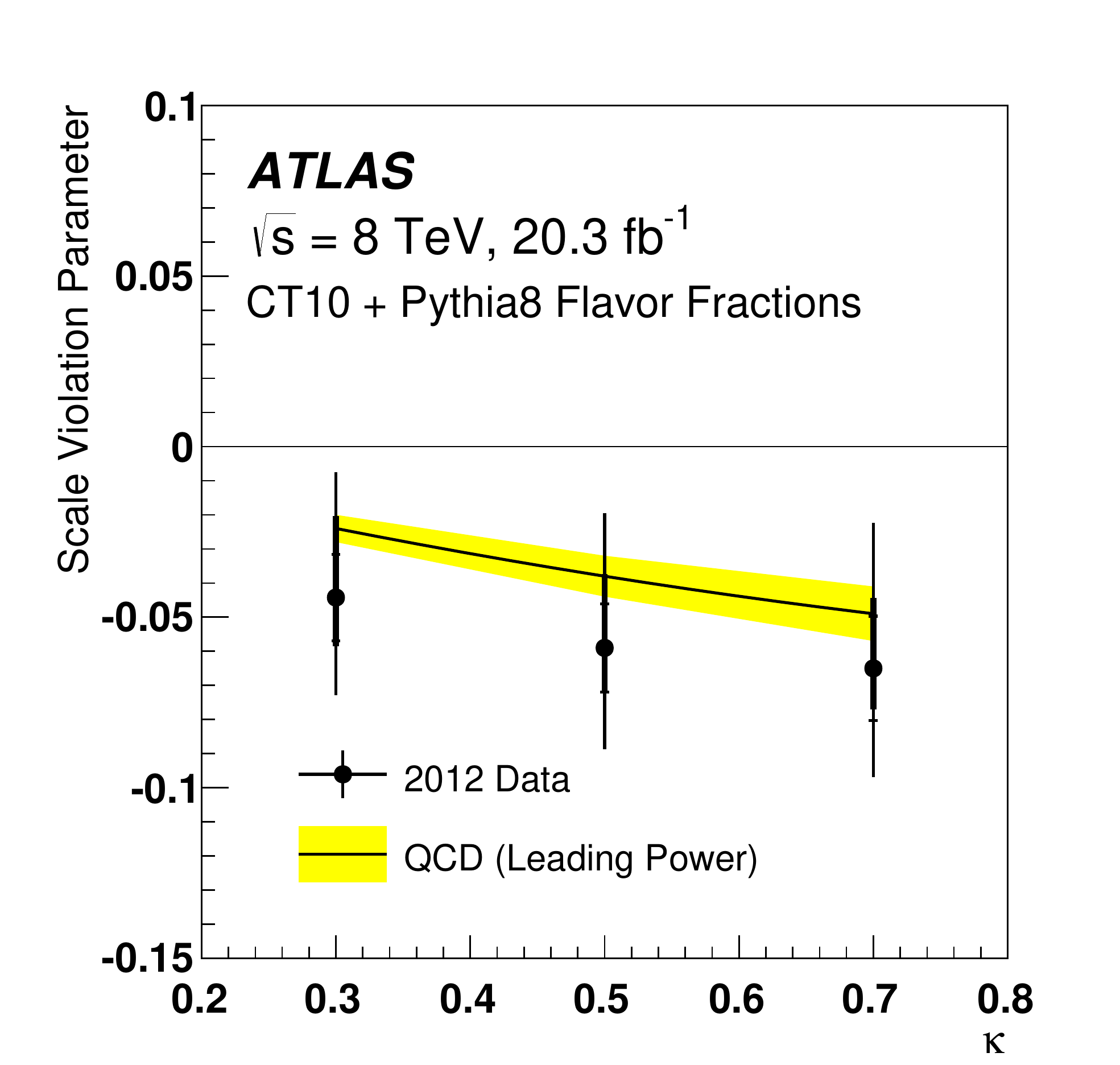}
\caption{The measured and predicted value of the average jet charge scale violation parameter $c_\kappa$. Reproduced from Ref.~\cite{Aad:2015cua}.}
\label{fig:ATLASMeasurements:jetchargescaleviolation}
\end{figure}

In addition to studying the sensitivity to various non-perturbative aspects of hadronization and parton distribution functions, the jet charge measurement by ATLAS includes the first direct comparison of a jet substructure quantity with a perturbative calculation at the LHC. As it is not collinear safe, the average jet charge is not calculable.  However, the $\pt$ dependence for a particular jet type has been calculated~\cite{Waalewijn:2012sv,Krohn:2012fg}. A new technique was introduced in Ref.~\cite{Aad:2015cua} to separately extract the average up and down quark jet charge.  For a fixed $\pt$, the more forward of the two dijets has a higher energy and is therefore more likely to be the scattering parton with a higher momentum fraction of the proton.  In turn, the higher momentum fraction parton is most likely to be a valence quark.  Therefore, the fraction of up quark jets is higher for the more forward dijet than the more central dijet. Assuming further that the jet charge is entirely determined by the jet $\pt$ and parton origin, one can then solve a system of equations to extract the average up and down quark jet charge in each bin of jet $\pt$:
\begin{align}
\nonumber
\langle Q_J^f\rangle&=f_u^f\langle Q_J^u\rangle+f_d^f\langle Q_J^d\rangle\\\label{eq:system}
\langle Q_J^c\rangle&=f_u^c\langle Q_J^u\rangle+f_d^c\langle Q_J^d\rangle,
\end{align}
where $f=\text{forward}$, $c=\text{central}$, $u=\text{up}$ and $d=\text{down}$.  As expected (though not an input), the average up quark charge is positive and the average down quark charge is negative; furthermore, the latter is roughly half the former in absolute value.  The $\pt$ dependence of $\langle Q_J^{u,d}\rangle$ are fit with a logarithmic scale violating term $c$: $\langle Q\rangle_i =\langle Q\rangle_0(1+c_{\kappa}\ln(p_{\text{T},i}/p_\text{T,$0$}))$, where $i$ represents the $\pt$ bin.  Figure~\ref{fig:ATLASMeasurements:jetchargescaleviolation} shows the measured and predicted values of $c_\kappa$.  The uncertainties are large, but there is an indication that $c<0$ and $\partial c/\partial\kappa < 0$, as predicted.


\subsubsection{Other Jet Substructure Observables}

The ATLAS and CMS Collaborations have performed further precision
measurements of hadronic jet substructure in $pp$ collisions,
correcting for acceptance and resolution such as jet and event
shapes~\cite{Aad:2011kq, Aad:2012np, Aad:2013fba, Chatrchyan:2012mec,Aaboud:2019aii,Sirunyan:2018asm},
charged particle
multiplicities~\cite{Aad:2011gn,Aad:2016oit,Chatrchyan:2012mec}, the
jet fragmentation functions~\cite{Aad:2011sc,Aaboud:2018uiu}, color
flow~\cite{Aad:2015lxa} and \kt splitting scales, $N$-subjettiness
ratios as well as further substructure variables such as Planar Flow
and angularity~\cite{ATLAS:2012am, Aad:2013ueu}. 

\subsection{Measurements with Jet Substructure\label{sec:meas2}}

While measurements of jet substructure observables such as jet mass,
jet charge and event shape variables have been discussed in section~\ref{sec:substructuremethodsobservables}, the following sections present measurements of other
quantities through the exploitation of  jet substructure techniques
such as top tagging. 


\subsubsection{Differential \ttbar Cross Section Measurements}
\label{sec:atlas:ttbar}

\begin{figure}[tbp]
\centering
\includegraphics[width=0.43\textwidth]{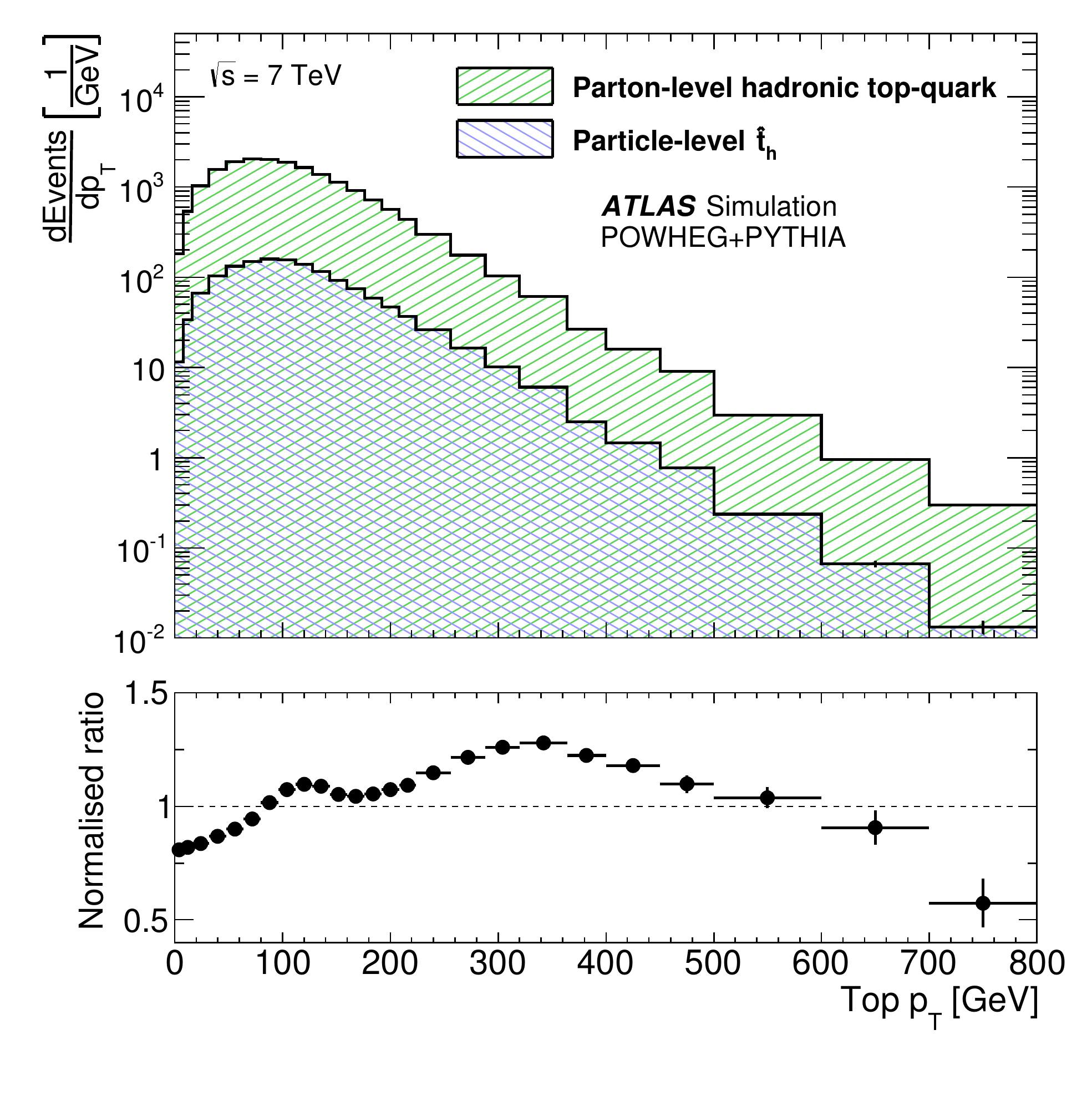}
\caption{The drop in fiducial efficiency  at top  $\pt> 600$ \GeV when reconstructing top quarks with individual \antiktfour jets (\textit{resolved} reconstruction). Adapted from ~\cite{Aad:2015eia}. }
\label{fig:atlas:ttbar1}
\end{figure}

\begin{figure}[tbp]
\centering
\includegraphics[width=0.43\textwidth]{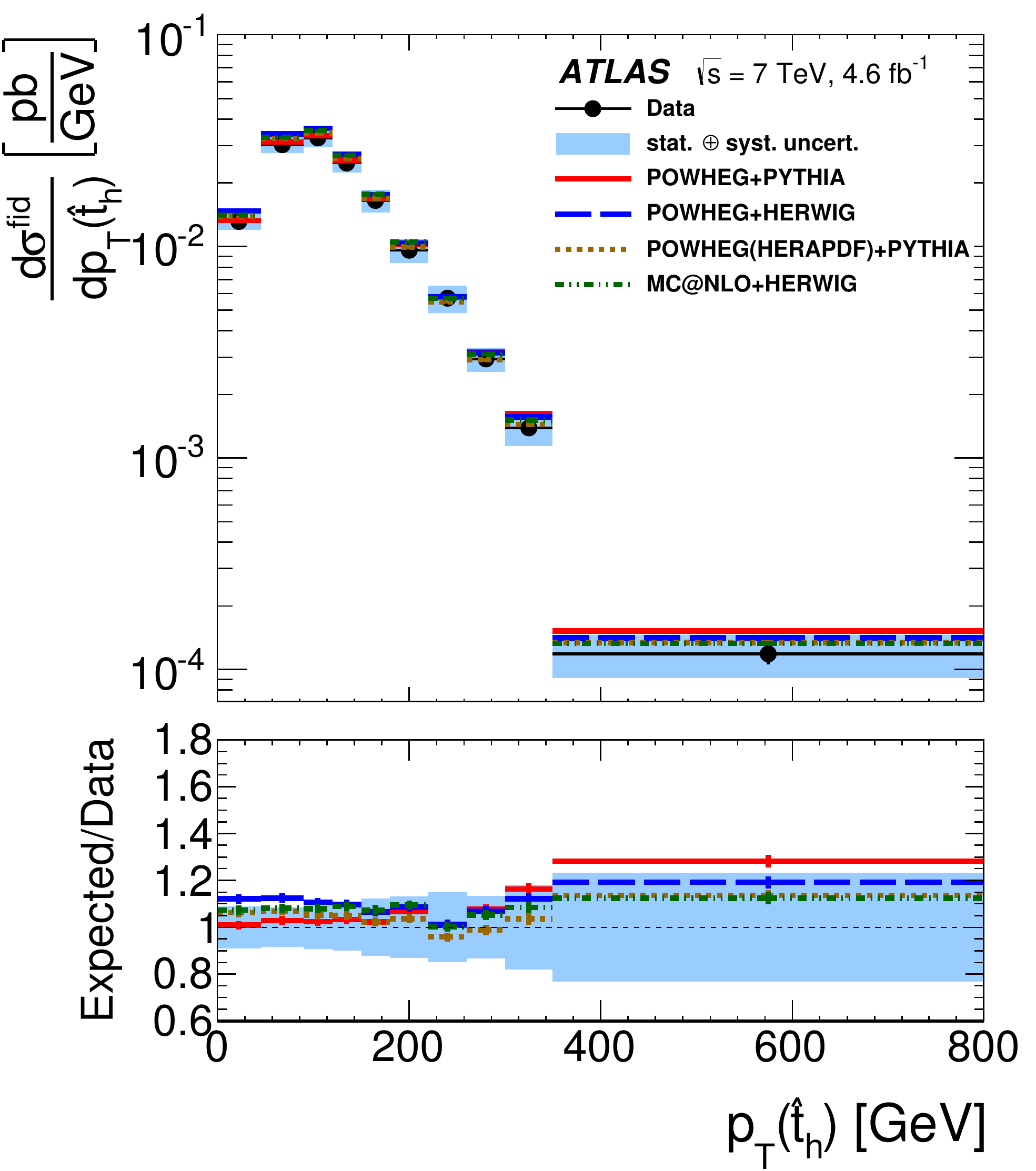}
\caption{The small number of top jets identified at high-\pt results in very coarse cross section measurement when using the \textit{resolved} reconstruction technique. Adapted from ~\cite{Aad:2015eia}. }
\label{fig:atlas:ttbar2}
\end{figure}

The selection cuts applied in traditional \ttbar cross section measurements~\cite{Chatrchyan:2012saa, Khachatryan:2015oqa, Khachatryan:2016mnb, Aad:2015eia, Aad:2015mbv, Aaboud:2016syx, Aaboud:2017fha} are chosen to maximize the acceptance and minimize the associated uncertainties on the fiducial and total cross section measurements. The fiducial region is such that events with top $\pt$ below 100\GeV and above 600\GeV are under-represented, with the former caused by trigger and reconstruction efficiencies and the latter by collimated decays from large Lorentz boosts. This is evident from figure~\ref{fig:atlas:ttbar1}, where a drop in selection efficiency below 100\GeV and above 600\GeV is apparent. This results in a small number of events being selected with high top quark $\pt$, as seen in the ATLAS \runone (7 \TeV) measurement shown in figure~\ref{fig:atlas:ttbar2}.  This means that a very interesting region in terms of new physics is the least well-measured. Despite often having similar signal efficiencies to resolved reconstructed techniques, boosted top tagging techniques allow for more precise measurements at high \pt due to their higher background rejection. 

\begin{figure}[tb]
\centering
\includegraphics[width=0.49\textwidth]{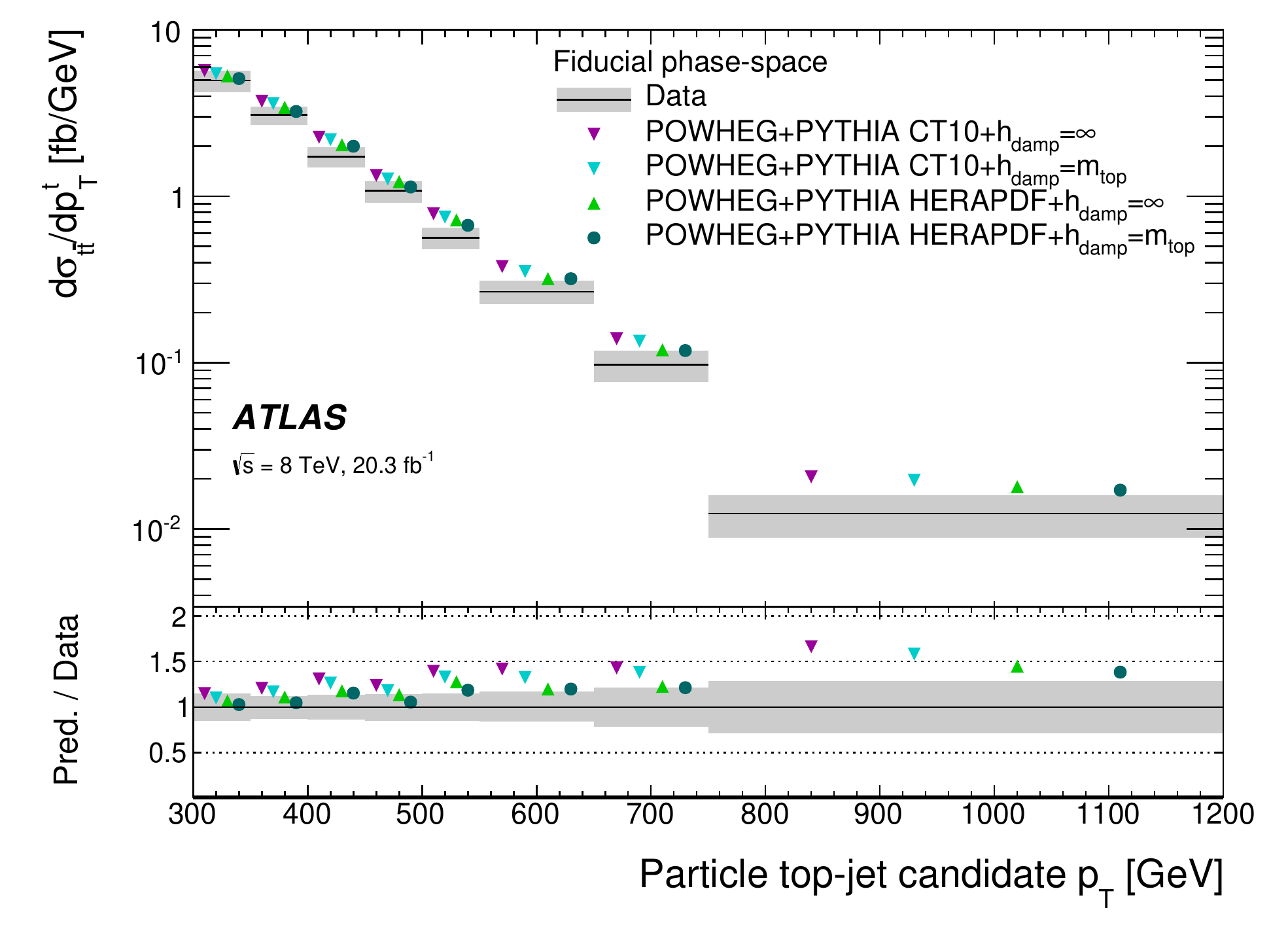}
\caption{Particle-level differential $t\bar{t}$ cross section measurement for two different PDF sets and choices of the $h_{\rm damp}$ parameters. Taken from Ref.~\cite{Aad:2015hna}.}
\label{fig:atlas:ttbar3}
\end{figure}

The ATLAS Collaboration performed a measurement of the boosted $\ttbar$ differential cross section as a function of the top quark $\pt$ in the lepton+jets channel~\cite{Aad:2015hna}. A least one \antikt jet, trimmed with $R_{\rm sub}=0.3$ and $f_{\rm cut}=0.05$ is required with $|\eta|<2$ and $\pt >300~\GeV$. To select events with boosted top quarks, the large-$R$ jet is required to have a mass larger than 100\GeV and $\sqrt{d_{12}}> 40\GeV$ (Tagger III, see section~\ref{sec:TopTagging}). The reconstructed $\pt$ distribution of the \antikt $R = 1.0$ trimmed jet is unfolded to the parton and particle-level. The measured particle-level differential cross section is compared in figure~\ref{fig:atlas:ttbar3} to the predictions of several MC generators normalized to the NNLO+NNLL inclusive cross section. Overall good agreement is observed, but a harder $\pt$ spectrum is predicted by the simulation than observed in data with larger discrepancies at high $\pt$. The differential cross section measurement is also compared to predictions from Powheg+Pythia using either the HERAPDF~\cite{Aaron:2009aa} or CT10~\cite{Lai:2010vv} PDF set and two different values for the resummation damping factor $h_{\rm damp}$, $h_{\rm damp} = m_{\rm top}$ and $h_{\rm damp} = \infty$. The best data/MC agreement is observed when using the HERAPDF set and $h_{\rm damp} = m_{\rm top}$. For each of the settings, the trend of a harder $\pt$ spectrum in simulation compared to data persists.

\begin{figure}[tbp]
\centering
\includegraphics[width=0.43\textwidth]{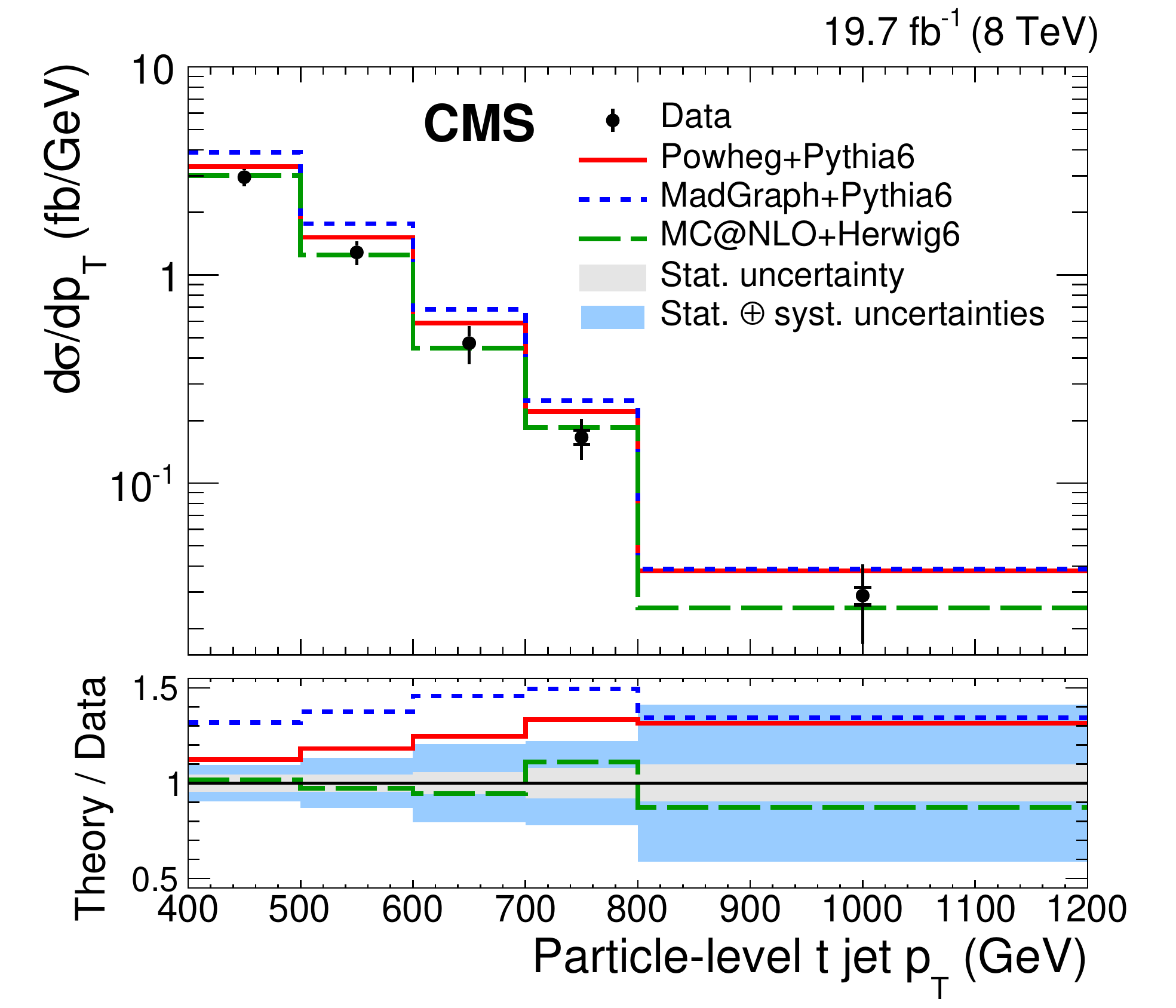}
\caption{Comparison of the particle-level differential $t\bar{t}$ cross section as a function of the jet \pt\ to three different MC generators. Taken from Ref.~\cite{Khachatryan:2016gxp}.}\label{fig:cms:ttbar_pt}
\end{figure}

A similar measurement by the CMS Collaboration based on 8\TeV data~\cite{Khachatryan:2016gxp} uses the CMSTT algorithm to reconstruct boosted top quarks. The unfolded results are in agreement with the ATLAS measurement and show a similar trend between data and simulation, as shown in figure~\ref{fig:cms:ttbar_pt}. 

These measurements extend up to a top quark $\pt$ of 1.2~\TeV, allowing for higher precision thanks to the usage of jet substructure techniques. The largest uncertainties at the highest values of $\pt$ in ATLAS and CMS come from the large-$R$ jet energy scale and the extrapolation of the $b$-jet calibration to high \pt. 

The parton-level differential cross section in top quark $\pt$ has also been measured in the all-hadronic final state by the CMS Collaboration using 8\TeV data~\cite{CMS-PAS-TOP-16-018}. This measurement relies on pruned jets with an $N$-subjettiness and subjet-$b$ tagging requirement to suppress the huge amount of background from QCD dijet production. The cross section is determined from a maximum likelihood fit to the jet mass distributions for signal-enriched and signal depleted regions. This allows for a simultaneous extraction of the $t\bar{t}$ cross section and the QCD background. The measurement is in agreement with the results from the lepton+jets final states, but has somewhat larger statistical uncertainties of up to about 40\% in the highest $\pt$ bin with $0.8<\pt<1.2\TeV$.

The increased $\sqrt{s}$ at Run 2 of the LHC offers the possibility for more precise differential $t\bar{t}$ cross section measurements in the highly-boosted regime. The $t\bar{t}$ production cross section increased by more than a factor of ten for top quark $\pt>400\GeV$ when going from $\sqrt{s}=8\TeV$ to $13\TeV$.

\begin{figure}[tb]
\centering
\includegraphics[width=0.43\textwidth]{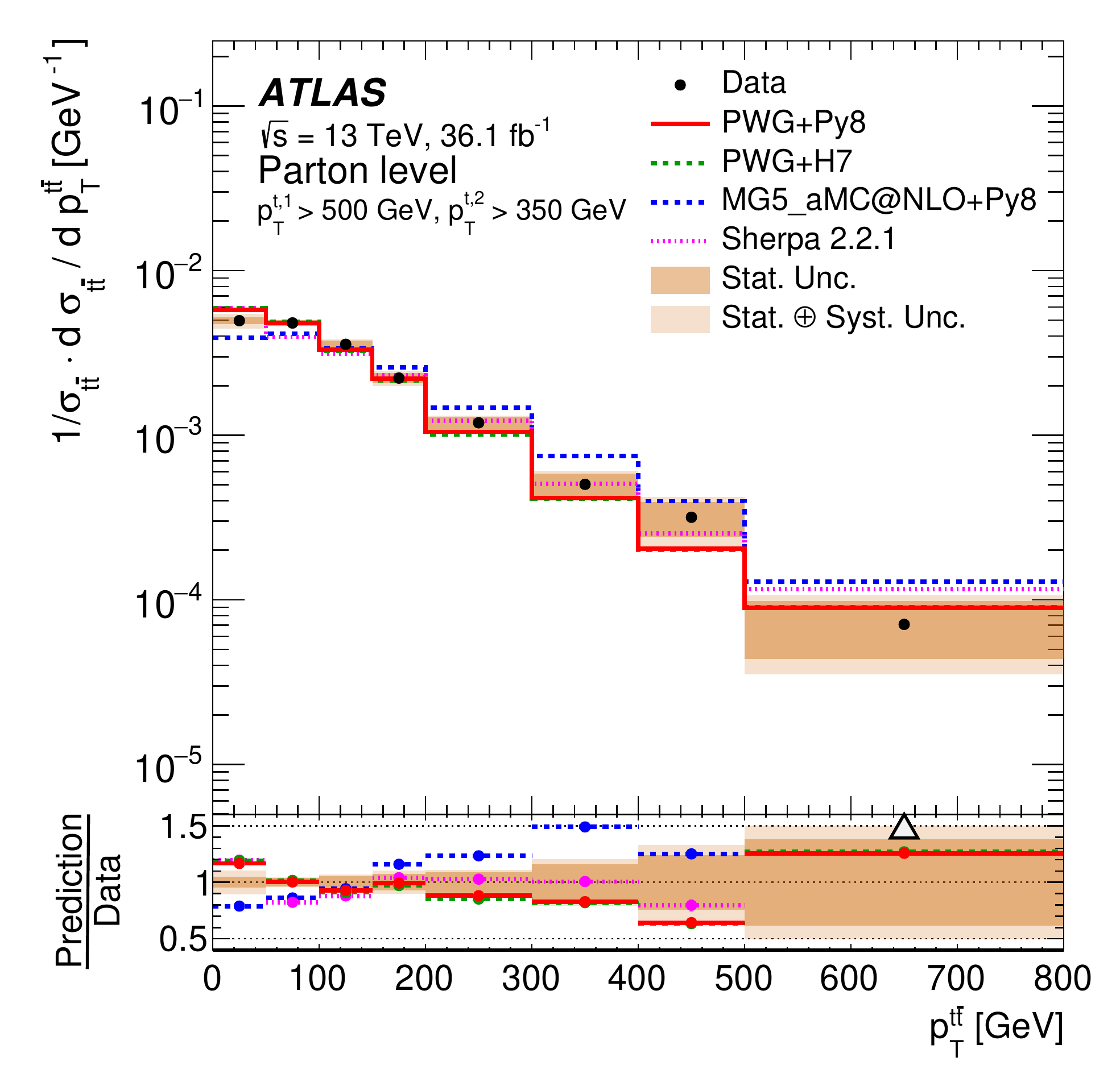}
\caption{The normalized differential cross section as a function of the $t\bar{t}$ $\pt$ as measured by ATLAS in the all-hadronic channel at 13\TeV. Taken from Ref.~\cite{Aaboud:2018eqg}.}
\label{fig:atlas:ttbar4}
\end{figure}

A first measurement based on 3.2~$\fbinv$ of 13\TeV data in the lepton+jets channel has been performed by ATLAS~\cite{Aaboud:2017fha}. The measurement extends to $\pt$ of 1.5\TeV and a similar trend as at 8\TeV is observed between the data and the simulation at high $\pt$. 
A newer measurement of the $t\bar{t}$ differential cross section in the all-hadronic channel is performed by the ATLAS Collaboration with $36.1\fbinv$ of 13\TeV data~\cite{Aaboud:2018eqg}. The measurement uses trimmed \antikt $R = 1.0$ jets with $R_{\rm sub}=0.2$ and $f_{\rm cut}=0.05$. To obtain a flat signal efficiency of 50\% and a quark/gluon rejection of approximately 17 (10) for $\pt = 500\ (1000)\GeV$, $\pt$ dependent criteria are applied on the jet mass and $\toptau$. Furthermore the two top-tagged large-$R$ jets are required to have a $b$ tagged small-$R$ jet within $\Delta R < 1.0$. The event selection results in a signal-to-background ratio of approximately 3:1. The measured fiducial phase-space cross section is $\sigma = 292 \pm 7\ (\rm stat) \pm 76\ (\rm sys)$~fb compared to the Powheg+Pythia8 prediction of $384 \pm 36$~fb at NNLO+NNLL. The measured normalized differential cross section as a function of the top jet \pt and rapidity is in good agreement with the different MC predictions. Larger discrepancies are observed for the \pt of the $t\bar{t}$ system as shown in figure~\ref{fig:atlas:ttbar4}. The measurement is dominated by the systematic uncertainties on the jet energy, mass and substructure scale of the large-$R$ jets, alternative parton shower model and the uncertainties on the $b$ jet identification. 

\subsubsection{$W/Z/H$ Cross Sections}
\label{sec:wzcrosssections}

\begin{figure}[tb]
\centering
\includegraphics[width=0.43\textwidth]{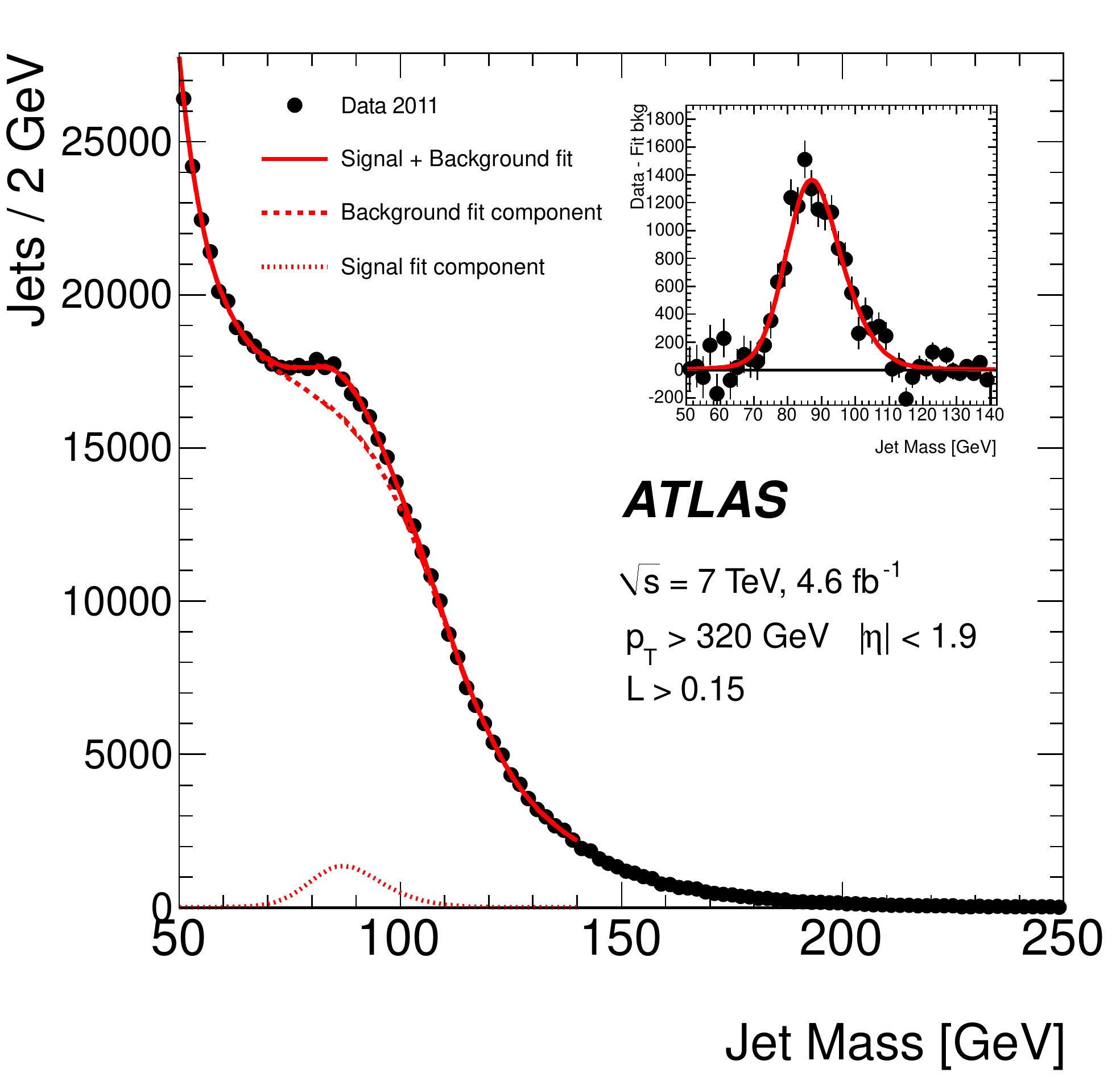}
\caption{Binned maximum likelihood fit to the jet mass distribution in data for selected $W/Z$ events reconstructed as one single ungroomed \antikt $R = 0.6$ jet. Taken from Ref.~\cite{Aad:2014haa}.}
\label{fig:WZXsection}
\end{figure}

The cross section of boosted $W$ and $Z$ boson production was measured by ATLAS in 4.6~fb$^{-1}$ of 7\TeV $pp$ collisions~\cite{Aad:2014haa}. The hadronically decaying $W$ and $Z$ bosons are reconstructed as one single ungroomed \antikt $R = 0.6$ jet with $\pt > 320$~\GeV, $|\eta| < 1.9$ and masses ranging between 50 and 140~\GeV. The $W$ and $Z$ signal is enhanced over the dominating QCD background by constructing a likelihood discriminant from three substructure variables; thrust minor~\cite{Brandt:1964sa,PhysRevLett.39.1587}, sphericity~\cite{Bjorken:1969wi} and aplanarity~\cite{Bjorken:1969wi}, resulting in a signal efficiency of 56\% and a background rejection of 89\%. The jet mass distribution after subtracting the expected background from $t\bar{t}$ events is shown in figure~\ref{fig:WZXsection}. A binned maximum likelihood fit to the jet mass distribution is used to extract the $W/Z$ jet signal yield and to calculate the inclusive cross section. Only the combined $W+Z$ cross section measurement is performed in this analysis due to the limited jet mass resolution. The combined $W+Z$ cross section is measured to be $\sigma_{W+Z}= 8.5 \pm 0.8$~(stat.)~$\pm 1.5$~(syst.)~pb and is in agreement with the Standard Model prediction of $\sigma_{W+Z} = 5.1 \pm 0.5$~pb within 2 standard deviations. The dominating systematic uncertainties are the jet mass resolution and the choice of the QCD background PDF. The signal significance was furthermore studied when using groomed jets instead of ungroomed jets. Without an optimization of the analysis for groomed jets, similar significances were observed for groomed and ungroomed jets as expected due to the low number of pile-up vertices in the 7\TeV dataset. 

\begin{figure}[tb]
\centering
\includegraphics[width=0.43\textwidth]{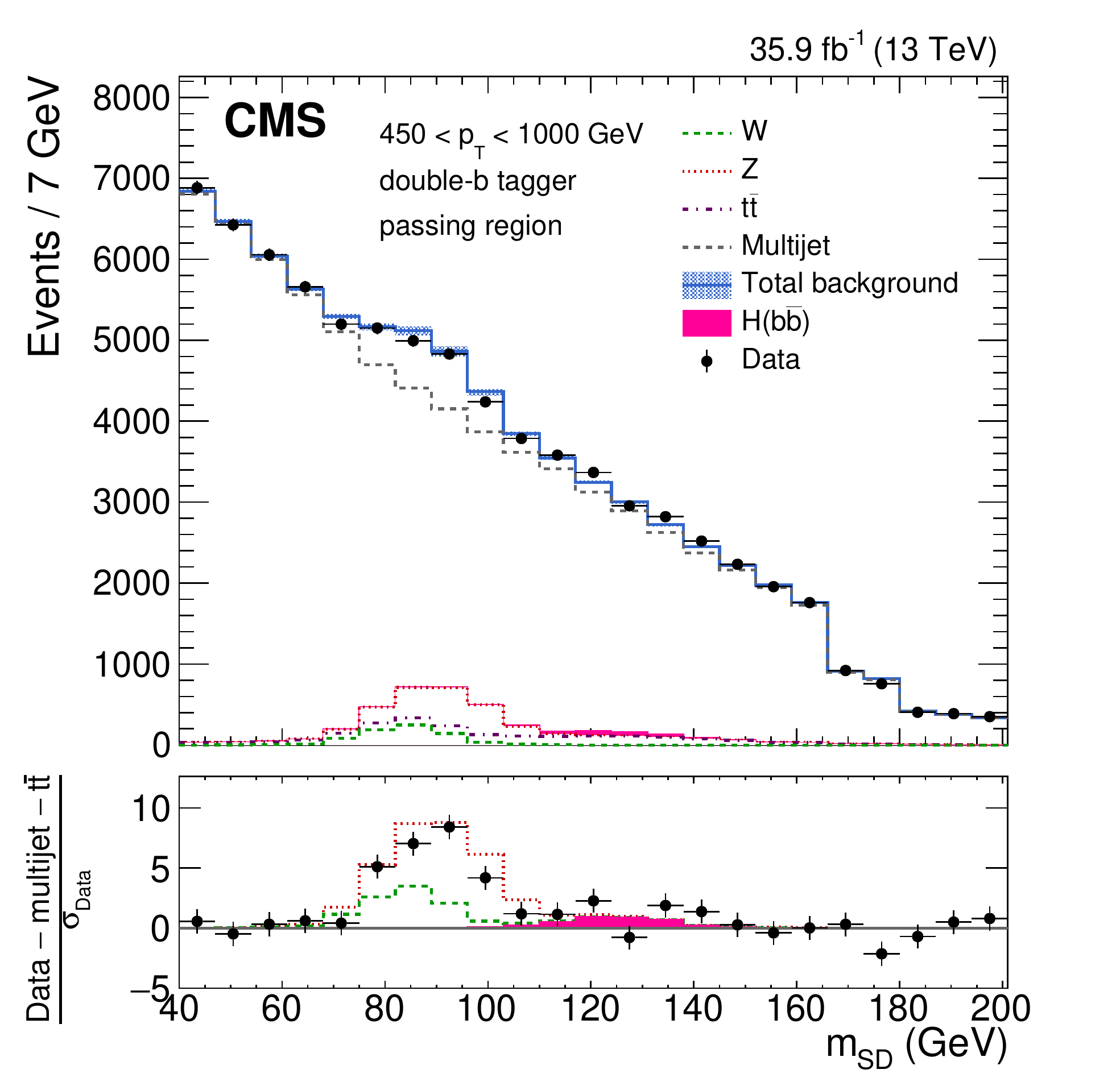}
\caption{Soft drop jet mass $\rm m_{SD}$ of \antikt $R = 0.8$ jets in data and for
  the dominating background processes; multijet production and
  $W/Z$+jets events. Jets are required to pass criteria on $N^1_2$ and
  to be identified as double-$b$ jets by the double-$b$ tagger introduced
  in section \ref{sec:btagging}. Taken from Ref.~\cite{Sirunyan:2017dgc}.}
\label{fig:mass6}
\end{figure}

As discussed in section \ref{sec:btagging} the SM Higgs boson decays
with approximately 58\% into $b\bar{b}$. However the $H\rightarrow
b\bar{b}$ decay in the resolved channel can only be studied in 
associated production with either a vector boson ($W/Z$)
\cite{Aaboud:2017xsd,Sirunyan:2017elk}, top quarks, or via the vector-boson-fusion production mechanism due 
to the overwhelming multijet background. To search for $H\rightarrow
b\bar{b}$ in the gluon-gluon fusion production mode with an additional high-\pt jet, 
jet substructure techniques can be employed to suppress the
enormous multijet background. The CMS Collaboration performed a search
for the SM Higgs boson using a dijet topology with 35.9~fb$^{-1}$
of 13\TeV $pp$ collisions~\cite{Sirunyan:2017dgc}. The analysis uses
\antikt $R = 0.8$ jets corrected with the PUPPI algorithm to reduce the effects from 
pile-up, and modified with the soft drop algorithm
($\beta = 0$, $\zsoftdrop = 0.1$) to mitigate the effects from the underlying event and 
soft/wide-angle radiation. At least one large-$R$ jet with
$\pt > 450\GeV$ and $|\eta| < 2.5$ is required. To distinguish the
two prong structure of a jet containing the full $H\rightarrow
b\bar{b}$ decay from quark- or gluon-initiated jets, the $N_{2}^{1}$
variable, calculated from the generalized energy correlation
functions, is exploited. To ensure a flat QCD background rejection of
26\% over the considered mass and \pt range, a decorrelation procedure
\cite{Dolen:2016kst} is applied to $N_{2}^{1}$.  The multijet
background is further suppressed by utilizing the double-$b$ tagger. 
The $W/Z$+jets background is estimated from MC simulation and the shape of the
multijet background is determined in a validation region in data with
lower values of the double-$b$ tagger discriminator. The soft drop
mass distribution of the leading jet is shown in figure~\ref{fig:mass6}
with a clear resonant structure at the mass of the $W$ and $Z$
boson. The SM background processes and the potential signal from SM
$H\rightarrow b\bar{b}$ production are estimated simultaneously. The observed (expected) significance for the $H\rightarrow b\bar{b}$ process is 1.5(0.7)$\sigma$.  
The measured cross section for the $Z$+jets process is $0.85\pm 0.16$~(stat.)~$^{+1.0}_{-0.4}$~(syst.)~pb
which is in agreement with the SM prediction of $1.09 \pm 0.11$~pb. This is the
first observation of $Z\rightarrow b\bar{b}$  in the single jet
topology.  

The ATLAS Collaboration also measured the high \pt $Z\rightarrow b\bar{b}$ cross section using two nearby $b$ tagged \antikt $R = 0.4$ jets (instead 
of one large-radius jet) in 19.5~fb$^{-1}$ of 8~\TeV $pp$ collisions~\cite{Aad:2014bla}. The measured fiducial cross section was determined to be 
$\sigma_{Z\rightarrow b\bar{b}} = 2.02 \pm 0.33$~pb which is in excellent agreement with the next-to-leading-order theoretical predictions.


\section{Searches for New Physics}\label{sec:searches}

Jet substructure methods have been successfully applied in a large
variety of searches for physics beyond the SM. The respective
exclusion limits are substantially improved through the application of
these methods. In some cases the decay signature of heavy BSM
particles would not be accessible without the application of jet
substructure methods. 

As the number of such BSM searches is very large, only a small subset
of the published results can be discussed here. The following sections
give an overview of a selection of searches for \ttbar resonances~\cite{Chatrchyan:2012ku,Khachatryan:2015sma,Aad:2013nca,Aad:2012ans,Aad:2015fna,ATLAS-CONF-2016-014,Sirunyan:2017uhk}, diboson resonances~\cite{Chatrchyan:2012ypy,Khachatryan:2014hpa,Khachatryan:2014gha,Khachatryan:2015ywa,Khachatryan:2015bma,Khachatryan:2016yji,Khachatryan:2016cfa,Aad:2015ipg,Aaboud:2017fgj,Aaboud:2017itg,Aaboud:2016okv,Aaboud:2017cxo,Aaboud:2017eta,Aaboud:2017ahz,Aad:2015uka,Aaboud:2016xco,Khachatryan:2016cfx,Sirunyan:2017nrt,Sirunyan:2017acf,Sirunyan:2016cao,Sirunyan:2017wto,CMS-PAS-B2G-16-026}, vector-like quarks~\cite{Chatrchyan:2013uxa,Khachatryan:2015axa,Khachatryan:2015oba,Khachatryan:2015gza,Khachatryan:2016vph,Sirunyan:2017ynj,Sirunyan:2016ipo,Sirunyan:2017ezy,Sirunyan:2017usq,Sirunyan:2017bfa,Sirunyan:2017pks,Sirunyan:2017lzl,Aad:2015voa,Aad:2016qpo,Aaboud:2017zfn,Aaboud:2017qpr} and leptophobic $Z'$~\cite{Aaboud:2018zba,Sirunyan:2017nvi}. Further searches using jet substructure techniques can be found in Refs.~\cite{Khachatryan:2015wza,Khachatryan:2016zcu,Khachatryan:2016oia,Khachatryan:2017rhw,Sirunyan:2017wif,Sirunyan:2017pjw,Sirunyan:2017bsh,Khachatryan:2016mdm,Sirunyan:2016wqt,Sirunyan:2017hci,Sirunyan:2017hnk,Sirunyan:2017jix,Sirunyan:2017hsb,Sirunyan:2018gka,Khachatryan:2015edz,Khachatryan:2015mta,Sirunyan:2017ukk,Sirunyan:2017djm,Aad:2015dva,Aad:2013oja,Aaboud:2016qgg,Aaboud:2016trl,Aaboud:2017ecz,Aad:2014xra,Aaboud:2018juj,Aaboud:2017vwy,Aaboud:2017hrg,Aaboud:2017ayj,Aaboud:2017aeu,Aaboud:2016lwz}.

\subsection{Diboson Resonances}\label{sec:dibosonsearch}

Several new physics models  predict resonances coupling strongly to vector bosons to play a role in the cancellation of 
large corrections to the Higgs mass. These models include extensions of the SM Higgs doublet, where the simplest 
realizations are two-Higgs-doublet models~\cite{Branco:2011iw} with heavy, neutral Higgs bosons, 
which can have large branching fractions to top quarks and $W$/$Z$/$H$ bosons. 
Alternatives are composite Higgs models~\cite{Kaplan:1983fs,Kaplan:1983sm,Georgi:1984ef,Banks:1984gj,Georgi:1984af,Dugan:1984hq,Georgi:1985hf,Bellazzini:2014yua} or Randall-Sundrum Kaluza-Klein models~\cite{Randall:1999ee,Agashe:2003zs,Davoudiasl:1999tf,Pomarol:1999ad}. 

Searches for new resonances generally focus at high masses with $m > 1\TeV$ such that the SM bosons receive high Lorentz boosts. 
In more than 60\% of the cases, $W$/$Z$/$H$ bosons decay into a quark
anti-quark pair, which makes the reconstruction of such decays with jet substructure techniques an
essential ingredient for these searches.
In the following, the analysis strategies and results from CMS and ATLAS using $pp$ collision data with $\sqrt{s}=13\TeV$ are discussed.

The searches for diboson resonances are performed  in 
semi-leptonic~\cite{Aaboud:2016okv,Sirunyan:2017nrt} 
and fully hadronic final states~\cite{Aaboud:2017ahz, Aaboud:2017eta, Sirunyan:2017acf, Sirunyan:2016cao}. As the methods of jet substructure analyses exhibit their full strength in hadronic
final states, the following discussion gives a summary and comparison of
the ATLAS and CMS results in the search for $W$/$Z$ resonances in hadronic
final states only.

In an analysis performed by the CMS Collaboration~\cite{Sirunyan:2017acf} 
events with two \antikt jets with $R=0.8$, corrected with the PUPPI algorithm, and 
$65<m_{\rm soft\ drop}<105\GeV$ are selected. The jet is considered to be a $W$ 
boson candidate if the mass is in the range 65--85\GeV, while it is
 a $Z$ boson candidate if the mass is in the range 85--105\GeV.
This leads to the three signal categories $WW$, $ZZ$ and $WZ$. The jets are further
categorized according to $\tau_{21}$ into  high purity (HP, $\tau_{21} < 0.35$) 
and low purity (LP, $0.35<\tau_{21}<0.75$). 
Events are always required to have one HP $V$ jet, and are
divided into HP and LP events, depending on whether the other $V$ jet is
of high or low purity. To further suppress the large QCD multijet background a requirement on
the dijet kinematics $|\eta_{1}-\eta_{2}|<1.3$ is applied. 

\begin{figure}[tb]
\centering
\includegraphics[width=0.45\textwidth]{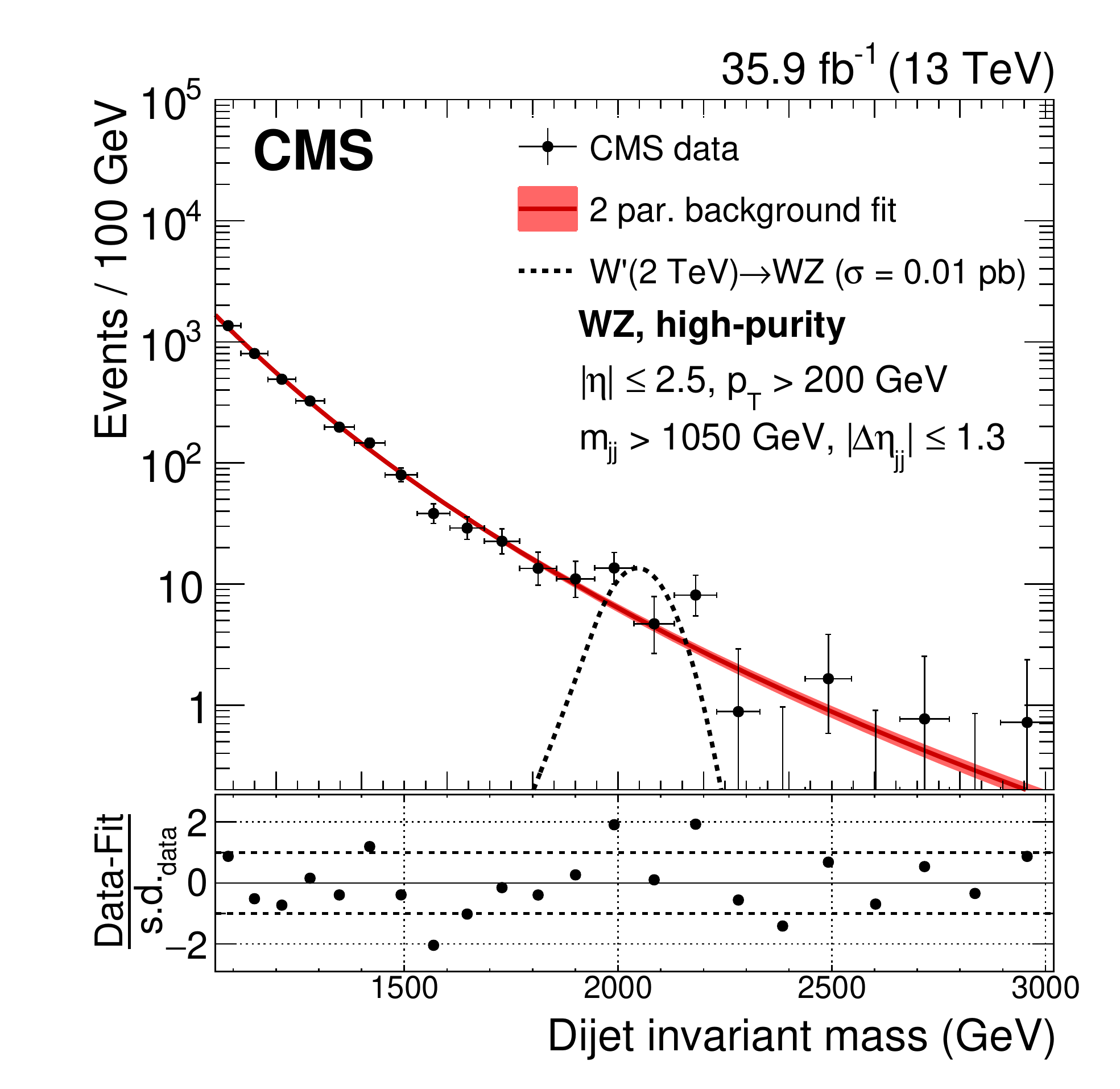}
\caption{\label{fig:vvCMS} Dijet invariant mass distribution in the high purity $WZ$ category of the fully hadronic $WW$/$WZ$/$ZZ$ resonance search. The fit under the background-only hypothesis is overlayed. Taken from Ref.~\cite{Sirunyan:2017acf}.}
\end{figure}

The background is estimated from a signal+background fit with the function
$\frac{dN}{dm_{jj}}=\frac{P_0}{(m_{jj}/\sqrt{s})^{P_1}}$, where
$P_0$ is a normalization parameter and $P_1$ is a parameter describing
the shape. This parametrization has been tested and validated on simulated 
events and on data in a control region. 
As shown in figure~\ref{fig:vvCMS} the data in the signal region is well described by
the fit function.
Figure~\ref{fig:vvCMS} also shows that no excess over the background-only hypothesis is 
observed. 

A similar analysis has been performed by the ATLAS Collaboration~\cite{Aaboud:2017eta}.  
In this analysis events are required to have at least two large-$R$ jets with 
$\pt>200\GeV$ in the pseudo-rapidity range $|\eta|<2.0$.
These jets are reconstructed with the \antikt algorithm with a radius parameter $R = 1.0$.  
The trimming algorithm is applied using $\kt$ subjets with $R=0.2$.  
The rapidity separation between the two leading jets has to
satisfy $|\Delta y_{12}|<1.2$.

The large-$R$ jet mass is computed from the Combined Mass (see section~\ref{sec:jetrec-mass}), 
and is required to be within a window of the expected $W$ or $Z$ mass value. The window width varies
from 22 to 40\GeV depending on the jet \pt. In addition, the $D_2^{\beta = 1}$ 
variable is used to select jets with a two-prong structure.

Similar as in the CMS analysis, the background is estimated by fitting the dijet invariant
mass distribution with the parametric form $\frac{dn}{dx}= p_1(1-x)^{p2 + \xi p_3} x^{p3}$, 
where $n$ is the number of events, $x$ is a dimensionless variable related to the dijet mass $m_{\rm JJ}$, 
$x = m_{\rm JJ}/\sqrt{s}$, $p_1$ is a normalization factor,
$p_2$ and $p_3$ are dimensionless shape parameters, and
$\xi$ is a constant chosen to remove the correlation between $p_2$ and $p_3$ in the fit.

\begin{figure}[tb]
\centering
\includegraphics[width=0.43\textwidth]{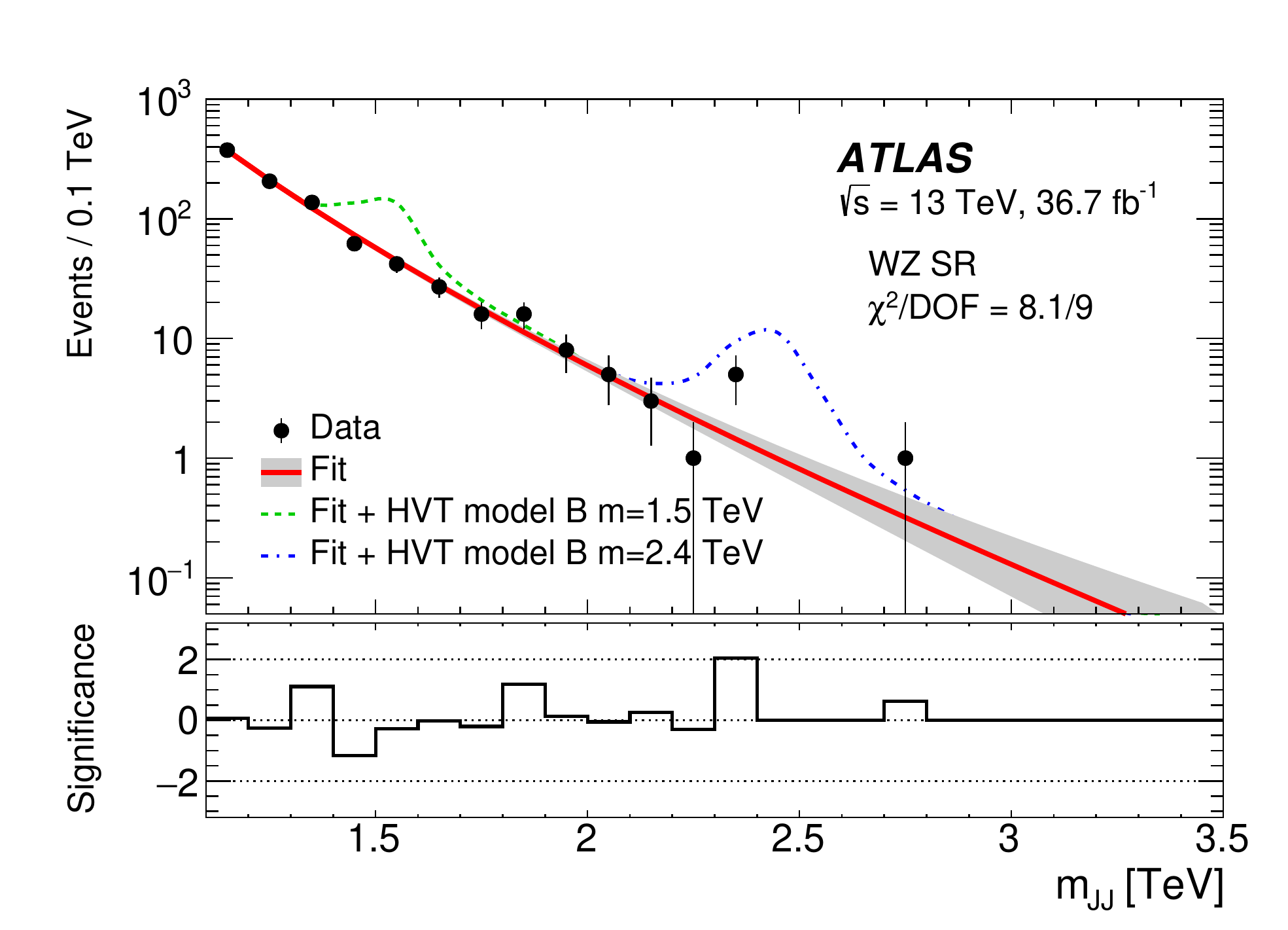}
\caption{The observed data in the signal region of the $WZ$ category. Also shown is the fitted background prediction. The gray region represents the uncertainty in the background estimate. Taken from Ref.~\cite{Aaboud:2017eta}.\label{fig:atlas:massdist}}
\end{figure}

The dijet invariant mass distributions for these events are shown in figure~\ref{fig:atlas:massdist}, 
where good agreement is found between data and the expectations from the background fit. 

In case of boosted $H$ bosons, different reconstruction
methods have to be used to benefit from the presence of $b$ quarks in
$H \to b\bar{b}$ decays (see section~\ref{sec:btagging}). Results have been published on 
the search for for $WH$/$ZH$ final states~\cite{Aaboud:2017ahz,Aaboud:2017cxo,Sirunyan:2017wto} 
as well as for $HH$ final states~\cite{Aaboud:2016xco,Aad:2015uka,CMS-PAS-B2G-16-026}. 

\subsection{$t\bar{t}$ Resonances \label{sec:ttresonancesearch}}

The models of new physics mentioned in the previous section also predict resonances decaying to pairs of top quarks. 
An example for an alternative model is the topcolor model which contains a $Z'$ boson~\cite{Hill:1994hp}, with exclusive decays to top quarks. 

In case of boosted $t\to bW$ events with leptonic $W$ boson decays,  the lepton  may overlap with the associated $b$ quark jet.  Therefore, the usual lepton-isolation criteria, which are used to mitigate the contamination with QCD multijet background,  are relaxed. The CMS and ATLAS Collaborations follow different strategies for this purpose. In  CMS~\cite{Khachatryan:2015sma,Sirunyan:2017uhk}, the lepton must have a large angular separation from the associated $b$ jet candidate of $\Delta R({\rm lepton, jet}) > 0.5$ or it must have a transverse momentum relative to the jet axis $\pt^{\rm rel}$ above $25\GeV$.  This  requirement removes background contributions from semi-leptonic $B$ hadron decays. In  ATLAS~\cite{Aad:2015fna,Aaboud:2018mjh}, the lepton isolation is achieved by a variable isolation cone that changes as a function of the transverse momentum~\cite{Rehermann:2010vq}. Interestingly, studies performed in CMS for 13\TeV show that the CMS implementation of such a variable isolation criterion is not as powerful as the selection based on 
$\Delta R({\rm lepton, jet})$ and $\pt^{\rm rel}$~\cite{CMS-PAS-B2G-15-002}.

To reconstruct the boosted hadronic top decay, the presence of a single high-momentum, large-$R$, top-tagged jet is required. In CMS (ATLAS) the large-$R$ jet is reconstructed with the CA (\antikt) algorithm with a size parameter of $R= 0.8$ ($1.0$). The selection requirement on the transverse momentum is $\pt > 400 (300)\GeV$.  ATLAS applies trimming to  the large-$R$ jets  with the parameters $f_{\rm cut} = 0.05$ and $R_{\rm sub} = 0.3$ and the jets are required to have a mass $m_{\rm jet} > 100\GeV$ and $\sqrt{d_{12}}> 40\GeV$.  The strategy followed by CMS is to apply the CMSTT algorithm  (as defined in section~\ref{sec:TopTagging}), where the mass of the jet has to satisfy $140 < m_{\rm jet} < 250\GeV$. In addition, the $N$-subjettiness ratio $\toptau$ must be smaller than $0.7$.

The variable of interest is the invariant mass $m_{t\bar{t}}$ of the $t\bar{t}$ system. It is reconstructed from the top-tagged large-$R$ jet, a $b$ tagged small-$R$ jet as well as the lepton and the missing energy.  Once the top-pair system is reconstructed, events are further divided into categories based on the lepton flavor and the number of $b$-tagged and top-tagged jets.  This gives several analysis categories with different background compositions:  the  top-tagged and $b$-tagged events are dominated by the SM $t\bar{t}$ background, while events without top tags and $b$ tags are mostly composed of $W$+jets events. 

Similar methods are applied in case both $W$ bosons decay hadronically~\cite{Khachatryan:2015sma}. To access the region with jets of lower momenta with $ 200< \pt < 400\GeV$  a dedicated algorithm with a larger jet size parameter of $R= 1.5$ (CA15 jets) is applied in CMS. The larger jet size extends the analysis coverage to the case of intermediate or smaller Lorentz boosts. These low-\pt jets are required to be identified by the HEPTopTagger algorithm (as described in section~\ref{sec:TopTagging}). This approach improves the sensitivity for smaller masses of the hypothetical \ttbar resonance. 

Even with the requirement of two top-tagged jets, the event sample is dominated by QCD dijet events.  This background is estimated using a data-driven technique, where an anti-tag and probe method is used. The $\toptau$ requirement is reversed on one jet to select a sample dominated by QCD events.  The opposite jet is then used to measure the misidentification rate for the top-tagging requirements.  The measured misidentification rate ranges from $5$ to $10\%$, depending on the jet momentum, $\toptau$ and the $b$ tagging requirements applied.  This differential rate is used in a sample of single top-tagged events to predict the double top-tagged event contribution from QCD processes in each individual event category. Closure tests performed in data and simulation are performed to validate the background estimation for each of the signal regions.

\begin{figure}[t]
\centering
\includegraphics[width=0.45\textwidth]{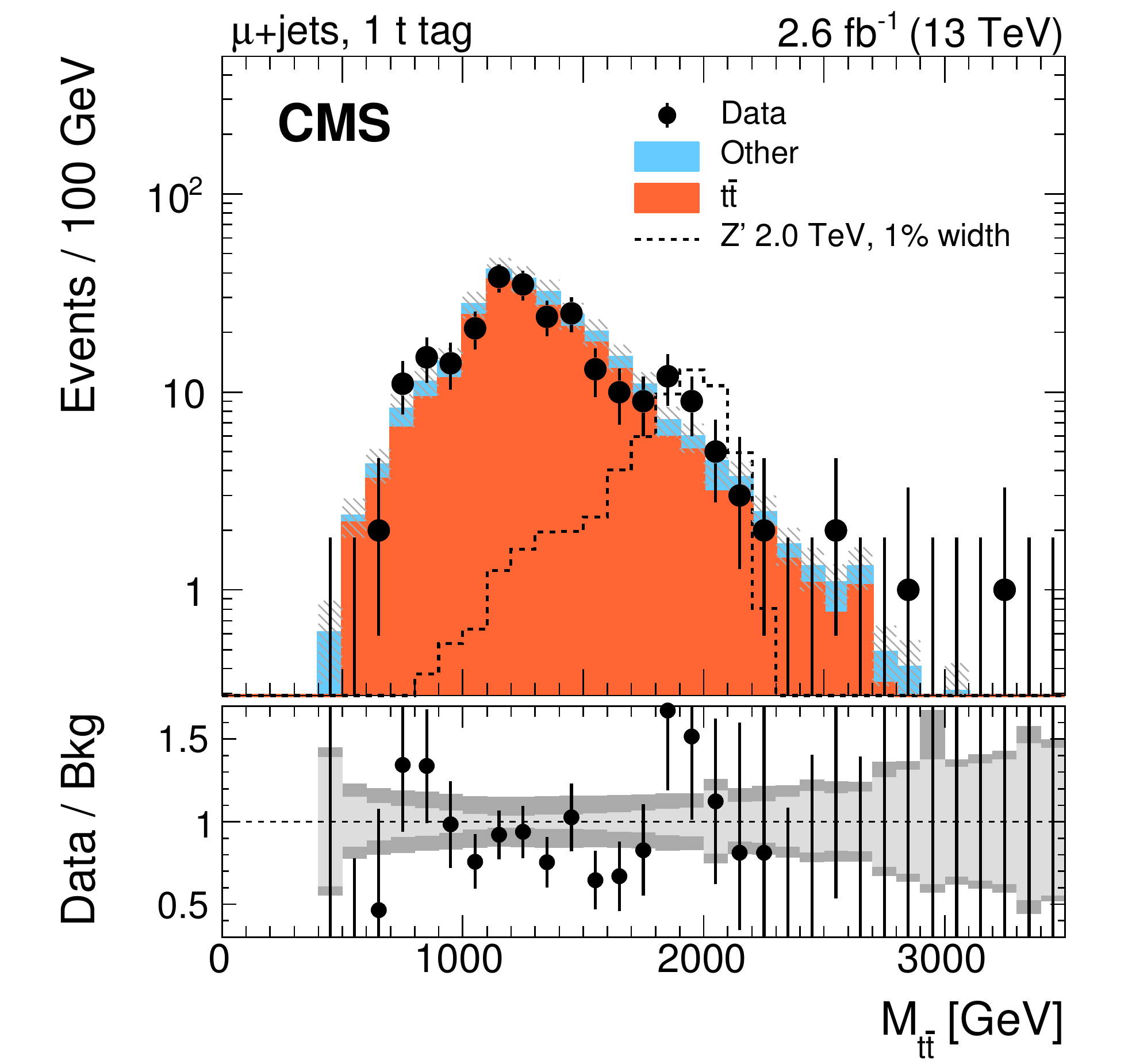} 
\caption{Invariant mass of the reconstructed $t\bar{t}$-pair in data and simulation for the lepton+jets channel in the category with one top-tagged jet, taken from Ref.~\cite{Sirunyan:2017uhk}.} 
\label{fig:ttResCMS}
\end{figure}

No significant excess above the predicted background is observed in the measured $t\bar{t}$ invariant mass spectrum. Figure~\ref{fig:ttResCMS} shows the $m_{t\bar{t}}$ spectrum in the analysis category with the highest $S/B$ fraction. 

Depending on the model, narrow $t\bar{t}$ resonances are excluded for masses less than approximately 4 \TeV.  The exclusion limits are weaker for scenarios with large width of the resonance.

\subsection{Vector-like Quarks}

Vector-like quarks (VLQs) are predicted by a variety of theories introducing a mechanism that stabilizes the mass of the Higgs particle. Such theories include little 
Higgs models~\cite{ArkaniHamed:2001nc,Schmaltz:2005ky}, 
models with extra dimensions~\cite{Antoniadis:2001cv,Hosotani:2004wv}, and
composite Higgs models~\cite{Antoniadis:2001cv,Hosotani:2004wv,Agashe:2004rs}. 
As VLQs are expected to have large masses and have top quarks and vector-bosons as decay products, jet substructure analyses have been applied in  many searches for VLQs. 

The first search for VLQs using jet substructure methods was an inclusive search for pair-produced 
$T$ quarks~\cite{Chatrchyan:2013uxa}. As VLQs may have many decay modes ($T\to bW$, $T\to tZ$, $T\to tH$, $B\to tW$, $B\to bZ$, $B\to bH$),  a large variety of final states needs to be explored. For this reason, an inclusive search has been performed without the attempt to reconstruct a specific decay chain.  The CA algorithm was used with a distance parameter $R=0.8$ (CA8 jets). Boosted $W$ jets are identified based on the mass of the CA8 jet while boosted top jets are identified with the CMSTT, 
described in section~\ref{sec:ttresonancesearch}. 

\begin{figure}[tb]
\centering
\includegraphics[width=0.95\columnwidth]{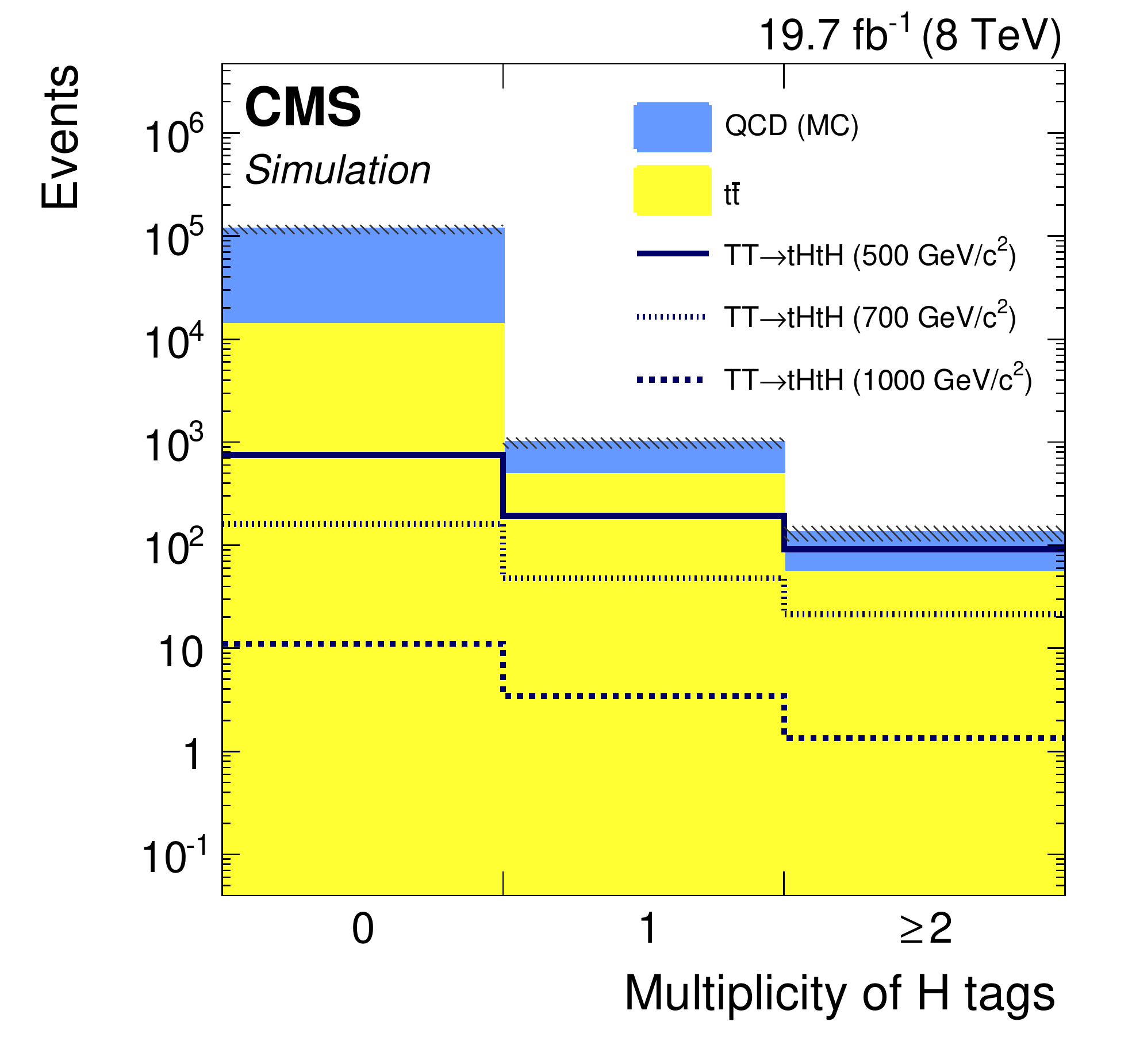}
\caption{Multiplicity of CA15 jets which fulfill the Higgs tagging criteria. The solid histograms represent the simulated background processes (\ttbar and QCD multijet). The hatched error bands show the statistical uncertainty of the simulated events. Taken from Ref.~\cite{Khachatryan:2015axa}.\label{fig:VLQHiggstagger}} 
\end{figure}

The first search for VLQs in the all-hadronic final state \cite{Khachatryan:2015axa} targeted the $T\to tH$ decay mode. The CA algorithm with a large size parameter of $R=1.5$ was applied to cluster top quarks and Higgs bosons in single large jets. To identify the origin of the large CA jets a top tagging algorithm (HEPTopTagger) and a Higgs tagging algorithm based on subjet-$b$ tagging (see section~\ref{sec:btagging}) are used. 
This was the first time these two algorithms have  been applied in a data analysis by the CMS Collaboration. 
Two subjets must be $b$ tagged and their invariant mass must be greater than 60\GeV to fulfill the Higgs tagging requirement. The multiplicity of these Higgs tags is shown in figure~\ref{fig:VLQHiggstagger} which demonstrates that both the QCD multijet and the \ttbar backgrounds can be suppressed by several orders of magnitudes. 

Extensive use of substructure methods has also been made by the ATLAS Collaboration, in particular for the search for single production of VLQs. The single production modes may have higher cross sections than pair production depending on the VLQ mass and the coupling parameters \cite{Aguilar-Saavedra:2013qpa}. 
ATLAS performed an analysis~\cite{Aad:2015voa} where the VLQ is searched for in the decay mode with a $W$ boson and a top quark ($B \to tW$). Final states with at least one lepton are considered, where either the $W$ boson or the top quark appear in a boosted configuration. They are identified by the application of a jet mass requirement ($m>50\GeV$) on a trimmed large-$R$ \antikt jet with a distance parameter $R=1.0$. 

A different strategy is followed in another ATLAS search~\cite{Aad:2016qpo}, where the decay into the $bW$ final state is investigated ($T/Y \to bW$). As the $W$ boson is assumed to decay leptonically, no boosted hadronic $W$ or top quark decays are present. Therefore, the analysis uses a veto on the presence of massive ($m>70\GeV$), trimmed large-$R$ \antikt jets with $R=1.0$, to suppress the dominant \ttbar background. 

Today, jet substructure methods are widely employed in almost all VLQ searches  published by the LHC Collaborations, see e.g.\ Refs.~\cite{Aaboud:2017zfn,Khachatryan:2016vph,Sirunyan:2017ynj,Sirunyan:2016ipo,Sirunyan:2017ezy,Sirunyan:2017usq,Sirunyan:2017bfa}.  The excluded VLQ masses are exceeding 
1\TeV for all branching fractions,  thanks to jet substructure techniques.

\subsection{Leptophobic $Z'$}

Besides resonaces coupling to heavy SM particles, there exist predictions for resonances that couple to quarks and gluons~\cite{Baur:1987ga,Hewett:1988xc,Baur:1989kv,Langacker:2008yv}, including simplified Dark Matter (DM) models in which resonances couple only to quarks and DM particles~\cite{An:2012ue,Rajaraman:2011wf,Goodman:2010ku}.  When the new particle (such as a $Z'$) is sufficiently light ($m_{Z'} \ll 1$\TeV), it can be boosted when produced in association with initial-state radiation and thus entirely captured by a single large-radius jet~\cite{Sirunyan:2017nvi,Aaboud:2018zba}.   Searching in this mode can significantly extend the sensitivity of the existing search program, where resolved low-mass resonance searches typically degrade due to high trigger thresholds and the enormous QCD multijet background.

\begin{figure}[tb]                                                                                                                                                                 
\includegraphics[width=0.43\textwidth]{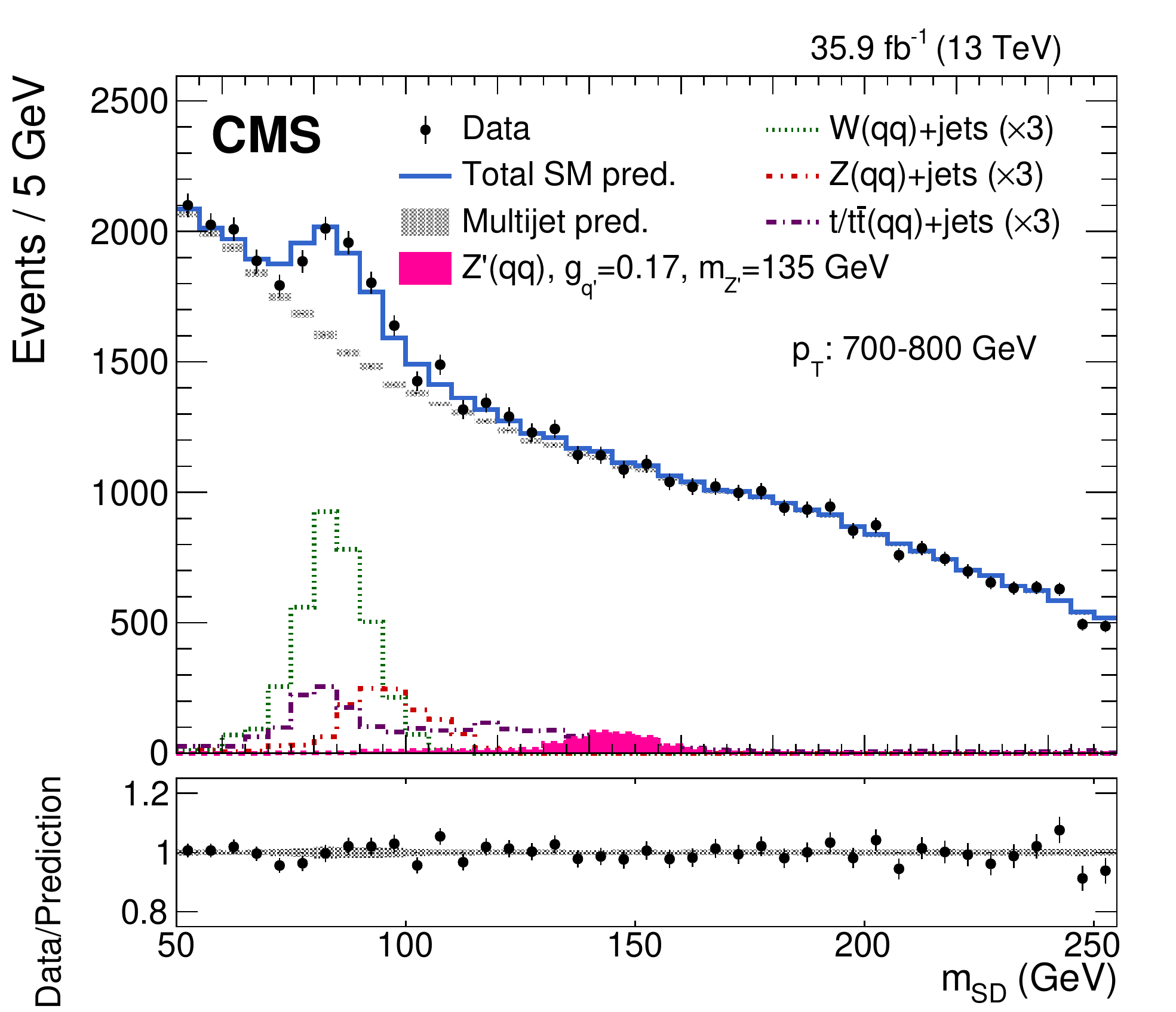}
\caption{Soft drop jet mass of anti-$k_t$ $R = 0.8$ jets in data and for the dominating background processes; multijet production and $W/Z$+jets events. Taken from Ref.~\cite{Sirunyan:2017nvi}.}
\label{fig:Zprime_CMS}
\end{figure}

\begin{figure}[tb]
\includegraphics[width=0.43\textwidth]{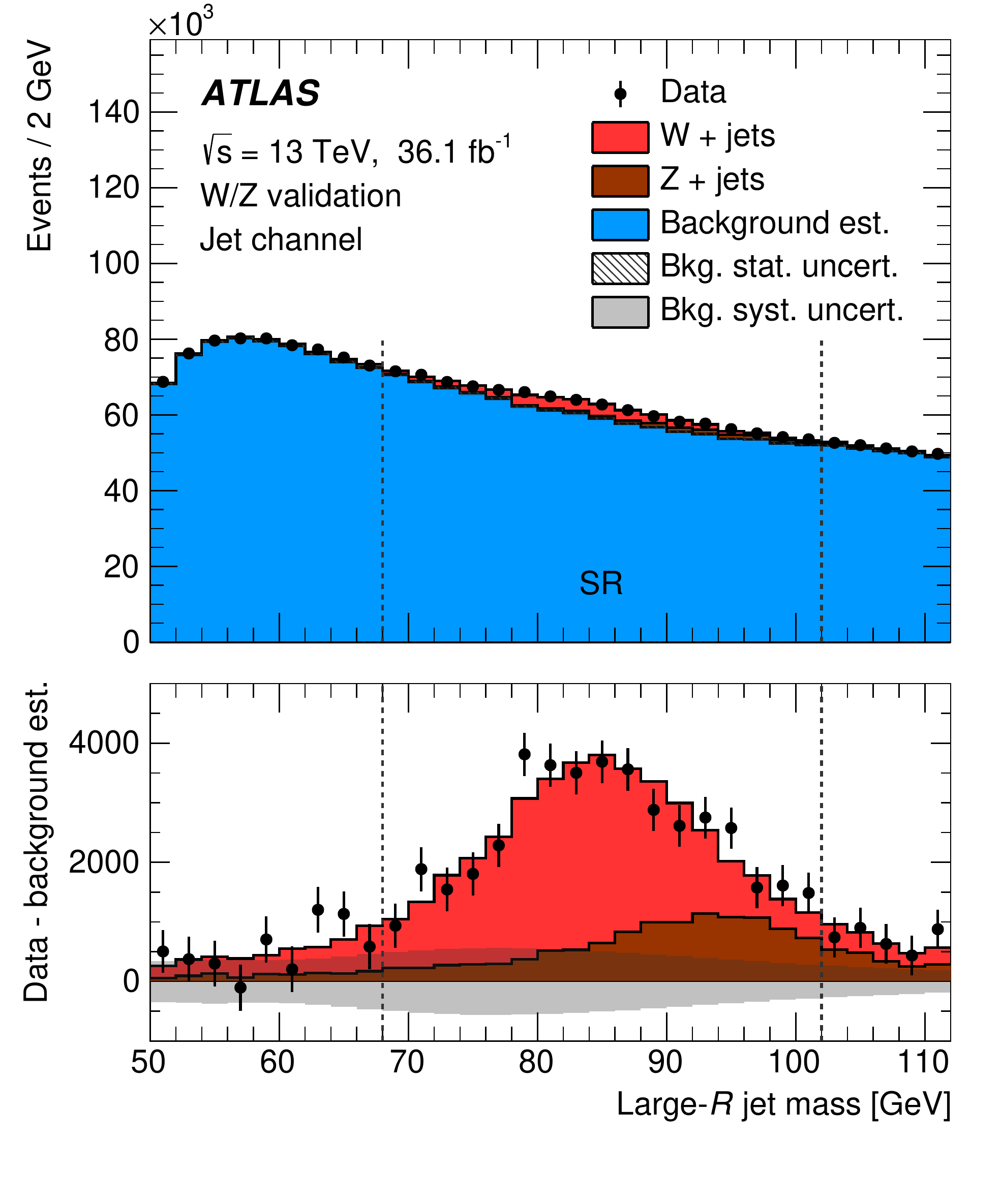}
\caption{Trimmed jet mass distribution \antikt $R = 1.0$ jets in data and for the dominating background processes. Taken from Ref.~\cite{Aaboud:2018zba}.}
\label{fig:Zprime_ATLAS}
\end{figure}

Both ATLAS and CMS have used this strategy to look for boosted $Z'$ jets.  Jets in the CMS analysis are reconstructed with the \antikt algorithm with $R = 0.8$ and corrected for effects from pile-up and the underlying event with PUPPI and the soft drop algorithm ($\beta = 0$, \zsoftdrop = 0.1) whereas \antikt $R = 1.0$ jets, trimmed with $\rsub = 0.2$ and $\fcut = 5\%$ are used in ATLAS. To suppress the dominating QCD multijet background, CMS applies criteria on $N^{1}_{2}$~\cite{Moult:2016cvt} and ATLAS chooses $\tau_{21}$ as discriminator. To avoid distortions of the jet mass spectrum due to large correlation between the jet mass and substructure variables, a decorrelation  with the DDT method is applied. Data-driven techniques are used to determine the dominating background from QCD multijet production.  Subdominant processes such as $W/Z$+jets events are estimated from MC simulation. The jet mass distributions of the large-$R$ jet is shown in figure~\ref{fig:Zprime_CMS} and \ref{fig:Zprime_ATLAS} for the CMS and ATLAS analyses, respectively. No evidence for a resonant structure on top of the SM background is observed.

\section{Conclusions}
\label{sec:conclusions}

Jet substructure is the term used to describe the calculations, algorithms, and analysis techniques developed over the last decade and reviewed in this article. These methods are used to exploit the details of hadronic activity detectable by modern particle detectors such as ATLAS and CMS, and precision Standard Model measurements and searches for physics beyond the Standard Model at both these experiments increasingly rely on one or more of the tools developed by the jet substructure community. With increasingly sophisticated hardware and software capabilities, jet substructure techniques of the future will grow in complexity and utility, further empowering the exploration of the subnuclear properties of nature.

\section*{Acknowledgments}

Much of the work in this field in recent years has been galvanized by the Boost Workshop Series~\cite{boost11,boost12,boost13}, which continues to inspire fruitful collaborations between experimentalists and theorists.  

The editors thank CERN and the ATLAS and CMS Collaborations, the participants and organizers of the Boost Workshops held in Zurich 2016~\cite{boostzurich} and Buffalo 2017~\cite{boostbuffalo} for discussions and input, and Jon Butterworth for suggesting this jet substructure review article.  We also thank Andrew Larkoski and Ian Moult for the collaboration on the theoretical review.  

\bibliography{experiment}

\end{document}